\documentclass{jfm}
  \paperheight\pdfpageheight 
  \paperwidth\pdfpagewidth   
  \usepackage[Symbolsmallscale]{upgreek}
  \usepackage{graphicx}
  \usepackage{amsmath}
  \usepackage{mathtools}
  \usepackage[T1]{fontenc}
  \usepackage{newtxtext}
  \usepackage{newtxmath}

  \author{Freddie Jensen\aff{1}
          \and Edward James Brambley\aff{1,2}\corresp{\email{E.J.Brambley@warwick.ac.uk}}}

  \affiliation{\aff{1}Mathematics Institute, University of Warwick, Coventry CV4~7AL, UK
  \aff{2}WMG, University of Warwick, Coventry CV4~7AL, UK}
  \newcommand{\mat}[1]{\mathsfbi{#1}} 
  \newcommand{\gmat}[1]{\boldsymbol{#1}} 
  \newcommand{\pvect}[1]{\boldsymbol{#1}} 
  \newcommand{\mvect}[1]{\boldsymbol{#1}} 

\usepackage[longnamesfirst]{natbib}
\usepackage{hyperref}
\hypersetup{
    colorlinks = true,
    urlcolor   = blue,
    citecolor  = black,}

\newcommand{\D}{\mathrm{D}} 
\newcommand{\I}{\mathrm{i}} 
\newcommand{\e}{\mathrm{e}} 
\newcommand{\J}{\mathrm{J}} 
\newcommand{\Ib}{\mathrm{I}} 
\newcommand{\K}{\mathrm{K}} 
\newcommand{\intd}{\mathrm{d}} 
\newcommand{\cnon}{\upbeta_0} 
\newcommand{\T}{\mathrm{T}} 

\newcommand{\sgn}{\mathrm{sgn}}

\def\clap#1{\hbox to 0pt{\hss#1\hss}}

\title{Multimodal nonlinear acoustics in\\two- and three-dimensional curved ducts}

\begin{document}
\maketitle

\begin{abstract}
We develop a weakly nonlinear model of duct acoustics in two and three dimensions (without flow).  The work extends the previous work of \citet[][\emph{J.~Fluid Mech. 875, pp.~411--447}]{mctavish+brambley-2019} to three dimensions and significantly improves the numerical efficiency.  The model allows for general curvature and width variation in two-dimensional ducts, and general curvature and torsion with radial width variation in three-dimensional  ducts.  The equations of gas dynamics are perturbed and expanded to second order, allowing for wave steepening and the formation of weak shocks. The resulting equations are then expanded temporally in a Fourier series and spatially in terms of straight duct modes, and a multi-modal method is applied, resulting in an infinite set of coupled ODEs for the modal coefficients.  A linear matrix admittance and its weakly-nonlinear generalization to a tensor convolution are first solved throughout the duct, and then used to solve for the acoustic pressures and velocities.  The admittance is useful in its own right, as it encodes the acoustic and weakly-nonlinear properties of the duct independently from the specific wave source used.  After validation, a number of numerical examples are presented that compare two- and three-dimensional results, the effects of torsion, curvature and width variation, acoustic leakage due to curvature and nonlinearity, and the variation in effective duct length of a curved duct due to varying the acoustic amplitude.  The model has potential future applications to sound in brass instruments.  \textsc{Matlab} source code is provided in the supplementary material.
\end{abstract}


\section{Introduction}

Brass instruments are understood to sound brassy due to nonlinear wave-steepening within the instrument~\citep{hirschberg}. This has been observed experimentally for a trombone~\citep{hirschberg,rendon2010} and for a trumpet~\citep{pandya2003,rendon2013}, and is supported by relatively simple physical models~\citep[e.g.][]{gilbert2008}.  In these studies, attention is often given to the width, or bore, of an instrument, and its variation along its length.  Varying the bore so that it progressively gets wider lowers the amplitude of the outward-propagating wave, resulting in less nonlinear steepening and a less brassy sound.  Conversely, a cylindrical bore that does not widen (such as is necessary for a trombone slide to be able to slide) allows nonlinear steepening to occur, exciting higher harmonics and leading to a more brassy sound.  \Citet{arnold2012} even proposed a dimensionless brassiness parameter $B$ based on the bore variation in order to quantitatively classify musical instruments by their brassiness.  Most of the modelling in this area assume plane-wave propagation and straight ducts, although~\citet{fernando} allowed for non-plane-wave propagation by introducing a cross-duct modal expansion while retaining the straight duct approximation.

However, very few brassy-sounding musical instruments are straight; for example, a traditional B$\flat$ trumpet has a duct that, if straightened, would be about three times as long as the instrument.  In order to investigate the acoustics of curved ducts, \citet{felix1} generalized the multi-modal method previously developed by~\citet{pagneux1996} for ducts with width variation.  While initially developed in two dimensions, it was subsequently extended to three~\citep{felix2}.  The multi-modal method involves the projection of the curved-duct acoustics equations for pressure and velocity onto a basis of straight-duct modes, converting the governing PDEs into an infinite coupled set of ODEs for the amplitude of each mode.  The ratio of the velocity and pressure, known as the admittance, is then taken, and a Riccati-style equation for the admittance is solved.  As well as being computationally tractable, this formulation has the potential to provide better physical intuition than, for example, direct numerical simulation, as the admittance encodes the downstream effects of the duct geometry independently of the form of the acoustic wave introduced upstream.  This will be seen below, as we will follow a similar procedure here.  However, this multi-modal method depends on the linearity of the acoustic equations, and so is not able to model the weak nonlinearity necessary for investigating brassiness.

The tuning of a wind instrument is given by the effective length of the duct, and acoustic modes travelling around the outside of a duct bend would see a different effective length compared with acoustic modes travelling around the inside of a duct bend, resulting in a different resonant pitch.  Brass instruments shift their pitch slightly when played louder. Could perhaps nonlinear steepening affect the effective length of a duct bend?  Could instruments be designed for pitch stability as the sound amplitude is varied?  Or, if curved ducts behave the same as straight ducts, is it possible to say why and to quantify to what extent?  Such an analysis would require a framework capable of analysing nonlinear acoustics in curved ducts.

Recently, \citet{mctavish+brambley-2019} combined the curved-duct linear multi-modal method of~\citet{felix1} with the straight-duct nonlinear multimodal method of~\citet{fernando} to describe the weakly-nonlinear acoustics of curved ducts in two dimensions.  Even restricted to two dimensions, the algebraic complexity of the resulting system was prohibitive, and so a linear matrix and nonlinear tensor convolution notation was developed.  However, because the matrices and tensors involved in the governing equations varied with as the duct geometry varied, the technique was computationally inefficient.  Nonetheless, it was possible to analyze a range of curved geometries in both the linear and nonlinear regimes.

The aim of the present work presented here was originally to extend the two-dimensional analysis of~\citet{mctavish+brambley-2019} to three dimensions.  In so doing, a modified mathematical framework was developed that combines both two- and three-dimensional cases, is computationally more efficient, and is easier to understand in terms of the interplay between the nonlinearity and the duct geometry (consisting of width variation, curvature, and, in three-dimensions, torsion).  Section~\ref{sec:gov-equ} derives this mathematical framework from the governing fluid mechanics equations, resulting in the multimodal set of governing ODEs valid in both two and three dimensions given in section~\ref{spatprodandnot}.  These equations are then solved using the multi-modal method in section~\ref{sec:solve}, first by introducing the admittance (\S\ref{sec:admittance}) and then by truncating and numerically solving (\S\ref{sec:numerics}).  A number of numerical results are then presented in section~\ref{sec:results}, including validation for a straight duct (\S\ref{sec:results-straight}), a constant curvature bend in two and three dimensions (\S\ref{sec:results-bend} and~\S\ref{sec:results-bend3D}), and an exponential horn (\S\ref{sec:results-horn}), along with more interesting cases involving combinations of curvature, width variation, torsion, and nonlinearity.  In particular, the leakage of acoustics from a duct owing to either curvature or nonlinearity is demonstrated in section~\ref{sec:results-curvature-and-width}, a direct comparison of two-dimensional results against three-dimensional results made possible by the combined mathematical framework is given in section~\ref{sec:results-3Dvs3D}, and an investigation of the effective acoustic length of a bend is given in section~\ref{sec:results-bend-length}.  The results are summarized and discussed in the concluding section~\ref{sec:conclusion}, together with possibilities for future work.  \textsc{Matlab} source code that produces the results shown here is provided in the supplementary material, along with videos animating some of the figures.

\section{Formulation of the Governing Equations}\label{sec:gov-equ}

In this section, we derive the governing equations in the weakly nonlinear limit that will be solved in subsequent sections.  The derivation follows that of \citet{mctavish+brambley-2019}, although here we do not limit ourselves to only two dimensions.

\subsection{Derivation of the weakly nonlinear equations}
\label{weaknonlinderivation}
We start with the equations of mass and momentum conservation for an inviscid fluid,
\begin{align}
\label{conseqns}
    \frac{\partial\hat{\rho}}{\partial\hat{t}} + \pvect{\hat\nabla\cdot}(\hat{\rho}\pvect{\hat{u}}) = 0, &&        \hat{\rho}\left(\frac{\partial\hat{\pvect{u}}}{\partial \hat{t}} + \pvect{\hat{u}\cdot\hat{\nabla}\hat{u}}\right) = -\hat{\pvect{\nabla}}\hat{p}.
\end{align}
Dimensional quantities are written with hats here. In this paper, we work in either two or three dimensions; note that in $n$ dimensions, these give us $n+1$ equations in $n+2$ variables (one density, $n$ velocities, and one pressure).  We therefore close the system of equations using the equation of state in combination with the thermal energy equation for an adiabatic gas (consistent with an inviscid non-heat-conducting fluid),
\begin{align}
    \hat{p}=\hat{p}(\hat{S},\hat{\rho}), &&\frac{\D\hat{S}}{\D\hat{t}} = 0.
\end{align}
Since we are interested in acoustics, we expand these equations for a small perturbation about an ambient state of rest. To do so, we first need some notion of size, so we introduce reference quantities and non-dimensionalise. Let the ambient density, pressure and entropy be denoted by $\hat{\rho}_0$, $\hat{p}_0$ and $\hat{S}_0$ respectively.  The ambient speed of sound $\hat{c}_0$ is then given by
\begin{equation}
\label{soundspeeddef}
    \hat{c}_0^2 = \frac{\partial \hat{p}}{\partial\hat{\rho}}\bigg|_{\hat{S}}(\hat{\rho}_0,\hat{S}_0),
\end{equation}
where the subscript zero here and elsewhere in this section denotes evaluation at the ambient state.
Using $\hat{\rho}_0$ and $\hat{c}_0$, non-dimensional variables are defined by setting
\begin{align}\label{equ:nondim}
    &\hat{\rho} = \hat{\rho}_0\rho, & &\hat{\pvect{u}} = \hat{c}_0\pvect{u}, & &\hat{p} = \hat{\rho}_0 \hat{c}_0^2 p.
\end{align}
We also need spatial and temporal reference scales. The spatial reference lengthscale is denoted $\hat{\ell}_0$, typically given by the inlet radius of the duct. The timescale is then $\hat{c}_0/\hat{\ell}_0$, so the nondimensional operators are
\begin{align}\label{equ:nondim-deriv}
    &\hat{\pvect{\nabla}} = \frac{1}{\hat{\ell}_0}\pvect{\nabla} &
    &\text{and}&
    & \frac{\partial}{\partial\hat{t}} = \frac{\hat{c}_0}{\hat{\ell}_0}\frac{\partial}{\partial t}.
\end{align}
We now perturb about an ambient state of rest,
\begin{align}\label{equ:perturb}
    &\rho = 1 + \rho', & & \pvect{u} = \pvect{0} + \pvect{u}', & & p = p_0 + p',
\end{align}
taking $\rho' \sim p' \sim |\pvect{u}'| \sim M$, where $M\ll1$ is the perturbation Mach number.  Since we are interested in weak nonlinearity, we will neglect terms of order $O(M^3)$ or smaller but retain both the linear acoustic $O(M)$ and weakly nonlinear $O(M^2)$ terms.  Using the nondimensionalizations~(\ref{equ:nondim}, \ref{equ:nondim-deriv}) in the governing equations~\eqref{conseqns} and substituting the perturbed quantities~\eqref{equ:perturb} then results in
\begin{subequations}\begin{align}
    \label{perturbedmass}
        \frac{\partial\rho'}{\partial t} + \pvect{\nabla\cdot u}'
         &= -\rho'\pvect{\nabla\cdot u}' - \pvect{u}'\pvect{\cdot\nabla}\rho'
         = \rho'\frac{\partial\rho'}{\partial t} - \pvect{u}'\pvect{\cdot\nabla}\rho' + O(M^3),
\\
\label{perturbedmom}
        \frac{\partial\pvect{u}'}{\partial t} + \pvect{\nabla} p' &= -\pvect{u}'\pvect{\cdot}\pvect{\nabla}\pvect{u}' - \rho'\frac{\partial\pvect{u}'}{\partial t} + O(M^3).
    \end{align}\end{subequations}%
Here, and in what follows, we write the linear acoustic $O(M)$ quantities on the left hand side of equations and the nonlinear $O(M^2)$ terms on the right hand side.  Note that in~\eqref{perturbedmass} the linear expression $\pvect{\nabla\cdot u}' = -\partial\rho'/\partial t + O(M^2)$ was substituted in the right hand side, resulting in an expression that is still correct to order $O(M^3)$.  This is a technique we will make further use of below.

Since we are primarily interested in acoustic perturbations, we will neglect entropy perturbations, so that $S' \equiv 0$ and so $\hat{S} \equiv \hat{S}_0$ everywhere.  This means all perturbations are adiabatic, and so we may eliminate either the pressure or density perturbation using the expanded equation of state:
\begin{align}\label{equ:prho}
p' &= \rho' + \frac{1}{2}\frac{\partial^2 p}{\partial \rho^2}\bigg|_{S}\rho'^2 + O(M^3) &
&\Leftrightarrow&
\rho' &= p' - \frac{1}{2}\frac{\partial^2 p}{\partial \rho^2}\bigg|_{S}p'^2 + O(M^3),
\end{align}
where the second-order partial derivative is evaluated at the ambient state $(\rho_0, S_0)$, and the identity $p' = \rho' + O(M^2)$ has been used to rearrange between the left and right expressions.
If we consider a perfect gas, a consequence of adiabacity is that $(\hat{p}/\hat{p}_0)=(\hat{\rho}/\hat{\rho}_0)^{\upgamma}$, where the adiabatic index $\upgamma$ is the ratio of specific heats.  Consequently, for a perfect gas
\begin{align}
    \hat{c}_0^2 &= \frac{\partial \hat{p}}{\partial\hat{\rho}}\bigg|_{\hat{S}} = \frac{\upgamma\hat{p}_0}{\hat{\rho}_0}, &
    &\text{and}&
    \frac{\partial^2 \hat{p}}{\partial\hat{\rho}^2}\bigg|_{\hat{S}} &= \upgamma(\upgamma-1)\frac{\hat{p}_0}{\hat{\rho}_0^2} = (\upgamma-1)\frac{\hat{c}_0^2}{\hat{\rho}_0},
\end{align}
so that $p_0 = 1/\upgamma$ and $\partial^2 p/\partial \rho^2|_S = (\ugamma-1)$ when $\rho_0=1$ and $c_0=1$.  In what follows, for simplicity we will use this perfect gas notation, although the derivation is general provided it is taken that $\upgamma = 1+\partial^2 p/\partial \rho^2|_S$ and $p_0$ is not assumed to be $1/\upgamma$.

We now eliminate the density perturbation  $\rho'$ using~\eqref{equ:prho}, and, correct to $O(M^2)$ and dropping the $O(M^3)$ from each equation for brevity, this finally results in
\begin{subequations}\label{perturbedelimrho}\begin{align}
        \frac{\partial p'}{\partial t} + \pvect{\nabla\cdot u}'
         &= \upgamma p'\frac{\partial p'}{\partial t} - \pvect{u}'\pvect{\cdot\nabla}p'
          = \frac{1}{2}\frac{\partial}{\partial t}\left(\upgamma p'^2 + |\pvect{u}'|^2\right),
\label{equ:perturbedelimrhoa}\\
        \frac{\partial\pvect{u}'}{\partial t} + \pvect{\nabla} p'
        &= -\pvect{u}'\pvect{\cdot}\pvect{\nabla}\pvect{u}' - p'\frac{\partial\pvect{u}'}{\partial t}
         = \frac{1}{2}\pvect{\nabla}\left(p'^2\right) - \pvect{u}'\pvect{\cdot}\pvect{\nabla}\pvect{u}',\label{equ:perturbedmom}
\end{align}\end{subequations}
where again the $O(M)$ identity $\partial\pvect{u'}/\partial t = -\pvect{\nabla}p' + O(M^2)$ from the left hand side of~\eqref{equ:perturbedmom} has been used to rearrange the right-hand sides of both equations.

\subsection{Further assumptions}

Equations~\eqref{perturbedelimrho} may be simplified further.  Taking $\pvect{\nabla}$ of both sides of~\eqref{equ:perturbedmom}, we apply symmetry of mixed partial derivatives on $\pvect{\nabla}\pvect{\nabla} p'$ and find that the spin tensor of the velocity is linearly constant in time, i.e.
\begin{equation}
\label{constantAdef}
    \frac{\partial}{\partial t}\bigg(\pvect{\nabla}\pvect{u}'- (\pvect{\nabla}\pvect{u}')^\T\bigg) = \mat{0} + O(M^2).
\end{equation}
If we assume all acoustic variables to be periodic, we can define a time-average $\langle\bullet\rangle$, and then deduce that
\begin{equation}\label{equ:spin}
    \pvect{\nabla}\pvect{u}' - (\pvect{\nabla}\pvect{u}')^\T = \bigg\langle\pvect{\nabla}\pvect{u}' - (\pvect{\nabla}\pvect{u}')^\T\bigg\rangle + O(M^2).
\end{equation}
Having worked to a relatively high degree of generality so far, we now outline a hierarchy of physically-justified assumptions, ordered by strength, that will simplify the coming calculations.
\begin{itemize}
    \vspace{1ex}\item \textbf{Assumption 1 (weakest):} Taking the time average of~\eqref{equ:perturbedelimrhoa} shows that $\pvect{\nabla}\pvect{\cdot}\langle\pvect{u}'\rangle = O(M^3)$, meaning $\langle\pvect{u}'\rangle$ corresponds to a steady, viscosity-free incompressible mean flow to the order considered here. Under these circumstances, we would expect this mean flow to be vorticity-free, implying $\pvect{\nabla \langle u'\rangle} - (\pvect{\nabla\langle u'\rangle})^T = O(M^2)$, and consequently in light of~\eqref{equ:spin} that~\eqref{equ:perturbedmom} may be equivalently written as
    \begin{equation}
    \label{finalnonfouriermom}
        \frac{\partial\pvect{u}'}{\partial t} + \pvect{\nabla} p' = \frac{1}{2}\pvect{\nabla}\left(p'^2 - |\pvect{u}'|^2\right).
    \end{equation}
    
    \vspace{1ex}\item \textbf{Assumption 2a (stronger):} On top of assumption~1, by taking a time average of \eqref{finalnonfouriermom}  we see that the quantity $\langle p'\rangle - \frac{1}{2}\bigg\langle p'^2 - |\pvect{u}'|^2\bigg\rangle$ is spatially constant, where the constant corresponds to an $O(M)$ change in the far-field ambient pressure $p_0$.  By suitably choosing $p_0$ we may therefore set this change in ambient pressure to be zero, and consequently $\langle p'\rangle = O(M^2)$.
    
    \vspace{1ex}\item \textbf{Assumption 2b (also stronger):} On top of assumption 1, we may assume the mean flow $\langle\pvect{u'}\rangle$ to be identically zero. This is justified by imposing an inlet condition with no mean flow, and noting that no mean flow is induced by the nonzero frequency modes. Thus, $\langle\pvect{u}'\rangle = \pvect{0}$.
\end{itemize}\vspace{1ex}

Henceforth we will take all of these assumptions to be true. Defining the coefficient of nonlinearity $\cnon := (\upgamma + 1)/2$, and introducing the $O(M^2)$ quantity $Q' := (p'^2 - |\pvect{u}'|^2)/2$ (which could arguably be thought of as a Lagrangian of the perturbation), we have
\begin{align}
    &&\frac{\partial p'}{\partial t} + \pvect{\nabla}\pvect{\cdot}\pvect{u}' = \frac{\partial}{\partial t}\left(\cnon p'^2 - Q'\right), &&\frac{\partial\pvect{u}'}{\partial t} + \pvect{\nabla} p' = \pvect{\nabla} Q'.
\end{align}
Note that $\cnon$ has been written with a zero subscript in order to avoid confusion with the spatial mode enumerator $\beta$ used later on.

\subsection{Fourier modal decomposition}
At this point, because we are interested in tonal acoustics, we assume a base frequency $\omega$ and decompose pressure and velocity into Fourier modes by writing
\begin{align}
    & p'(\pvect{x},t) = \sum_{a = -\infty}^{\infty}P^a(\pvect{x})\e^{-\I a\omega t}, & & \pvect{u}'(\pvect{x},t) = \sum_{a = -\infty}^{\infty}\pvect{U}^a(\pvect{x})\e^{-\I a\omega t}.
\end{align}
Here $\omega$ is a dimensionless frequency, representing the Helmholtz number. To ensure that $p'$ and $\pvect{u}'$ are real, we must have $P^{-a} = P^{a*}$ and $\pvect{U}^{-a} = \pvect{U}^{a*}$, where an asterisk denotes the complex conjugate. Our equations for $p'$ and $\pvect{u}'$ (with $a=0$ modes discounted due to assumptions~2a and~2b above) become
\begin{subequations}\label{equ:coordfree}%
    \begin{equation}
    \label{coordfreemass}
        -\I a\omega P^a + \pvect{\nabla}\pvect{\cdot}\pvect{U}^a = -\I a\omega\left(- Q^a + \sum_{b \neq 0,a}\cnon P^{a - b}P^b \right),
    \end{equation}
    \begin{equation}
    \label{coordfreemom}
        -\I a\omega\pvect{U}^a + \pvect{\nabla} P^a = \pvect{\nabla} Q^a,
    \end{equation}
\end{subequations}
where $Q'$ has a Fourier series coefficient given by
\begin{equation}
    Q^a = \frac{1}{2}\sum_{b \neq 0,a}P^{a - b}P^b - \pvect{U^{a - b}\cdot U^b}.
\end{equation}
For brevity, we will now assume $P^0 \equiv U^0 \equiv 0$ and simply sum over all $b$.

\subsection{Duct coordinate system}
\subsubsection{The coordinate system in two dimensions}
\begin{figure}
\includegraphics{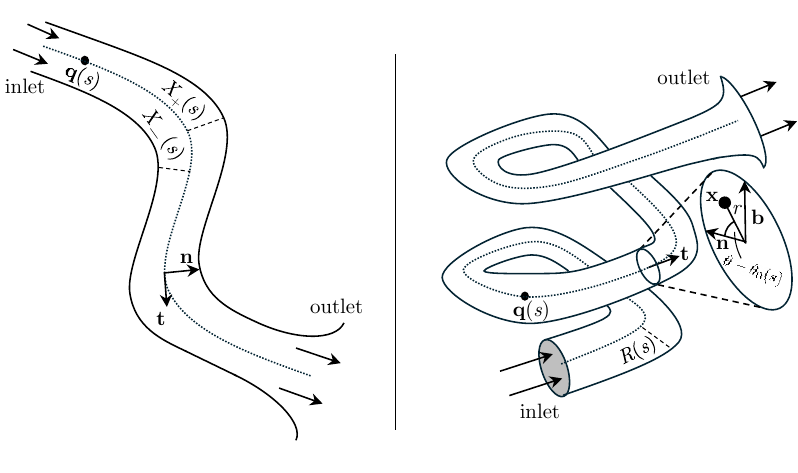}
\caption{The duct geometry in two dimensions (left) and three dimensions (right). A Frenet-Serret frame is employed in each case, with centreline $\pvect{q}(s)$, tangent $\pvect{t}(s)$ and normal $\pvect{n}(s)$; additionally in three dimensions we have a binormal $\pvect{b}(s)$. The two-dimensional duct has independently varying wall widths $X_+(s)$ and $X_-(s)$, whereas the three-dimensional duct has a single axisymmetrically-varying wall width $R(s)$.}
\label{figure:duct}
\end{figure}
As shown in figure~\ref{figure:duct} (left), the two-dimensional duct is defined by a centreline $\pvect{q}(s)$ (where $s$ is the arc-length parameterising distance along the duct), a total width at each point $X(s)$, and independently varying wall widths $X_-(s) < 0$ and $X_+(s) > 0$, defined such that $X(s) = X_+(s) - X_-(s)$. The direction along the duct is the \emph{longitudinal} direction, and the direction perpendicular to this is the \emph{transverse} direction. The longitudinal direction has coordinate $s$, and for the transverse direction we introduce a coordinate $x \in [X_-(s), X_+(s)]$. This coordinate basis arises from the Frenet--Serret frame defined by the centreline $\pvect{q}(s)$, i.e.\ the tangent and normal unit vectors $\pvect{t}$ and $\pvect{n}$ satisfying
\begin{align}
    &\frac{\intd \pvect{q}}{\intd s} = \pvect{t},& & \frac{\intd \pvect{t}}{\intd s} = \kappa\pvect{n},& & \frac{\intd \pvect{n}}{\intd s} = -\kappa\pvect{t},
\end{align}
with the scalar $\kappa(s)$ representing the curvature of the duct. Any point in the duct then has coordinates $(s,x)$ and position vector
\begin{equation}
    \pvect{x} = \pvect{q}(s) + x\pvect{n}(s).
\end{equation}
A differential $\intd \pvect{x}$ and a metric $\intd \pvect{x}\pvect{\cdot}\intd \pvect{x}$ may then be calculated, from which we may read off the Lam\'e coefficients and $(s,x)$ basis vectors
\begin{align}
    & h_s = 1 - \kappa x, & & h_x = 1,& & \pvect{e}_s = \pvect{t},& & \pvect{e}_x = \pvect{n}.
\end{align}
This is all of the machinery necessary to project the governing equations~\eqref{equ:coordfree}. Setting $\pvect{U}^a = U^a\pvect{e}_s + V^a\pvect{e}_x$, we get
\begin{subequations}
    \begin{gather}
        -\I a\omega P^a + \frac{1}{h_s}\frac{\partial U^a}{\partial s} + \frac{1}{h_s}\frac{\partial(h_sV^a)}{\partial x} = -\I a\omega\left(\sum_b\cnon P^{a - b}P^b - Q^a\right),
    \\
        -\I a\omega U^a + \frac{1}{h_s}\frac{\partial P^a}{\partial s} = \frac{1}{h_s}\frac{\partial Q^a}{\partial s}, 
        \qquad\qquad\qquad\qquad
        -\I a\omega V^a + \frac{\partial P^a}{\partial x} = \frac{\partial Q^a}{\partial x}.
    \end{gather}
\end{subequations}
We now proceed to eliminate the transverse velocities. Since we are neglecting $O(M^3)$ or smaller terms, we can form an expression for $V^a$ in terms of the other acoustic variables:
\begin{align}
\label{2DVexp}
    &V^a = \frac{1}{\I a\omega}\frac{\partial}{\partial x}\left(P^a - Q^a\right), &&\text{with} && Q^a = \frac{1}{2}\sum_b\left(P^{a - b}P^b - U^{a - b}U^b + \frac{\partial_xP^{a - b}\partial_xP^b}{(a - b)b\omega^2}\right),
\end{align}
where $\partial_x$ denotes $\partial/\partial x$ and the error in this expression for $Q^a$ is $O(M^3)$.  Since our eventual goal is to obtain ODEs in $s$ for spatial-mode coefficients of $P^a$ and $U^a$, these will be simplest to implement if we have as few $s$-derivatives present as possible. Therefore, we also wish to eliminate $s$-derivatives from the right-hand sides of our equations, using $O(M)$ substitutions of the form
\begin{equation}
    \frac{\partial P^a}{\partial s} = \I a\omega h_sU^a + O(M^2).
\end{equation}
Once we have eliminated $V^a$ and all $s$-derivatives at second order, we are left with
\begin{subequations}\label{equ:2Dgov}%
    \begin{align}
    \label{2Delimmass}
            \frac{\partial U^a}{\partial s} -& \I a\omega\left[h_s\left(1 + \frac{1}{a^2\omega^2}\frac{\partial^2}{\partial x^2}\right) - \frac{\kappa}{a^2\omega^2}\frac{\partial}{\partial x}\right]P^a \\\notag
            &= \I a\omega\Bigg\{-\cnon h_s\sum_bP^{a - b}P^b + \bigg[h_s\left(1 - \frac{1}{a^2\omega^2}\frac{\partial^2}{\partial x^2}\right) + \frac{\kappa}{a^2\omega^2}\frac{\partial}{\partial x}\bigg]Q^a\Bigg\},
\\\label{2Delimmom}
            \frac{\partial P^a}{\partial s} -& \I a\omega h_sU^a = \I\omega h_s\sum_b\Bigg\{U^{a - b}\left[(a - b) - b\left(1 + \frac{1}{b^2\omega^2}\frac{\partial^2}{\partial x^2}\right)\right]P^b + \frac{\partial_xU^{a - b}\partial_xP^b}{b\omega^2}\Bigg\}.
    \end{align}\end{subequations}%
The other piece of information that needs expression in the coordinate system is the hard-walled boundary condition $\pvect{u}\pvect{\cdot}\boldsymbol{\nu}_\pm = 0$ at $x = X_\pm$, where $\boldsymbol{\nu}_\pm$ is the normal to each duct wall (distinct from the Frenet-Serret normal $\pvect{n}$). The duct walls are defined by the equations $x - X_\pm(s) = 0$, so the normals are $\boldsymbol{\nu}_\pm = \pvect{\nabla}(x - X_\pm) = \pvect{e}_x - (\partial_sX_\pm/h_s)\pvect{e}_s$. The boundary condition is then given by the equation 
\begin{equation}
\label{2Dnopenetration}
     U^a\frac{\partial X_\pm}{\partial s} = h_sV^a = \frac{h_s}{\I a\omega}\frac{\partial}{\partial x}\left(P^a - Q^a\right)
     \qquad\text{at}\quad x = X_\pm(s).
\end{equation}

\subsubsection{The coordinate system in three dimensions}

Later on we will be projecting the equations onto a basis of functions representing the modes of a straight duct. The construction of such a basis is much easier and cleaner with separable boundary conditions in three dimensions, so here we drop the asymmetry about the centreline to focus on three-dimensional ducts with circular cross-sections; a constraint most brass instruments adhere to. However, as well as bending in the plane, the duct may now twist out of it, complicating the Frenet--Serret frame. So relative to two dimensions, the geometry in three dimensions is in one way simpler and in another more complex.

As shown in figure~\ref{figure:duct} (right), we begin once more with a centreline $\pvect{q}(s)$, but this time have a single radius function $R(s)$. The longitudinal coordinate is $s$ as before, and two more variables span the plane perpendicular to the centreline tangent. The Frenet--Serret frame now has an extra vector and a corresponding extra scalar
\begin{align}
    &\frac{\intd \pvect{q}}{\intd s} = \pvect{t}, && \frac{\intd \pvect{t}}{\intd s} = \kappa\pvect{n},&& \frac{\intd \pvect{n}}{\intd s} = -\kappa\pvect{t} + \tau\pvect{b}, && \frac{\intd \pvect{b}}{\intd s} = -\tau\pvect{n}.
\end{align}
Here $\pvect{b}$ is the \textit{binormal} (satisfying $\pvect{b} = \pvect{t} \times \pvect{n}$) and $\tau$ is the \textit{torsion}. We now introduce polar coordinates $r \in [0,R(s)]$ and $\theta \in [0,2\uppi)$ in the transverse plane, but rather than rotating this polar frame in line with the rotation of the basis vectors $\pvect{n}$ and $\pvect{b}$, we leave an extra degree of freedom here by also introducing a phase shift, or twist, $\theta_0(s)$, that may vary along the duct~\citep[following the work of][]{germano1982}, so that any point $\pvect{x}$ is given in these coordinates by
\begin{equation}
    \pvect{x} = \pvect{q}(s) + r\cos(\theta - \theta_0)\pvect{n} + r\sin(\theta - \theta_0)\pvect{b}.
\end{equation}
The corresponding metric $\intd \pvect{x}\pvect{\cdot}\intd \pvect{x}$ is then
\begin{equation}
    \intd \pvect{x}\pvect{\cdot}\intd \pvect{x} = \intd s^2\bigg[1 - \kappa r\cos(\theta - \theta_0) + r^2(\theta_0' - \tau)^2\bigg] + \intd r^2 + r^2\intd \theta^2 - 2\intd s\intd \theta r^2(\theta_0' - \tau).
\end{equation}
An orthogonal coordinate system requires that there be no cross-term differentials in the metric: to achieve this, following~\citet{germano1982}, we utilise the extra degree of freedom and take $\theta_0' = \tau$. The Lam\'e coefficients and $(s,r,\theta)$ basis vectors can then be written as
\begin{subequations}\begin{gather}
h_s = 1 - \kappa r\cos(\theta-\theta_0),
\qquad\qquad
h_r = 1,
\qquad\qquad
h_\theta = r,
\qquad\qquad
\pvect{e}_s = \pvect{t},
\\
\pvect{e}_r = \pvect{n}\cos(\theta-\theta_0) + \pvect{b}\sin(\theta-\theta_0),
\qquad\qquad
\pvect{e}_\theta = -\pvect{n}\sin(\theta-\theta_0) + \pvect{b}\cos(\theta-\theta_0).
\end{gather}\end{subequations}%
We decompose the velocity in the same way as before, but now with an extra coordinate, i.e. $\pvect{U}^a = U^a\pvect{e}_s + V^a\pvect{e}_r +  W^a\pvect{e}_\theta$. The mass and momentum equations~\eqref{equ:coordfree} become
\begin{subequations}\begin{gather}
            -\I a\omega P^a + \frac{1}{h_s}\frac{\partial U^a}{\partial s} + \frac{1}{rh_s}\frac{\partial(rh_sV^a)}{\partial r} + \frac{1}{rh_s}\frac{\partial(h_sW^a)}{\partial\theta} = \I a\omega\left(-\cnon\sum_bP^{a - b}P^b + Q^a\right),
\\
        -\I a\omega U^a + \frac{1}{h_s}\frac{\partial P^a}{\partial s} = \frac{1}{h_s}\frac{\partial Q^a}{\partial s},
        \qquad
        -\I a\omega V^a + \frac{\partial P^a}{\partial r} = \frac{\partial Q^a}{\partial r},
        \qquad
        -\I a\omega W^a + \frac{1}{r}\frac{\partial P^a}{\partial\theta} = \frac{1}{r}\frac{\partial Q^a}{\partial\theta}.
    \end{gather}\end{subequations}%
Proceeding as before, we find expressions for $V^a$ and $W^a$ in terms of $P^a$ and $U^a$:
\begin{align}
\label{3DVandWexp}
    &V^a = \frac{1}{\I a\omega}\frac{\partial}{\partial r}(P^a - Q^a), && W^a = \frac{1}{\I a\omega r}\frac{\partial}{\partial\theta}(P^a - Q^a),
\end{align}
with
\begin{equation}
    Q^a = \frac{1}{2}\sum_b\left(P^{a - b}P^b - U^{a - b}U^b + \frac{\pvect{\nabla}_\mathrm{t}P^{a - b}\pvect{\cdot}\pvect{\nabla}_\mathrm{t}P^b}{(a - b)b\omega^2}\right),
\end{equation}
where we have introduced the \textit{transverse gradient} $\pvect{\nabla}_\mathrm{t} = \pvect{e_r}\partial_r + \tfrac{1}{r}\pvect{e_\theta}\partial_\theta$ which acts only in the $r,\theta$ plane normal to the centreline. We also define, for any function $f$, the \textit{Frenet--Serret normal-derivative} (FSND)
\begin{equation}
    \frac{\partial f}{\partial\pvect{n}} = \pvect{n}\pvect{\cdot}\pvect{\nabla} f = \cos(\theta-\theta_0)\frac{\partial f}{\partial r} - \frac{\sin(\theta-\theta_0)}{r}\frac{\partial f}{\partial\theta}.
\end{equation}
This is a useful quantity, since it allows us to commute the transverse gradient with the longitudinal scale factor $h_s$, as in,
\begin{align}
        &\pvect{\nabla}_\mathrm{t}\pvect{\cdot}\left(h_s\pvect{\nabla}_\mathrm{t}f\right) = \left(h_s\pvect{\nabla}_\mathrm{t}^2 - \kappa\partial_{\pvect{n}}\right)f &&\text{and}&&\pvect{\nabla}_\mathrm{t}^2(h_sf) = \left(h_s\pvect{\nabla}_\mathrm{t}^2 - 2\kappa\partial_{\pvect{n}}\right)f.
\end{align}
This contrasts with the two-dimensional derivation, where both the transverse gradient and the FSND collapse onto the $x$-derivative (the former as a vector in the $\pvect{e}_x$ direction, the latter as a scalar). Once we have eliminated the $s$-derivatives as before, we then have
\begin{subequations}\label{equ:3Dgov}\begin{gather}
    \label{3Delimmass}\mathtoolsset{multlined-width=0.85\displaywidth}
        \begin{multlined}
            \frac{\partial U^a}{\partial s} - \I a\omega\left[h_s\left(1 + \frac{\pvect{\nabla}_\mathrm{t}^2}{a^2\omega^2}\right) - \frac{\kappa}{a^2\omega^2}\frac{\partial}{\partial\pvect{n}}\right]P^a \\
            = \I a\omega\Bigg\{-\cnon h_s\sum_bP^{a - b}P^b + \bigg[h_s\left(1 - \frac{\pvect{\nabla}_\mathrm{t}^2}{a^2\omega^2}\right) + \frac{\kappa}{a^2\omega^2}\frac{\partial}{\partial\pvect{n}}\bigg]Q^a\Bigg\},
        \end{multlined}
\\
    \label{3Delimmom}
            \frac{\partial P^a}{\partial s} - \I a\omega h_sU^a = \I\omega h_s\sum_b\Bigg\{U^{a - b}\left[(a - b) - b\left(1 + \frac{\pvect{\nabla}_\mathrm{t}^2}{b^2\omega^2}\right)\right]P^b + \frac{\pvect{\nabla}_\mathrm{t}U^{a - b}\pvect{\cdot}\pvect{\nabla}_\mathrm{t}P^b}{b\omega^2}\Bigg\}.
\end{gather}\end{subequations}%
Finally, the hard-walled boundary condition in three dimensions requires the normal $\boldsymbol{\nu}$, which we calculate similarly to earlier as $\boldsymbol{\nu} = \pvect{\nabla}(r - R)|_{r = R} = (\pvect{e}_r - R'\pvect{e}_s/h_s)|_{r = R}$. Plugging this into $\pvect{u}\pvect{\cdot}\boldsymbol{\nu}|_{r = R} = 0$ gives
\begin{equation}
\label{3Dnopenetration}
    R'U^a|_{r = R} = (h_sV^a)|_{r = R} = \left[\frac{h_s}{\I a\omega}\frac{\partial}{\partial r}\left(P^a - Q^a\right)\right]\bigg|_{r = R}.
\end{equation}

\subsection{Spatial modal decomposition}
At this point, we expand each of $P^a$ and $U^a$ in terms of a basis of straight-duct modes. We will ultimately be solving for the coefficients of the series expansions, which will only depend on $s$ with all transverse variation being contained within the basis functions.
\subsubsection{The spatial modes in two dimensions}
$P^a$ and $U^a$ are expanded as
\begin{align}
\label{2Dspatialmodedefs}
    &P^a = \sum_{\alpha = 0}^\infty P_\alpha^a(s)\psi_\alpha(s,x),&& U^a = \sum_{\alpha = 0}^\infty U_\alpha^a(s)\psi_\alpha(s,x),
\end{align}
with every $\psi_\alpha$ satisfying
\begin{itemize}
    \item Helmholtz's equation with eigenvalue $\lambda_\alpha$ in two dimensions (scaling out the eigenvalue's $X$-dependence for convenience)
    \begin{equation}
    \label{2DHelmholtzeq}
        \frac{\partial^2\psi_\alpha}{\partial x^2} + \frac{\lambda_\alpha^2}{X^2}\psi_\alpha = 0,
    \end{equation}
    \item a normalisation condition
    \begin{equation}
    \label{2Dmodenorm}
         \langle\psi_\alpha,\psi_\beta\rangle = \delta_{\alpha\beta},\text{ where }\langle\psi_\alpha,\psi_\beta\rangle := \int_{X_-}^{X_+}\psi_\alpha\psi_\beta~\intd x,
    \end{equation}
    \item and a Neumann condition on the duct walls (which ensures that the requisite Sturm-Liouville properties hold, while maintaining consistency with the  no-penetration condition)
    \begin{equation}
    \label{2DNeumann}
        \frac{\partial\psi_\alpha}{\partial x}\bigg|_{x = X_\pm} = 0.
    \end{equation}
\end{itemize}
Solutions to the Helmholtz equation are quantised by the boundary conditions, giving solutions
\begin{align}
\label{2Dmodedef}
    &\psi_\alpha = \frac{C_\alpha}{\sqrt{X}} \cos\left[\frac{\lambda_\alpha (x - X_-)}{X}\right],&& \lambda_\alpha = \alpha\uppi,
\end{align}
for $\alpha \in \mathbb{N}_0$, with $C_\alpha$ to be $\sqrt{2 - \delta_{\alpha0}}$ ensuring orthonormality.

\subsubsection{The spatial modes in three dimensions}
In three dimensions, the expansions are similarly written as
\begin{align}
\label{3Dspatialmodedefs}
    &P^a = \sum_{\alpha = 0}^\infty P_\alpha^a(s)\psi_\alpha(s,r,\theta),&& U^a = \sum_{\alpha = 0}^\infty U_\alpha^a(s)\psi_\alpha(s,r,\theta),
\end{align}
with every $\psi_\alpha$ satisfying the Helmholtz equation in the circular cross-section with Neumann boundary conditions. This gives
\begin{equation}
\label{modedef}
    \psi_\alpha = \frac{C_\alpha}{\sqrt{\uppi}R} \J_p(\lambda_\alpha r/R)\cos\left(p_\alpha(\theta-\theta_0) - \frac{\xi_\alpha\uppi}{2}\right),
\end{equation}
where $C_\alpha$ ensures orthonormalization, $\lambda_\alpha$ is the eigenvalue, $p_\alpha\in\mathbb{N}_0$ gives the azimuthal order and $\xi_\alpha \in\{0,1\}$ effectively provides both sine and cosine solutions.  Defining $j_{pq}'$ to be the $q^\text{th}$ zero of $\J_p'(x)$, we have $\lambda_\alpha = j_{p_\alpha q_\alpha}'$ for $q_\alpha \in\mathbb{N}_0$ independent of $\xi_\alpha$. Hence, modes are given by a bijection between $\alpha\in\mathbb{N}_0$ and $(p_\alpha,q_\alpha,\xi_\alpha)\in\mathbb{N}_0\times\mathbb{N}_0\times\{0,1\}$, although modes that are identically zero are skipped (such as $(0,0,1)$, $(2,0,0)$, $(2,0,1)$, $(4,0,0)$, etc). For orthonormality, we have 
\begin{align}
\langle\psi_\alpha,\psi_\beta\rangle &= \!\!\int_{0}^{R}\!\!\!\!\int_0^{2\uppi}\!\!\!\!\psi_\alpha\psi_\beta~r\intd r\intd\theta&
&\Rightarrow&
    C_\alpha &= 
    \begin{cases}
        \big|\J_{0}(j_{0 q_\alpha}')\big|^{-1},&p_\alpha = 0,\\
        \bigg(\sqrt{\frac{1}{2}\left[1 - \frac{{p_\alpha}^2}{{j_{p_\alpha q_\alpha}'}^2}\right]} ~\big|\J_{p_\alpha}(j_{p_\alpha q_\alpha})\big|\bigg)^{-1},&p_\alpha \neq 0.
    \end{cases}
\end{align}

\subsection{Spatial projection and notation}
\label{spatprodandnot}
Now that we have defined the spatial basis functions, we follow \citet{mctavish+brambley-2019} in projecting the governing equations (\ref{equ:2Dgov},~\ref{2Dnopenetration}) or (\ref{equ:3Dgov},~\ref{3Dnopenetration}) onto the spatial modal basis~\eqref{2Dmodedef} or~\eqref{modedef} to form an infinite set of ODEs for the coefficients $U^a_\alpha$ and $P^a_\alpha$.  This is achieved by multiplying each of the governing equations by the spatial basis functions $\psi_\beta$ and integrating across a duct cross-section, using the orthogonality of the spatial basis functions in the process. The result is simplified greatly using a compact notation, which we now introduce.

Let each $a\in\mathbb{N}$, let $\mvect{p}^a$ denote the vector of coefficients $P^a_\alpha$ for $\alpha\in\mathbb{N}_0$, and similarly let $\mvect{u}^a$ denote the vector of coefficients $U^a_\alpha$.  We use Roman letters as superscripts for the temporal Fourier modal decomposition, and Greek letters as subscripts for the spatial modal decomposition.  For each of these vectors, matrices $\mat{M}$ with coefficients $\mat{M}_{\alpha\beta}$ act in the normal way,
\begin{equation}
\big(\mat{M}\mvect{y}\big)_\alpha = \sum_{\beta=0}^\infty \mat{M}_{\alpha\beta}y_\beta.
\end{equation}
For the weakly nonlinear terms, we define a quadratic operator $\mathcal{M}$ with coefficients $\mathcal{M}_{\alpha\beta\gamma}$, which acts on vectors $\mvect{y}$ and $\mvect{z}$ as
\begin{equation}
\big(\mathcal{M}\langle\mvect{y},\mvect{z}\rangle\big)_\alpha
= \sum_{\beta=0}^\infty\sum_{\gamma=0}^\infty \mathcal{M}_{\alpha\beta\gamma}y_\beta z_\gamma.
\end{equation}
We also use the operator shorthand
\begin{equation}
\Big(\mathcal{M} + \mathcal{N}\langle\mat{I},\mat{A}\rangle\Big)\langle\mvect{y},\mvect{z}\rangle
= \mathcal{M}\langle\mvect{y},\mvect{z}\rangle + \mathcal{N}\langle\mvect{y},\mat{A}\mvect{z}\rangle.
\end{equation}
Unlike in~\citet{mctavish+brambley-2019}, here we choose a decomposition into matrix and quadratic operators that depends only on the modal basis and not on the variable $s$ or any physical parameters.  This allows for a unified approach to both the two- and three-dimensional cases listed above, and, since physical quantities such as curvature $\kappa$ occur explicitly, the effects of the physical quantities can be isolated and better understood.  The algebra behind the manipulations needed to get to the governing equations is given in appendix~\ref{spatprodapp}.  In two dimensions, this results in
\begin{subequations}\label{equ:2Dvector}\begin{gather}
    \label{2Dvectormass}
        \begin{aligned}
            \Bigg[\frac{\intd }{\intd s} &+ \frac{X'(s)}{2X(s)}\mat{W} + \frac{X_-'(s)}{X(s)}\widetilde{\mat{A}}\Bigg]\mvect{u}^a(s) \\
            &- \I a\omega\Bigg[\left(\mat{I} - \frac{\gmat{\Lambda}^2}{a^2\omega^2X(s)^2}\right)\bigg((1 - \kappa(s) X_-(s))\mat{I} - \kappa(s) X(s)\mat{A}\bigg) - \frac{\kappa(s)\widetilde{\mat{A}}}{a^2\omega^2X(s)}\Bigg]\mvect{p}^a(s) \\
            =& \frac{\I a\omega}{\sqrt{X(s)}}\sum_b\Bigg\{- \cnon\bigg((1 - \kappa(s) X_-(s))\mathcal{I} - \kappa X(s)\mathcal{A}\bigg)\langle\mvect{p}^{a - b}(s),\mvect{p}^b(s)\rangle\\&
            +\Bigg[\left(\mat{I} + \frac{\gmat{\Lambda}^2}{a^2\omega^2X(s)^2}\right)\bigg((1 - \kappa(s) X_-(s))\mathcal{I} - \kappa(s) X(s)\mathcal{A}\bigg)\\&
            \qquad\qquad\qquad\qquad\qquad\qquad\qquad + \frac{\kappa(s)\widetilde{\mathcal{A}}}{a^2\omega^2X(s)}\Bigg]\frac{\langle\mvect{p}^{a - b}(s),\mvect{p}^b(s)\rangle - \langle\mvect{u}^{a - b}(s),\mvect{u}^b(s)\rangle}{2} \\&
            + \left[\frac{\left(\mat{I} + \frac{\gmat{\Lambda}^2}{a^2\omega^2X(s)^2}\right)\bigg((1 - \kappa(s) X_-(s))\mathcal{I}^\lambda - \kappa(s) X(s)\mathcal{A}^\lambda\bigg) + \frac{\kappa(s)\widetilde{\mathcal{A}}^\lambda}{a^2\omega^2X(s)}}{2(a - b)b\omega^2X(s)^2} \right]\langle\mvect{p}^{a - b}(s),\mvect{p}^b(s)\rangle\Bigg\},
        \end{aligned}
\displaybreak[0]\\[1ex]
    \label{2Dvectormom}
        \begin{aligned}
            \Bigg[\frac{\intd }{\intd s} - &\frac{X'(s)}{2X(s)}\mat{W}^\T - \frac{X_-'(s)}{X(s)}\widetilde{\mat{A}}^\T\Bigg]\mvect{p}^a(s) - \I a\omega\Bigg[(1 - \kappa(s) X_-(s))\mat{I} - \kappa(s) X(s)\mat{A}\Bigg]\mvect{u}^a(s) \\
            =& \frac{1}{\sqrt{X(s)}}\sum_b\Bigg\{\bigg(\frac{X_+'(s)}{X(s)}\overline{\mathcal{W}}^+ - \frac{X_-'(s)}{X(s)}\overline{\mathcal{W}}^-\bigg)\langle\mvect{u}^{a - b}(s),\mvect{u}^b(s)\rangle
            \\&
            + \I\omega\Bigg[\bigg((1 - \kappa(s) X_-(s))\mathcal{I} - \kappa(s) X(s)\mathcal{A}\bigg)
            \left\langle\mat{I},(a - b)\mat{I} - b\left(\mat{I} - \frac{\gmat{\Lambda}^2}{b^2\omega^2X(s)^2}\right)\right\rangle
            \\&\qquad\qquad\qquad\qquad
            + \frac{(1 - \kappa(s) X_-(s))\mathcal{I}^\lambda - \kappa(s) X(s)\mathcal{A}^\lambda}{b\omega^2X(s)^2}\Bigg]\langle\mvect{u}^{a - b}(s),\mvect{p}^b(s)\rangle\Bigg\},
        \end{aligned}
    \end{gather}\end{subequations}%
where the matrices and tensors are defined in appendix~\ref{app:A1} and are independent of $s$.

Likewise, in three dimensions, using the derivation in appendix~\ref{app:A2}, we arrive at
\begin{subequations}\label{equ:3Dvector}
    \begin{gather}\begin{aligned}
        &\left(\frac{\intd }{\intd s} + \frac{R'(s)}{R(s)}\mat{W} + \tau(s)\mat{H}\right)\mvect{u}^a(s) - \I a\omega\left[\left(\mat{I} - \frac{\gmat{\Lambda}^2}{a^2\omega^2R(s)^2}\right)\left(\mat{I} - \kappa(s) R(s)\mat{A}\right) - \frac{\kappa(s)\widetilde{\mat{A}}}{a^2\omega^2R(s)}\right]\mvect{p}^a(s) \\
        &\quad= \frac{\I a\omega}{\sqrt{\uppi}R(s)}\sum_b\Bigg\{\Bigg[\left(\mat{I} + \frac{\gmat{\Lambda}^2}{a^2\omega^2R(s)^2}\right)\left(\mathcal{I} - \kappa(s) R(s)\mathcal{A}\right)\\
        &\qquad\qquad\qquad\qquad\qquad\qquad\qquad\qquad  + \frac{\kappa(s)\widetilde{\mathcal{A}}}{a^2\omega^2 R(s)}\Bigg]\frac{\langle\mvect{p}^{a - b}(s),\mvect{p}^b(s)\rangle - \langle\mvect{u}^{a - b}(s),\mvect{u}^b(s)\rangle}{2} \\
        &\qquad+ \Bigg[\frac{\left(\mat{I} + \frac{\gmat{\Lambda}^2}{a^2\omega^2R(s)^2}\right)\left(\mathcal{I}^\lambda - \kappa(s) R(s)\mathcal{A}^\lambda\right) + \frac{\kappa(s)\widetilde{\mathcal{A}}^\lambda}{a^2\omega^2R(s)}}{2(a - b)b\omega^2R(s)^2} \\
        &\qquad\qquad\qquad\qquad\qquad\qquad\qquad\qquad\qquad\qquad - \cnon\left(\mathcal{I} - \kappa(s) R(s)\mathcal{A}\right)\Bigg]\langle\mvect{p}^{a - b}(s),\mvect{p}^b(s)\rangle\Bigg\},
    \end{aligned}
\displaybreak[0]\\[1ex]
    \begin{aligned}
        &\left(\frac{\intd }{\intd s} - \frac{R'(s)}{R(s)}\mat{W}^\T - \tau(s)\mat{H}^\T\right)\mvect{p}^a(s) - \I a\omega\bigg(\mat{I} - \kappa(s) R(s)\mat{A}\bigg)\mvect{u}^a(s) \\
        &\qquad= \frac{1}{\sqrt{\uppi}R(s)}\sum_b\Bigg\{\frac{R'(s)}{R(s)}\overline{\mathcal{W}}\langle\mvect{u}^{a - b}(s),\mvect{u}^b(s)\rangle + \I\omega\Bigg[\left(\mathcal{I} - \kappa(s) R(s)\mathcal{A}\right) \\
        &\qquad\qquad\times\bigg\langle\mat{I},(a - b)\mat{I} - b\left(\mat{I} - \frac{\gmat{\Lambda}^2}{b^2\omega^2R(s)^2}\right)\bigg\rangle + \frac{\mathcal{I}^\lambda - \kappa(s) R(s)\mathcal{A}^\lambda}{b\omega^2R(s)^2}\Bigg]\langle\mvect{u}^{a - b}(s),\mvect{p}^b(s)\rangle\Bigg\},
    \end{aligned}
    \end{gather}
\end{subequations}
where again the matrices and tensors are defined in appendix~\ref{app:A2} and are independent of $s$.

The equations have a structure common to both two and three dimensions, and can be written compactly (as detailed in appendix~\ref{blockdetailapp}) as
    \begin{equation}
    \label{LandNintro}
        \frac{\intd }{\intd s}\begin{pmatrix}
            \mvect{u}^a \\ \mvect{p}^a
        \end{pmatrix} = \mat{L}^a\begin{pmatrix}
            \mvect{u}^a \\ \mvect{p}^a
        \end{pmatrix} + \sum_b\mathcal{N}^{ab}\left\langle\begin{pmatrix}
            \mvect{u}^{a - b} \\ \mvect{p}^{a - b}
        \end{pmatrix},\begin{pmatrix}
            \mvect{u}^b \\ \mvect{p}^b
        \end{pmatrix}\right\rangle,
    \end{equation}
in terms of square blocks $\mat{L}^a =: \begin{pmatrix}
        \mat{L}_1^a &\mat{L}_2^a \\ \mat{L}_3^a &\mat{L}_4^a
    \end{pmatrix}$, and `cubic blocks'
\begin{equation}
\mathcal{N}^{ab}\left\langle\begin{pmatrix}
            \mvect{u}^{a - b} \\ \mvect{p}^{a - b}
        \end{pmatrix},\begin{pmatrix}
            \mvect{u}^b \\ \mvect{p}^b
        \end{pmatrix}\right\rangle
        = \begin{pmatrix}
        \mathcal{N}^{ab}_1\langle \mvect{u}^{a-b}, \mvect{u}^b\rangle
        + \mathcal{N}^{ab}_6\langle \mvect{p}^{a-b}, \mvect{p}^b\rangle
        \\
        \mathcal{N}^{ab}_3\langle \mvect{u}^{a-b}, \mvect{u}^b\rangle
        + \mathcal{N}^{ab}_7\langle \mvect{u}^{a-b}, \mvect{p}^b\rangle
         \end{pmatrix},
\end{equation}
where the equivalent terms $\mathcal{N}^{ab}_2 \equiv \mathcal{N}^{ab}_4 \equiv \mathcal{N}^{ab}_5 \equiv \mathcal{N}^{ab}_8 \equiv 0$.
We recall that $p_\alpha^{-a} = p_\alpha^{a*}$ and that $p_\alpha^0=0$ by our earlier assumptions of real variables and vanishing time-averages.  Note that, as expected, the weakly nonlinear terms will mix the effects of different frequencies. We will now work in terms of these quantities, so that everything we do will apply in both two and three dimensions.

In summary, the governing equations are \eqref{equ:2Dvector} in two-dimensions and \eqref{equ:3Dvector} in three dimensions, and can be combined and written as \eqref{LandNintro}. Greater numerical efficiency has been achieved through the transferral of $s$-dependence from matrices to the scalars multiplying them (since this allows the matrices to be pre-calculated rather than requiring recalculation at each step). Upper (Roman) indices are Fourier series wavenumbers while lower (Greek) indices are duct mode wavenumbers.

\section{Solution of the Governing Equations}\label{sec:solve}

Here, we solve the above governing equations using a multi-modal method, inspired by~\citet{felix1} and~\citet{mctavish+brambley-2019}.  We first truncate the number of spatial modes to $\alpha_\text{max}$, before introducing the concept of an admittance which relates velocities to pressures.

\subsection{Admittance}\label{sec:admittance}
The \emph{admittance} $Y$, and its inverse the \emph{impedance} $Z$, characterise the relationship between acoustic pressure and acoustic velocity.  The admittance and impedance are defined here as
\begin{align}
\label{admittanceintros}
    &\mvect{u}^a = \mat{Y}^a\mvect{p}^a + \sum_b\mathcal{Y}^{ab}\langle\mvect{p}^{a - b},\mvect{p}^b\rangle,&& \mvect{p}^a = \mat{Z}^a\mvect{u}^a + \sum_b\mathcal{Z}^{ab}\langle\mvect{u}^{a - b},\mvect{u}^b\rangle.
\end{align}
The first of these relations may be used to eliminate the velocity from the governing equations~\eqref{LandNintro}, which results in a Riccati-style equation for the $s$-evolution of the admittance,
\begin{subequations}\label{equ:Y}
    \begin{gather}
    \label{linadmeq}
            \frac{\intd \mat{Y}^a}{\intd s} =  -\mat{Y}^a\mat{L}_3^a\mat{Y}^a + \mat{L}_1^a\mat{Y}^a - \mat{Y}^a\mat{L}_4^a + \mat{L}_2^a,
\displaybreak[0]\\
    \label{nonlinadmeq}
        \begin{aligned}
            \frac{\intd \mathcal{Y}^{ab}}{\intd s} = &-\mat{Y}^a\mathcal{N}_3^{ab}\langle\mat{Y}^{a - b},\mat{Y}^b\rangle - \mat{Y}^a\mathcal{N}_7^{ab}\langle\mat{Y}^{a - b},\mat{I}\rangle + \mathcal{N}_1^{ab}\langle\mat{Y}^{a - b},\mat{Y}^b\rangle + \mathcal{N}_6^{ab} \\
            &- \mathcal{Y}^{ab}\Bigg[-\mat{L}_1^a + \mat{Y}^a\mat{L}_3^a + \left\langle\mat{L}_4^{a - b} + \mat{L}_3^{a - b}\mat{Y}^{a - b},\mat{I}\right\rangle + \left\langle\mat{I},\mat{L}_4^b + \mat{L}_3^{b}\mat{Y}^b\right\rangle\Bigg].
        \end{aligned}
    \end{gather}
\end{subequations}
These equations will be solved to find the admittance, which encodes the acoustic properties of the duct.  Once the admittance is known, the acoustic pressure may be solved for by substituting the admittance definition~\eqref{admittanceintros} back into the governing questions~\eqref{LandNintro}, giving
\begin{equation}
\label{peq}
    \begin{aligned}
        \frac{\intd\mvect{p}^a}{\intd s} = \left(\mat{L}_3^a\mat{Y}^a + \mat{L}_4^a\right)\mvect{p}^a + \sum_b\bigg(\mat{L}_3^a\mathcal{Y}^{ab} + \mathcal{N}_3^{ab}\langle\mat{Y}^{a - b},\mat{Y}^b\rangle + \mathcal{N}_7^{ab}\langle\mat{Y}^{a - b},\mat{I}\rangle\bigg)\langle\mvect{p}^{a - b},\mvect{p}^b\rangle,
    \end{aligned}
\end{equation}
with the velocities then given by the known admittance and pressure from the admittance definition~\eqref{admittanceintros}.  In order to solve the first order equation~\eqref{equ:Y}, a known value of the admittance must be given as a boundary condition at some point in the duct, for example, at the outlet.  We next investigate special values of the admittance that might be used as boundary conditions for~\eqref{equ:Y}.

\subsection{Invariant admittances}\label{sec:invariant-admittances}

We consider admittances that solve~\eqref{equ:Y} and are constant in $s$, therefore having a vanishing $s$-derivative. Physically, these invariant solutions represent the admittances of ducts for which which no $s$ position is distinguishable from any other.  Such a duct would have a constant radius, curvature and torsion, leaving the possibilities of: (a) an infinite straight duct; (b) an annulus (two dimensions) or torus (three dimensions); and (c) an infinite helical duct.  Because of the quadratic term in the evolution equation for $\mat{Y}^a$, there is more than one constant solution for a given geometry.  Here, a `$+$' solution will be constructed from the set of eigenvalues that represent waves either decaying or propagating in the positive $s$-direction, and a `$-$' solution will be constructed from the eigenvalues representing growth in the positive $s$-direction or propagation in the negative $s$-direction. These are respectively the \textit{positive} and \textit{negative characteristic admittances}.   We denote by $(\mvect{u}^{a\pm},\mvect{u}^{a\pm})$ disturbances propagating in only the positive $+$ or negative $-$ directions, and define
\begin{align}
    &\mvect{u}^{a\pm} = \mat{Y}^{a\pm}\mvect{p}^{a\pm} + \sum_b\mathcal{Y}^{ab\pm}\langle\mvect{p}^{(a - b)\pm},\mvect{p}^{b\pm}\rangle,&& \mvect{p}^{a\pm} = \mat{Z}^{a\pm}\mvect{u}^{a\pm} + \sum_b\mathcal{Z}^{ab\pm}\langle\mvect{u}^{(a - b)\pm},\mvect{u}^{b\pm}\rangle.
\end{align}
The linear characteristic impedances $\mat{Z}^{a\pm}$ are the inverses of $\mat{Y}^{a\pm}$; the nonlinear characteristic impedances are found by the weakly-nonlinear inversion rule $\mathcal{Z}^{ab\pm} = -\mat{Z}^{a\pm}\mathcal{Y}^{ab\pm}\langle\mat{Z}^{(a - b)\pm},\mat{Z}^{b\pm}\rangle$.

\subsubsection{Straight-duct characteristic admittances}
\label{straightductcharadmderivation}
This computation will be done in the general case in order to lay the groundwork for the curvature/torsion cases, where the diagonalisation is less trivial.

For a straight duct, the matrix $\mat{L}^a$ is constant, and given by setting $\kappa \equiv \tau \equiv X' \equiv R' \equiv 0$,
\begin{equation}
\label{straightductoperator}
    \overline{\mat{L}}^a = \begin{pmatrix}
        \mat{0} &\overline{\mat{L}}_2^a \\
        \overline{\mat{L}}_3^a &\mat{0}
    \end{pmatrix},
\end{equation}
where $\overline{\mat{L}}_2^a$ is diagonal and $\overline{\mat{L}}_3^a$ is proportional to the identity. We seek a solution to 
\begin{equation}
\label{straightductproblem}
    \frac{\intd}{\intd s}\begin{pmatrix}
        \mvect{u}^a \\ \mvect{p}^a
    \end{pmatrix} = \overline{\mat{L}}^a\begin{pmatrix}
        \mvect{u}^a \\ \mvect{p}^a
    \end{pmatrix},
\end{equation}
which takes the form $\mvect{p}^a = \overline{\mvect{c}}^a\exp(\overline{\gamma}^a s)$, with $\overline{\gamma}^a$ an eigenvalue and $\overline{\mvect{c}}^a$ the corresponding eigenvector of $\overline{\mat{L}}^a$. We use the fact that for invertible matrices $\mat{A}$ and $\mat{D}$,
\begin{equation}
\label{dettheorem}
    \det\begin{pmatrix}
        \mat{A} &\mat{B} \\ \mat{C} &\mat{D}
    \end{pmatrix} = \det\mat{A}\det\left(\mat{D} - \mat{C}\mat{A}^{-1}\mat{B}\right) = \det\mat{D}\det\left(\mat{A} - \mat{B}\mat{D}^{-1}\mat{C}\right).
\end{equation}
With this in mind, we know that the characteristic equation reads
\begin{equation}
\label{straightductchareq}
    0 = \det\left[\overline{\mat{L}}^a - \overline{\gamma}^a\begin{pmatrix}
        \mat{I} &\mat{0} \\ \mat{0} &\mat{I}
    \end{pmatrix}\right] = \det\left(\big(\overline{\gamma}^a\big)^2\mat{I} - \overline{\mat{L}}_3^a\overline{\mat{L}}_2^a\right).
\end{equation}
Because $\overline{\mat{L}}_3^a\overline{\mat{L}}_2^a$ is diagonal, we may directly read off the eigenvalues.  Since we have truncated to $\alpha_\mathrm{max}$ spatial modes, and therefore both $\overline{\mat{L}}_2^a$ and $\overline{\mat{L}}_3^a$ are $\alpha_\mathrm{max}\times\alpha_\mathrm{max}$, we see that there are precisely $\alpha_\text{max}$ solutions for  ${\overline{\gamma}^a}^2$, of which some are positive and some negative (depending on the frequency $\omega$, since this is determining whether the modes are cut-on or cut-off). Thus, there will be a set of distinct eigenvalues $\{\overline{\gamma}_\alpha^a\}_{\alpha = 0}^{\alpha_\text{max}}$ that are exclusively in $\mathbb{R}^-$ or $\I\mathbb{R}^+$, corresponding to forward-decaying or forward-propagating modes respectively. These will be used to build the positive characteristic admittance; each will meanwhile have a mirror image in either $\mathbb{R}^+$ or $\I\mathbb{R}^-$, used to construct the negative characteristic admittance.

We may now partition the eigenvectors into two sets, those corresponding to $\{\overline{\gamma}_\alpha^a\}_{\alpha = 0}^{\alpha_\text{max}}$ and those corresponding to $\{-\overline{\gamma}_\alpha^a\}_{\alpha = 0}^{\alpha_\text{max}}$, denoted $\overline{\mvect{c}}^{a\pm}$. Splitting them (as we have split $\overline{\mat{L}}^a$) into upper and lower vectors $(\overline{\mvect{c}}_{\alpha,1}^{a\pm},\overline{\mvect{c}}_{\alpha,2}^{a\pm})$, and eliminating the upper vectors, we get
\begin{equation}
    \overline{\mat{L}}_3^a\overline{\mat{L}}_2^a\overline{\mvect{c}}_{\alpha,2}^{a\pm} = {\overline{\gamma}_\alpha^a}^2\overline{\mvect{c}}_{\alpha,2}^{a\pm}.
\end{equation}
Because $\overline{\mvect{c}}_{\alpha,2}^{a+}$ and $\overline{\mvect{c}}_{\alpha,2}^{a-}$ satisfy the same equation, we can choose $\overline{\mvect{c}}_{\alpha,2}^{a+} = \overline{\mvect{c}}_{\alpha,2}^{a-} = \overline{\mvect{c}}_{\alpha,2}^{a}$; further to this, we may note that \eqref{straightductchareq} implies that $\overline{\mat{L}}_3^a\overline{\mat{L}}_2^a = {\overline{\gmat{\Gamma}}^a}^2$, where $\overline{\gmat{\Gamma}}^a$ is a diagonal matrix with entries given by the elements of $\{\overline{\gamma}_\alpha^a\}_{\alpha = 0}^{\alpha_\text{max}}$. Defining a matrix $\overline{\mat{C}}^a$ with each column being a different $\overline{\mvect{c}}_{\alpha,2}^a$ (arranged so as to match the ordering of $\overline{\gmat{\Gamma}}^a$), we then see that
\begin{equation}
    \overline{\mat{L}}_3^a\overline{\mat{L}}_2^a\overline{\mat{C}}^a = \overline{\mat{C}}^a{\overline{\gmat{\Gamma}}^a}^2 = \overline{\mat{C}}^a\overline{\mat{L}}_3^a\overline{\mat{L}}_2^a.
\end{equation}
Since $\overline{\mat{C}}^a$ commutes with a diagonal matrix with unique entries, it too must be diagonal, so having not yet normalised the eigenvectors, we can now simply choose $\overline{\mat{C}}^a = \mat{I}$. The admittance is then simple to construct from substituting it into the lower block of \eqref{straightductproblem}, getting $\overline{\mat{L}}_3^a\overline{\mat{Y}}^{a\pm} = \pm\overline{\gmat{\Gamma}}^a$. In both two and three dimensions, $\overline{\mat{L}}_3^a = \I a\omega\mat{I}$, so we have
\begin{equation}
    \overline{\mat{Y}}^{a\pm} = \pm\frac{1}{\I a\omega}\overline{\gmat{\Gamma}}^a =: \pm\overline{\mat{Y}}^a.
\end{equation}
For clarity, the admittances in two dimensions have the following explicit expression
\begin{equation}
\label{straightcharadm}
    \overline{\mat{Y}}_{\alpha\beta}^{a\pm} = \pm\I\delta_{\alpha\beta}\exp\left\{-\frac{\I\uppi}{4}\left[\sgn\left(1 - \frac{\lambda_\alpha^2}{a^2\omega^2X^2}\right) + 1\right]\right\}\sqrt{\left|1 - \frac{\lambda_\alpha^2}{a^2\omega^2X^2}\right|},
\end{equation}
and the three-dimensional case is identical, only with an $R$ in place of every $X$. We can then easily write down the definition of the straight-duct cut-off frequency for each mode in two and three dimensions
\begin{equation}
\label{cutofffreqdef}
    \overline{\omega}_\alpha^a(s) = \begin{cases}
        \frac{\lambda_\alpha}{aX(s)} \quad &2\text{D}, \\
        \frac{\lambda_\alpha}{aR(s)} \quad &3\text{D}, \\
    \end{cases}
\end{equation}
and note that $\overline{\omega}_\alpha^a$ increases with the spatial modenumber, while having the opposite relationship with the temporal modenumber.

In order to calculate the nonlinear characteristic admittances, we substitute the linear characteristic admittances into \eqref{nonlinadmeq}, and look for fixed points
\begin{equation}
    \begin{aligned}
        0 = &-\overline{\mat{Y}}^a\overline{\mathcal{N}}_7^{ab}\langle\overline{\mat{Y}}^{a - b},\mat{I}\rangle + \overline{\mathcal{N}}_1^{ab}\langle\overline{\mat{Y}}^{a - b},\overline{\mat{Y}}^b\rangle + \overline{\mathcal{N}}_6^{ab} \\
        &\mp \overline{\mathcal{Y}}^{ab\pm}\left[\overline{\mat{Y}}^a\overline{\mat{L}}_3^a + \langle\overline{\mat{L}}_3^{a - b}\overline{\mat{Y}}^{a - b},\mat{I}\rangle + \langle\mat{I},\overline{\mat{L}}_3^b\overline{\mat{Y}}^b\rangle\right].
    \end{aligned}
\end{equation}
If we note that the square-bracketed quantity is actually equal to $\overline{\gmat{\Gamma}}^a + \langle\overline{\gmat{\Gamma}}^{a - b},\mat{I}\rangle + \langle\mat{I},\overline{\gmat{\Gamma}}^b\rangle$, a sum of three diagonal matrices each acting on a different component, then we may simply divide through by this, pointwise, for each entry of $\overline{\mathcal{Y}}^{ab\pm}$, getting
\begin{equation}
    \overline{\mathcal{Y}}_{\alpha\beta\gamma}^{ab\pm} = \pm\frac{\left(\overline{\mathcal{N}}_1^{ab}\langle\overline{\mat{Y}}^{a - b},\overline{\mat{Y}}^b\rangle + \overline{\mathcal{N}}_6^{ab} - \overline{\mat{Y}}^a\overline{\mathcal{N}}_7^{ab}\langle\overline{\mat{Y}}^{a - b},\mat{I}\rangle\right)_{\alpha\beta\gamma}}{\overline{\gamma}_\alpha^a + \overline{\gamma}_\beta^{a - b} + \overline{\gamma}_\gamma^b} =: \pm\overline{\mathcal{Y}}_{\alpha\beta\gamma}^{ab}.
\end{equation}
Obviously we will run into problems if the denominator of this turns out to be zero: this would constitute a \textit{resonant triad}~\citep[see, e.g.]{bustamante}. Empirically, we do not seem to run into this problem in the straight-duct case, but for the more complicated characteristic admittances that follow there seems to be no such guarantee. We avoid the problem of singular entries in the characteristic admittance by excising them from the matrix when a resonant triad is found, but proper work on the avoidance/physical significance of them should at some point be conducted.

\subsubsection{Other characteristic admittances}
If instead we were to consider the characteristic admittance of an infinitely long duct with constant curvature, the eigenvalues would remain on the real/imaginary axes, but the eigenvector matrices would cease to be diagonal, resulting in a slightly more complicated expression for the corresponding characteristic admittance $\breve{\mat{Y}}^a$. Including torsion in addition would cause the eigenvalues to bifurcate into the complex plane: as a result the torsional-duct characteristic admittances $\widetilde{\mat{Y}}^{a\pm}$ would no longer satisfy $\widetilde{\mat{Y}}^{a+} = -\widetilde{\mat{Y}}^{a-}$.  For completeness, expressions for these quantities are derived explicitly in appendix~\ref{charadmapp}, although here we will only make use of the straight duct admittance above.

\subsection{Splitting operators}

By using the characteristic admittances defined above, we can split waves into forward- and backward-going parts,
\begin{align}
    &\mvect{u}^a = \mvect{u}^{a+} + \mvect{u}^{a-}, && \mvect{p}^{a} = \mvect{p}^{a+} + \mvect{p}^{a-}.
\end{align}
We may perform this decomposition from the total pressure by defining linear and nonlinear splitting operators $\mat{S}^{a\pm}$ and $\mathcal{S}^{ab\pm}$,
\begin{equation}
    \mvect{p}^{a\pm} = \mat{S}^{a\pm}\mvect{p}^a + \sum_b\mathcal{S}^{ab\pm}\langle\mvect{p}^{a - b},\mvect{p}^b\rangle;
\end{equation}
a similar procedure could be defined for the velocity.  Assuming we have picked a characteristic admittance, we now proceed to calculate these splitting operators in terms of the characteristic admittance.  Restricting ourselves to the linear case for a moment, we have
\begin{equation}
    \mvect{p}^{a\pm} = \mvect{p}^a - \mvect{p}^{a\mp} = \mvect{p}^a - \mat{Z}^{a\mp}(\mvect{u}^a - \mvect{u}^{a\pm}) = (\mat{I} - \mat{Z}^{a\mp}\mat{Y}^a)\mvect{p}^a  + \mat{Z}^{a\mp}\mat{Y}^{a\pm}\mvect{p}^{a\pm},
\end{equation}
so that
\begin{equation}
    \mvect{p}^{a\pm} = (\mat{I} - \mat{Z}^{a\mp}\mat{Y}^{a\pm})^{-1}(\mat{I} - \mat{Z}^{a\mp}\mat{Y}^{a})\mvect{p}^a = \mat{S}^{a\pm}\mvect{p}^a,
\end{equation}
which may equivalently be written more compactly as
\begin{equation}
    \mat{S}^{a\pm} = (\mat{Y}^{a\pm} - \mat{Y}^{a\mp})^{-1}(\mat{Y}^a - \mat{Y}^{a\mp}).
\end{equation}
If the above equations are expanded to second order in terms of $\mat{S}^{a\pm}$, we find $\mathcal{S}^{ab\pm}$ to be
\begin{equation}
    \mathcal{S}^{ab\pm} = (\mat{Y}^{a\pm} - \mat{Y}^{a\mp})^{-1}\bigg(\mathcal{Y}^{ab} - \mathcal{Y}^{ab+}\langle\mat{S}^{(a - b)+},\mat{S}^{b+}\rangle - \mathcal{Y}^{ab-}\langle\mat{S}^{(a - b)-},\mat{S}^{b-}\rangle\bigg).
\end{equation}
\subsection{Pressure Boundary Condition}
\label{pressurebc}
When we come to solve these equations, we prescribe the admittances at the duct's outlet, amounting to a radiation condition, before solving, in order, the equations \eqref{linadmeq} and \eqref{nonlinadmeq}. After this, we know everything about the global radiative properties of the duct, and correspondingly we have only a first-order ODE~\eqref{peq} to solve for the pressure. For musical instruments, the pressure perturbation is prescribed by the musician, who disturbs the internal air column of the instrument at the inlet, meaning that the pressure is specified at the entrance (taken here to be at $s = 0$) and solved forwards from there.

One matter that is not entirely clear is that of the musician's interaction with reflecting waves. Should the musician's playing amount to a prescription of the forward-going pressure only, meaning that waves reflected back impact without consequence on the source? Or should it rather be a prescription of the total pressure, meaning that the musician is adapting to the impact of the reflections they themselves generated?  If the former, we may specify the forward-going pressure at the inlet, $p^+(\pvect{x},t)|_{s = 0}$, with corresponding coefficient $\mvect{p}^{a+}(0)$, and convert this (for the purpose of solving the equation) to the total pressure with the inverse splitting operator,
\begin{equation}
    \mvect{p}^a(0) = (\mat{S}^{a+})^{-1}\left[\mvect{p}^{a+}(0) - \sum_b\mathcal{S}^{ab}\bigg\langle(\mat{S}^{(a - b)+})^{-1}\mvect{p}^{(a - b)+}(0),(\mat{S}^{b+})^{-1}\mvect{p}^{b+}(0)\bigg\rangle\right].
\end{equation}
The simplest boundary condition for the pressure is that of a sinusoidal piston source at $s = 0$, given by $p^+(\pvect{x},t)|_{s = 0} = M\sin(\omega t)$, and projected onto the coordinate basis as
\begin{equation}
\label{pistonprojection}
P_\alpha^{a+}(0) = \frac{M\sqrt{A_\text{cs}(0)}\sgn(a)\delta^{|a|,1}\delta_{\alpha0}}{2\I},
\end{equation}
where $A_\text{cs}(0) = X(0)$ in two dimensions and $\uppi R(0)^2$ in three. We can also consider non-plane modes. In two dimensions, for a variable-width straight duct (with width variation symmetric about the centreline), symmetric and antisymmetric modes are uncoupled, so there is a family of pressure distributions inaccessible to the plane-wave inlet condition. The `fundamental' member of this family is the first antisymmetric mode, so we can consider a variant on condition \eqref{pistonprojection} where the $\delta_{\alpha0}$ is replaced with a $\delta_{\alpha1}$.

This effect is amplified for any variable-width straight-duct in three dimensions, where modes of different azimuthal wavenumber are uncoupled, resulting in uncountably many mutually-inaccessible families of pressure distributions. The `fundamental' of each therefore represents another possible inlet condition, meaning we replace the $\delta_{\alpha0}$ with a $\delta_{p_\alpha n}\delta_{q_\alpha 1}$ for any $n$ of our choice.

\subsection{Numerical Method}\label{sec:numerics}
\subsubsection{Truncation}
When solving this infinite set of coupled ODEs numerically, we must truncate both the number of spatial and temporal modes used. If we pick maximum values of each, we then have spatial modes ranging from $0$ to $\alpha_\text{max}$ and temporal modes ranging from $-a_\text{max}$ to $a_\text{max}$. This would mean that we have $(\alpha_\text{max}+1)(2a_\text{max}+1)$ equations to solve for $P_\alpha^a$, $(\alpha_\text{max}+1)^2(2a_\text{max}+1)$ for $\mat{Y}_{\alpha\beta}^a$ and $(\alpha_\text{max}+1)^3(2a_\text{max}+1)^2$ for $\mathcal{Y}_{\alpha\beta\gamma}^{ab}$. However, by our previous assumptions we are neglecting zero modes, and all of our acoustic quantities being real implies that $p^{-a}_\alpha = p^{a*}_\alpha$ and $u^{-a}_\alpha = u^{a*}_\alpha$, resulting in $(\alpha_\text{max}+1)a_\text{max}$ independent equations for $P_\alpha^a$ and $(\alpha_\text{max}+1)^2a_\text{max}$ for $\mat{Y}_{\alpha\beta}^a$.

A slightly messier calculation at this point puts the number of independent entries of $\mathcal{Y}_{\alpha\beta\gamma}^{ab}$ at $(\alpha_\text{max}+1)^3(2a_\text{max}-1)a_\text{max}$. More entries may yet be discarded though, since discrete convolutions behave strangely upon truncation; if $a$ and $b$ range from $-a_\text{max}$ to $a_\text{max}$ then $a - b$ ranges from $-2a_\text{max}$ to $2a_\text{max}$, so as a consequence, at each timestep, for each possible triple $(\alpha,\beta,\gamma)$, there is an $ab$ matrix $\mathcal{Y}_{\alpha\beta\gamma}^{ab}$ of which the only calculable entries form a \textit{banded} matrix (with bandwidth $a_\text{max}$), resulting in $a_\text{max}(a_\text{max}+1)$ lost entries. Having exploited the conjugacy property already, this means a subtraction of $a_\text{max}(a_\text{max}+1)/2$ entries from the total, resulting in $3(\alpha_\text{max}+1)^3a_\text{max}(a_\text{max}-1)/2$ independent entries for the nonlinear admittance.

\subsubsection{Numerical solver}
We use a 4th order Runge--Kutta method to integrate the admittance and pressure equations in $s$. Unless otherwise specified, we use the adaptive-step solver \texttt{ode45} in \textsc{Matlab}, although for one case in section~\ref{sec:results-horn} we use a fixed-step solver, as that proves more accurate in the neighbourhood of a cusp in the duct diameter.

\subsubsection{Numerical Viscosity}\label{sec:numerical-viscosity}
One of the pitfalls of truncation is the pooling of energy at higher modes. This can be countered by employing a numerical viscosity, which we do by subtracting a new term $E^a$ from the right-hand-side of the pressure equation \eqref{peq}.  Here, we take this to be
\begin{equation}
\label{truncationerrorpostcalc}
    E^a = -\nu_0\frac{|a|\omega\cnon M}{1 + M\cnon\omega s}\log\left(1 - \frac{|a| - 1}{a_\text{max}}\right)P_0^a \sim \nu_0\frac{|a|^2\omega\cnon M}{a_\text{max}}P_0^a,
\end{equation}
where $\nu_0$ is a positive scalar determining how great an effect we want the viscosity to have.  The second approximation here is in the limit of small $s$ relative to the shock formation distance, and with $1 \ll |a| \ll a_\text{max}$ (where energy pooling starts to occur).  This choice of numerical viscosity is informed by the truncation error for a sawtooth plane wave in a one-dimensional duct, and is discussed further in section~\ref{sec:results-straight} below.

The physical molecular viscosity could also instead be considered as a starting point, and comparing the result of one-dimensional numerical sawtooth stabilisation with its physical counterpart, we find that the first-order correction term agrees with the $a^2$ factor from the second approximation, but also that the combined shear and bulk viscosities $4\mu/3 + \zeta$ must be approximately equal to $2\cnon M/a_\text{max}\omega$. This calculation may be found in appendix \ref{numviscapp}.

\subsubsection{Alternative numerical damping}
\label{altnumdamp}
Numerical instabilities can also arise in linear problems. If a node occurs in the pressure (scenarios involving this are discussed in section \ref{inverseexpresults}), this causes a singularity in the admittance. Such singularities may be avoided (or `dampened') by the addition of a small imaginary part to the frequency $\omega$, which corresponds to a sound source slowly exponentially growing in time, and therefore a wave pattern that slowly exponentially decays in space away from the source.  This technique is borrowed from stability analysis~\citep[e.g.][]{briggs-1964,bers-1983}, where adding an imaginary part to the frequency $\omega$ is used to guarantee a causal solution.

\subsection{Modal resolution}

The following test cases require different numbers of modes depending on what physics is being considered. For linear acoustics, spatial modes are the only limiting factor: when comparing with pre-existing linear work, we use the previous resolution as a benchmark (e.g. $\alpha_\text{max} = 30$ for figure \ref{FP2002_fig2}), but when comparing with  results, we pick a new standard (taking $\alpha_\text{max} = 50$ for figures \ref{mctavfig5}, \ref{webstercomp}, \ref{padmtile}, \ref{testcasecomp.pdf}). In all of these cases the absence of nonlinearity allows us to set $a_\text{max} = 1$, and unoptimised \textsc{Matlab} code run on a standard laptop finishes in the order of seconds.

Where nonlinearity comes in, memory requirements are likely to exceed that of a standard laptop, although all results here have been computed on a single desktop with $128\,\mathrm{GiB}$ of memory. For most of these computations (i.e.\ figures \ref{McTavBramb_fig7} and \ref{McTav_fig3.6.pdf}-\ref{curvaturemachtiles.pdf}) spatial resolution is necessary, so we take $a_\text{max} = 10$ and $\alpha_\text{max} = 10$, and computations take on the order of tens of minutes. For figure \ref{blackstocktiles.pdf}, we consider only plane waves in a straight duct, so we can take $\alpha_\text{max} = 0$ and $a_\text{max} = 100$; while this does not exceed the memory limit, the calculation now takes of the order of half a day to compute using a single core.

The numerical procedure converges as the number of temporal and spatial modes is increased, and the truncations above are sufficient for accurate results in all of the cases presented here.  A demonstration of the numerical convergence with the number of modes is provided for a constant-curvature bend in three dimensions in section~\ref{sec:results-bend3D} below.

\section{Results}\label{sec:results}
We now consider applying the procedure above to a number of test cases.  The numerical \textsc{Matlab} code to compute these cases is provided in the supplementary material.

\subsection{A straight duct of constant diameter}\label{sec:results-straight}
The simplest test geometry in either two or three dimensions is that of a straight duct with constant diameter. For a plane piston source, the equations are simplified greatly, both by the absence of curvature and width-variation terms, and also by the collapsing of the spatial mode vectors onto a single scalar component. The admittance equations are then solved trivially, since the boundary condition is a constant solution, giving 
\begin{align}
    &\mat{Y}_{00}^a(s) = 1,&& \mathcal{Y}_{000}^{ab}(s) = -\frac{1}{2\sqrt{A_\text{cs}}}\left(\cnon + \frac{a - 2b}{a}\right)\quad \forall s.
\end{align}
This holds in both two and three dimensions, and we write $A_\text{cs}$ for the area of the duct cross-section so that we can work in generality.
Plugging these into the pressure equation, we get (again in both cases, once we substitute the value of $\mathcal{I}_{000}$)
\begin{equation}
    \frac{\intd P_0^a}{\intd s} = \I a \omega P_0^a + \frac{\I\omega}{2\sqrt{A_\text{cs}}}\sum_b\bigg[-a\cnon + (a - 2b)\bigg]P_0^{a - b}P_0^b,
\end{equation}
but the second term in the square bracket here vanishes upon summation. Since the zero mode $\psi_0$ does not depend on $s$ for a straight duct, we can multiply through by it, getting an equation for Fourier coefficients
\begin{equation}
\label{fouriereqstraight}
    \frac{\intd P^a}{\intd s} = \I a\omega P^a - \frac{\I a\omega\cnon}{2}\sum_bP^{a - b}P^b.
\end{equation}
This equation was solved up to shock formation by \cite{fubini} (for any boundary conditions both periodic and odd in t), in the form of a sine series for the pressure
\begin{align}
    &\frac{p}{M} = \sum_aB_a\sin[a\omega(t - s)];&& B_a = \frac{2}{a\sigma}J_a(a\sigma),
\end{align}
where $\sigma = M\cnon\omega s$ is the arc-length normalised by the shock formation distance. \cite{fay} found a post-shock-formation solution in the form of a sawtooth wave,
\begin{equation}
    B_a = \frac{2}{a(1 + \sigma)},
\end{equation}
valid for $\sigma \gtrapprox 3$. \cite{blackstock} matched these two solutions with the following two terms
\begin{equation}
    B_a = \frac{2}{a\uppi}P_{\text{sh}} + \frac{2}{a\uppi\sigma}\int_{\Phi_\text{sh}}^\uppi\cos a(\Phi - \sigma\sin\Phi)\mathrm{d}\Phi,
\end{equation}
where the shock amplitude $P_\text{sh}$ and phase $\Phi_\text{sh}$ satisfy
\begin{align}
    &P_\text{sh} = \sin(\sigma P_\text{sh}),&& \Phi_\text{sh} = \sigma\sin\Phi_\text{sh}.
\end{align}
For $\sigma < 1$, the only solution for $(P_\text{sh},\Phi_\text{sh})$ is $(0,0)$, so the first term vanishes and the second becomes the Fubini solution. As $\sigma$ grows greater than 1, two more solutions appear either side of 0; the relevant solutions here are the positive ones. For $\sigma \gtrapprox 5\uppi/2$ more solutions appear; we remain interested from this point in the minimal positive solution for each of $P_\text{sh}$ and $\Phi_\text{sh}$. Since $\Phi_\text{sh}$ converges to $\uppi$ from below, the second term tends to 0, while the first term behaves like the Fay solution.

The numerical viscosity used here, described in section~\ref{sec:numerical-viscosity} above, is justified by considering the truncation error of a sawtooth wave when substituted into equation \eqref{fouriereqstraight}. The sawtooth wave as derived in \cite{fay} has Fourier coefficients given by
\begin{equation}
    P^a = \frac{\I Me^{\I a\omega s}}{a(1 + \sigma)}
\end{equation}
so the terms missing if we truncate the nonlinear term in equation~\eqref{fouriereqstraight} to a finite sum are then
\begin{equation}
\label{truncationerrorprecalc}
    \begin{aligned}
        E^a &= \frac{\I a\omega\cnon \sqrt{A_\text{cs}}}{2}\sum_{\substack{b = -\infty,\\b \neq 0,a}}^\infty P^{a - b}P^b - \frac{\I a\omega\cnon \sqrt{A_\text{cs}}}{2}\sum_{\substack{b = a - \sgn(a)a_{\text{max}},\\b \neq 0,a}}^{\sgn(a)a_{\text{max}}}P^{a - b}P^b \\
        &= -\frac{\I\omega\cnon \sqrt{A_\text{cs}}M^2e^{\I a\omega s}}{(1 + \sigma)^2}\Bigg(\sum_{\substack{b = -\infty,\\b \neq 0,a}}^\infty - \sum_{\substack{b = a - \sgn(a)a_{\text{max}},\\b \neq 0,a}}^{\sgn(a)a_{\text{max}}}\Bigg)\frac{a}{2(a - b)b}
    \end{aligned}
\end{equation}
where the limits on the second sum ensure that for both positive and negative $b$, $|a - b|$ may never exceed $a_{\text{max}}$. By carefully splitting the summands into partial fractions, which make the sums easier to compute, we gain a form which has integral bounds that we use to calculate the numerical viscosity. This is done in detail in appendix \ref{numviscapp}, and results (with the inclusion of the viscous scale factor $\nu_0$) in equation \eqref{truncationerrorpostcalc}.

\begin{figure}
    \centering
    \includegraphics[]{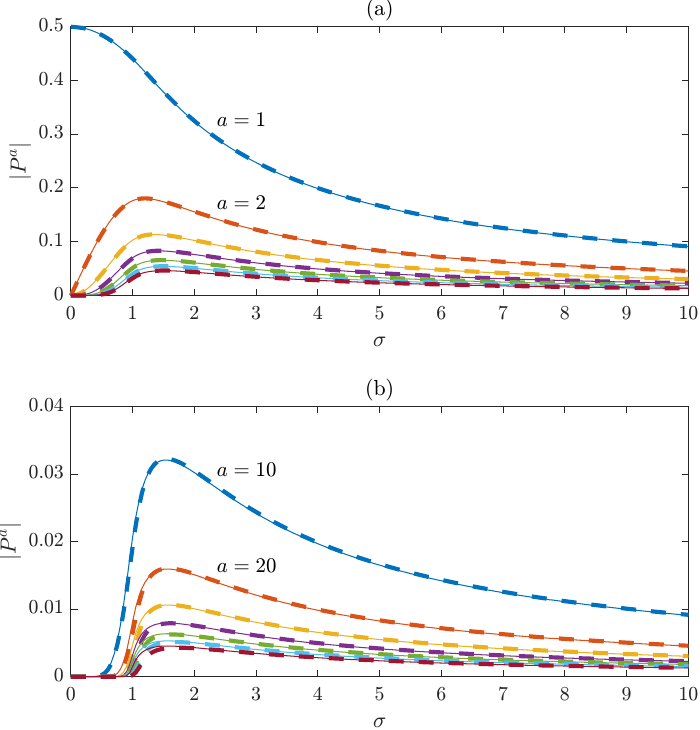}
    \caption{Comparison of numerically-calculated mode amplitudes (dashed) with those predicted by the Blackstock solution (solid), using the code in two dimensions. (a) shows modes 1 through 7, while (b) shows 10 through 70. Truncation was taken at $\alpha_\text{max} = 0$ and $a_\text{max} = 100$.}
    \label{blackstocktiles.pdf}
\end{figure}

We can compare our numerics with this. This can be done in both two and three dimensions; figure \ref{blackstocktiles.pdf} compares various modes in the two-dimensional case, for a simulation with 1 spatial mode (since no spatial coupling is induced for this geometry anyway) and 100 temporal modes. As one would expect, good matches are achieved for the lower (dominant) modes, while for higher modes the numerical solutions undershoot at first and over-compensate later on. We have achieved matches of equal quality with the code in three dimensions.

Eliminating spatial variation in favour of higher temporal modes is a useful special case because it allows for the calibration of the numerical viscosity. Figure \ref{blackstocktiles.pdf} was created with $\nu_0 = 1$, because higher scale factors than this place far too much damping on the higher modes, while lower factors allow instabilities to build up much more quickly. In light of this, all other nonlinear calculations take $\nu_0 = 1$ as well.

\subsection{Constant curvature in two dimensions}\label{sec:results-bend}
\begin{figure}
    \centering
    \includegraphics{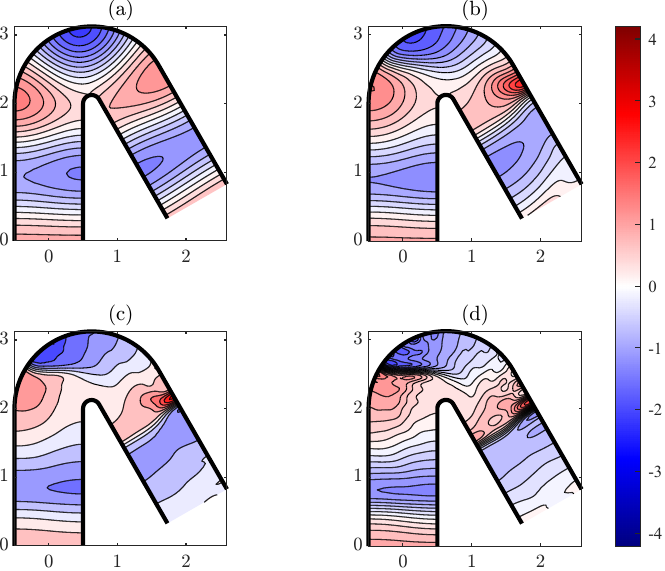}
    \caption{Pressure (normalised to the source amplitude) in a bend identical to the one used in \cite{mctavish+brambley-2019}, with a plane piston source of frequency $3/X$, for (a) linear, (b) $M$ = 0.05, (c) 0.10, and (d) 0.15. Truncation was taken at $\alpha_\text{max} = a_\text{max} = 10$. An animated version of this figure is available as Movie~1 in the supplementary material.}
    \label{McTavBramb_fig7}
\end{figure}

In two dimensions, \citet{mctavish+brambley-2019} considers linear and nonlinear propagation around a constant-width bend, first computed in the linear case by~\citet{felix1}.  This provides a test case for the combination of curvature and nonlinearity. Figure~\ref{McTavBramb_fig7} reproduces figure~7 of \citet{mctavish+brambley-2019}, with all of the same parameters, i.e.\ (for constant duct width $X$) a bend of curvature $8/5X$ located at a distance $2X$ downstream of a plane piston source of frequency $3/X$, with truncation taken at 10 spatial modes and 10 temporal modes, plotted in the linear regime and nonlinearly for $M = 0.05$, 0.10 and 0.15. Good agreement is observed.

\subsection{Constant curvature in three dimensions}\label{sec:results-bend3D}
\begin{figure}
    \centering
    \includegraphics{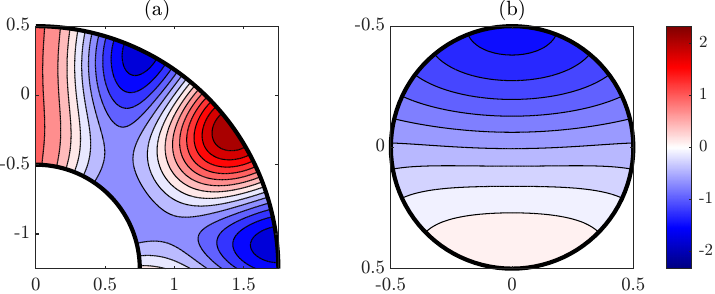}
    \caption{Pressure (normalised to the source amplitude) in a bend identical to the one used in \cite{felix2}, with a plane piston source of frequency $2.4/R$, plotted both through the midplane and across the duct outlet in the linear regime. Truncation was taken at $\alpha_\text{max} = 30$ and $a_\text{max} = 1$. An animated version of this figure is available as Movie~2 in the supplementary material.}
    \label{FP2002_fig2}
\end{figure}

In three dimensions, no published work exists that can be used to test the combined effects of curvature and nonlinearity, but in the linear case \cite{felix2} provides an example that may be tested against. Here, for constant duct radius $R$, a plane piston source of frequency $2.4/R$ is placed at the mouth of a bend of curvature $4/5R$, and a cross-section through the half-plane is plotted alongside the outlet's circular cross-section. Comparing figure \ref{FP2002_fig2} with figure 2 of \cite{felix2}, we again see good agreement. Truncation was taken at 30 spatial modes to match their calculation, with the usual 1 temporal mode in the absence of nonlinearity.

\begin{figure}
    \centering
    \includegraphics[]{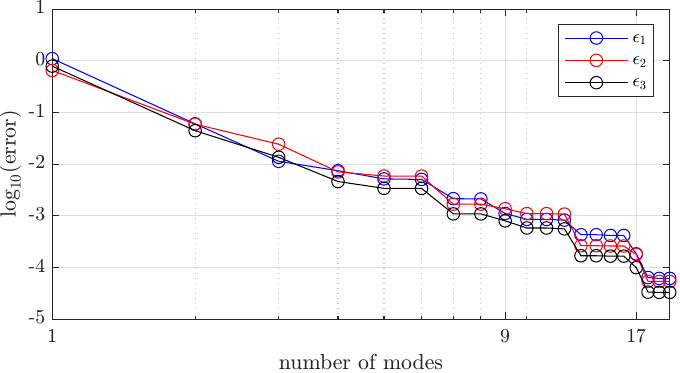}
    \caption{Numerical convergence in a three-dimensional planar bend for three definitions of the error, detailed in equation \eqref{errordefs}. $a_\mathrm{max}=1$, and so the number of modes is $\alpha_\text{max}+1$.}
    \label{convergenceplot.pdf}
\end{figure}

\begin{figure}
    \centering
    \includegraphics{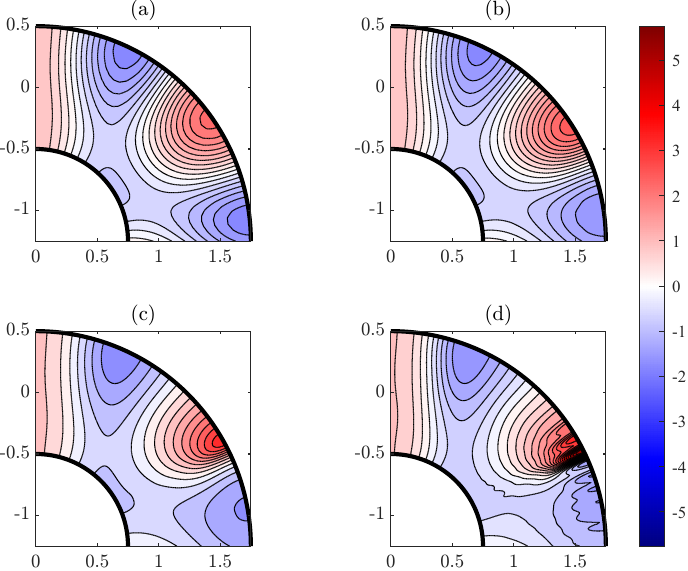}
    \caption{Pressure (normalised to the source amplitude) in a bend identical to the one used in \cite{felix2}, with a plane piston source of frequency $2.4/R$, for (a) linear, (b) $M$ = 0.02, (c) 0.05, and (d) 0.10. Truncation was taken at $\alpha_\text{max} = a_\text{max} = 10$. An animated version of this figure is available as Movie~3 in the supplementary material.}
    \label{McTav_fig3.6.pdf}
\end{figure}

For this geometry, we also follow \cite{felix2} in the inclusion of a convergence study for the spatial modes. As well as their two definitions of the error involving integration along a surface, we also calculate a volumetric error, which behaves similarly. $p_\text{ref}$ was taken here with $\alpha_\text{max} = 20$ and $a_\text{max} = 1$. Figure \ref{convergenceplot.pdf} shows the result. As with \cite{felix2}, there is uneven behaviour due to the different effects of higher azimuthal or radial resolution; nonetheless, the inclusion of more modes never causes the error to grow. The error definitions are
\begin{subequations}
\label{errordefs}
    \begin{equation}
        \epsilon_1 = \sqrt{\frac{\int_{t = 0}^{2\uppi/\omega}\int_{s = 0}^{s_d}\int_{r = 0}^R\left[h_s\|p - p_\text{ref}\|^2\right]_{\theta = 0}^{\uppi}\mathrm{d}r\mathrm{d}s\mathrm{d}t}{\int_{t = 0}^{2\uppi/\omega}\int_{s = 0}^{s_d}\int_{r = 0}^R\left[h_s\|p_\text{ref}\|^2\right]_{\theta = 0}^{\uppi}\mathrm{d}r\mathrm{d}s\mathrm{d}t}},
    \end{equation}
    \begin{equation}
        \epsilon_2 = \sqrt{\frac{\int_{t = 0}^{2\uppi/\omega}\int_{\theta = 0}^{2\uppi}\int_{r = 0}^Rr\|p - p_\text{ref}\|^2|_{s = s_d}\mathrm{d}r\mathrm{d}\theta\mathrm{d}t}{\int_{t = 0}^{2\uppi/\omega}\int_{\theta = 0}^{2\uppi}\int_{r = 0}^Rr\|p_\text{ref}\|^2|_{s = s_d}\mathrm{d}r\mathrm{d}\theta\mathrm{d}t}},
    \end{equation}
    \begin{equation}
        \epsilon_3 = \sqrt{\frac{\int_{t = 0}^{2\uppi/\omega}\int_{\theta = 0}^{2\uppi}\int_{r = 0}^R\int_{s = 0}^{s_d}rh_s\|p - p_\text{ref}\|^2|_{s = s_d}\mathrm{d}s\mathrm{d}r\mathrm{d}\theta\mathrm{d}t}{\int_{t = 0}^{2\uppi/\omega}\int_{\theta = 0}^{2\uppi}\int_{r = 0}^R\int_{s = 0}^{s_d}rh_s\|p_\text{ref}\|^2|_{s = s_d}\mathrm{d}r\mathrm{d}s\mathrm{d}\theta\mathrm{d}t}}.
    \end{equation}
\end{subequations}
In figure \ref{McTav_fig3.6.pdf} the same geometry has a plane piston source of frequency $\omega = 2.4/R$ prescribed in the total pressure at the top left, for various Mach numbers. We see good agreement both with the published linear work of \cite{felix2}, where evenly-spaced peaks and troughs travel around the outside of the bend, and also with the unpublished nonlinear work of \cite{mctavphd}, where the contours are deformed by the steepening of the peaks as the Mach number is increased. To match the latter calculation, 10 spatial modes and 10 temporal modes were used.

\subsection{An exponential horn}\label{sec:results-horn}
The exponential horn is a convenient geometry: if an approximation is made allowing only plane waves to propagate, an analytical solution \citep{webster} exists, with the growth rate of the horn acting to dampen oscillations. The geometry is best parametrised across both two and three dimensions by specifying the growth exponent of the horn's cross-sectional area, i.e. by setting
\begin{equation}
    A_\text{cs}(s) = \begin{cases}
        A_\text{cs}(0)\exp(2ms)\quad &s\in[0,s_\mathrm{o}],\\
        A_\text{cs}(0)\exp(2ms_\mathrm{o}) &s>s_\mathrm{o}.
    \end{cases}
\end{equation}
so that the radial growth factor in two dimensions is doubled relative to that in three. The section of duct with $s > s_\mathrm{o}$ forms the infinite straight duct providing the characteristic admittance boundary condition $\mat{Y}^a = \mat{Y}^{a+}$ at the outlet $s=s_0$. Since we are only interested in plane waves, the (scalar) linear admittance equation (in the variable radius region) becomes
\begin{equation}
    \frac{\intd Y^a}{\intd s} = -\I a\omega (Y^a)^2 - 2mY^a + \I a\omega,
\end{equation}
with the boundary condition at $s = s_\mathrm{o}$ being $Y^a = 1$ (which also holds for all $s > s_\mathrm{o}$). Solving this ODE, we find that
\begin{equation}
    Y^a = \frac{n_a\cos n_a(s_\mathrm{o} - s) - (\I a\omega - m)\sin n_a(s_\mathrm{o} - s)}{n_a\cos n_a(s_\mathrm{o} - s) - (\I a\omega + m)\sin n_a(s_\mathrm{o} - s)},
\end{equation}
for $s \leq s_\mathrm{o}$. Here $n_a$ is a quantity satisfying $n_a^2 = a^2\omega^2 - m^2$, which is real in the case of cut-on modes and imaginary otherwise. Having found $Y^a$, it may now be plugged into the equation for the pressure, which in the plane-wave case reduces to
\begin{equation}
    \frac{\mathrm{d}P_0^a}{\mathrm{d}s} = \begin{cases}
        (m + \I a\omega Y^a)P_0^a \quad &s \leq s_\mathrm{o}, \\
        \I a\omega P_0^a &s > s_\mathrm{o},
    \end{cases}
\end{equation}
with our initial condition being
\begin{equation}
\label{webstertotalplanewaveBC}
    P_0^a = \frac{M\sqrt{A_\text{cs}(0)}}{2\I}\delta^{|a|1}\sgn(a),
\end{equation}
or alternatively
\begin{equation}
    P_0^a = \frac{M\sqrt{A_\text{cs}(0)}}{2\I}\delta^{|a|1}\sgn(a)\frac{n_a\cos n_as_\text{o} - (\I a\omega + m)\sin n_as_\text{o}}{n_a\cos n_as_\text{o} - \I a\omega\sin n_as_\text{o}}
\end{equation}
if we wish to specify only the forward-going pressure. Solving this equation, with condition \eqref{webstertotalplanewaveBC}, the pressure is
\begin{equation}
    P_0^a(s) = \frac{M\sgn(a)\delta^{|a|,1}}{2\I}\sqrt{A_\text{cs}(0)}\frac{n_a\cos n_a(s_\text{o} - s) - (\I a\uomega + m)\sin n_a(s_\text{o} - s)}{n_a\cos n_as_\text{o} - (\I a\omega + m)\sin n_as_\text{o}},
\end{equation}
and if we specify the forward-going pressure instead, the $m$ disappears from the bracketed term in the denominator.

\begin{figure}
    \centering
    \includegraphics[]{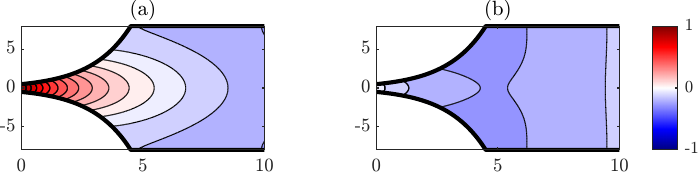}
    \caption{Pressure (normalised to the source amplitude) in an exponential horn identical to the one used in \cite{mctavish+brambley-2019}, with a plane piston source of frequency $0.95\overline{\omega}_1^1(s_\text{o}) = 0.95\uppi/X(s_\text{o})$, for (a) linear, (b) linear, quarter of a cycle later. Truncation was taken at $\alpha_\text{max} = 50$ and $a_\text{max} = 1$. An animated version of this figure is available as Movie~4 in the supplementary material.}
    \label{mctavfig5}
\end{figure}

To test our code on this geometry, we follow \cite{mctavish+brambley-2019} in considering a two-dimensional horn of inlet width $X_\text{i}$, length $4.5X_\text{i}$ and width increase ratio of 16, and a plane piston source of frequency of $0.95\overline{\omega}_1^1(s_\text{o}) = 0.95\uppi/X(s_\text{o})$ for the inlet condition on the total pressure (where we recall the definition of the cutoff frequency from equation \eqref{cutofffreqdef}). With 50 spatial modes and 1 temporal mode (since with no nonlinearity there is no temporal coupling) a good match is observed with their results in the linear case (figure \ref{mctavfig5}).

\begin{figure}
    \centering
    \includegraphics[]{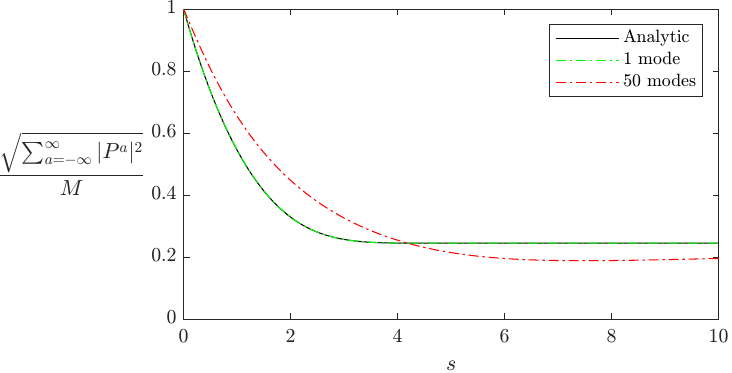}
    \caption{RMS pressure along the centreline of the exponential horn for two different modal resolutions, compared with the plane-wave approximation \citep{webster}. Mode coupling induced by the variation in duct width causes the plane-wave approximation to be an inaccurate one for this geometry.}
    \label{webstercomp}
\end{figure}

Further to this, a more direct comparison with Webster's analytic approximation is shown in figure \ref{webstercomp}, where the RMS pressure at the centreline is plotted (in contrast to \cite{mctavish+brambley-2019}, the analytic solution is calculated for the whole domain, including the straight-duct section post-outlet). We achieve a very good match by restricting our calculation to plane waves only: this is achieved with a fixed (and very small)-stepsize method, since the discontinuity in $X'$ at the outlet is severe enough to cause errors in a variable stepsize solver like \texttt{ode45}. In contrast, the 50-mode calculation disagrees very obviously with the Webster approximation, showing that the Webster approximation will necessarily fail to capture the spatial coupling effects of a duct bell.

\subsection{An inverse exponential horn in two-dimensions}
\label{inverseexpresults}
If we consider an `inverse' exponential horn, that is, one that narrows rather than widening, we introduce the possibility of nodes. This is because waves are now reflecting to a much greater degree, so cancellations can occur. This effect is most readily observed with antisymmetric waves (and in fact cannot occur in a straight duct when considering only plane waves). Since antisymmetric waves are uncoupled from symmetric ones in a straight duct, there are grounds to consider the derivation of an analogous Webster-like solution to the first antisymmetric mode. We will solve only the two-dimensional problem here, since the three-dimensional problem is complicated by a greater number of ways to break azimuthal symmetry. As with the Webster solution, this discussion also covers only linear acoustics in this geometry.

When the Webster solution is derived, we consider a single-mode truncation of equation \eqref{LandNintro} for straight ducts of variable width (whose centreline is always in the middle of the duct, so $X_+ = - X_- = X/2$) in the linear regime. As such, we have no $\mathcal{N}^{ab}$, and $\mat{L}^a$ (which for $a = 1$ we call $\mat{L}$) reads
\begin{equation}
    \mat{L} = \begin{pmatrix}
        -\frac{X'}{2X}\left(\mat{W} - \widetilde{\mat{A}}\right) &\I\omega\left(\mat{I} - \frac{\gmat{\Lambda}^2}{\omega^2X^2}\right) \\
        \I\omega\mat{I} &\frac{X'}{2X}\left(\mat{W}^T - \widetilde{\mat{A}}^T\right)
    \end{pmatrix}.
\end{equation}
The non-coupling between symmetric and antisymmetric waves is encoded here by the fact that the matrix $\mat{W} - \widetilde{\mat{A}}$ has non-zero entries only where the column index and row index are either both even (antisymmetric) or both odd (symmetric). In particular, this results in the following 2-mode truncation of equation \eqref{LandNintro}
\begin{equation}
\label{2modetruncation}
    \frac{\intd}{\intd s}\begin{pmatrix}
        u_0 \\ u_1 \\ p_0 \\ p_1
    \end{pmatrix} = \begin{pmatrix}
        -\frac{X'}{2X}\begin{pmatrix}
            1 &0 \\ 0 &2
        \end{pmatrix} &\I\omega\begin{pmatrix}
            1 &0 \\ 0 &1 - \frac{\uppi^2}{\omega^2X^2}
        \end{pmatrix} \\
        \I\omega\begin{pmatrix}
            1 &0 \\ 0 &1
        \end{pmatrix} &\frac{X'}{2X}\begin{pmatrix}
            1 &0 \\ 0 &2
        \end{pmatrix}
    \end{pmatrix}\begin{pmatrix}
        u_0 \\ u_1 \\ p_0 \\ p_1
    \end{pmatrix}.
\end{equation}
Solving the symmetric problem $(u_0,p_0)$ results in the Webster Horn Equation, while solving the antisymmetric problem $(u_1,p_1)$ results in the following equation for $p_1$ (from which we henceforth drop the subscript 1, as with $u_1$)
\begin{equation}
\label{antisympressureeq}
    \frac{\intd^2p}{\intd s^2} - \left[4m^2 - \omega^2 + \frac{\uppi^2}{X_\text{i}^2}\e^{4|m|s}\right]p = 0,
\end{equation}
where for inlet width $X_\text{i}$ we have now substituted in the two-dimensional inverse exponential horn definition $X := X_\text{i}\e^{-2|m|s}$. If we use the substitution $\sigma = \uppi\e^{2|m|s}/2|m|X_\text{i}$, this reduces to the modified Bessel equation
\begin{equation}
    \sigma^2\frac{\intd^2p}{\intd \sigma^2} + \sigma\frac{\intd p}{\intd\sigma} - (\nu^2 + \sigma^2)p = 0,
\end{equation}
where $\nu$ is defined by $\nu^2 = 1 - \omega^2/4m^2$, and takes purely real values or purely imaginary values, depending on whether $\omega$ exceeds $2|m|$. This equation has two solutions, $\Ib_\nu(\sigma)$ and $\K_\nu(\sigma)$. In appendix \ref{inverseexpapp} we derive the solution for the pressure in detail, finding that $\K_\nu(\sigma)$ dominates for most of the duct. Nodes are then found at roots of this function, i.e. roots $s_\text{node}$ of $\K_{\I\sqrt{\omega^2/4m^2 - 1}}(\uppi\e^{2|m|s}/2|m|X_\text{i})$.

\begin{figure}
    \centering
    \includegraphics[]{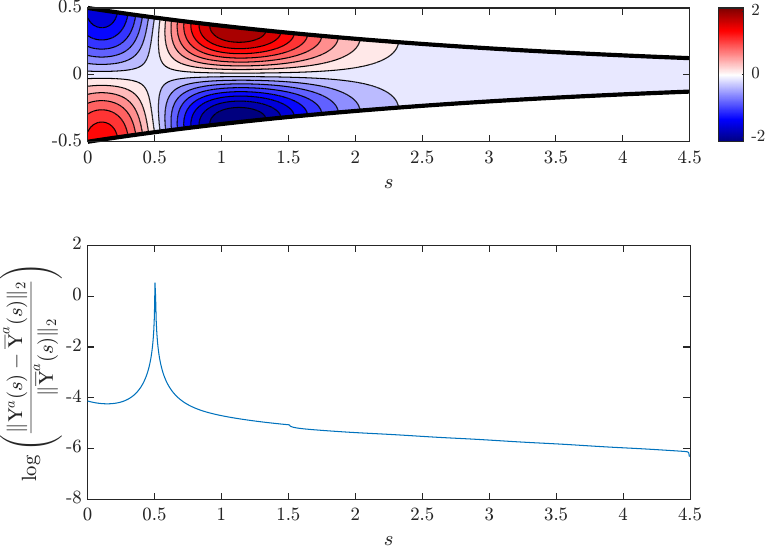}
    \caption{Pressure (top, normalised to the source amplitude) inside an inverse exponential horn, with an antisymmetric source of frequency $(5 + 0.01\I)/X_\text{i}$, together with the normalised deviation of the admittance from the characteristic admittance(bottom). Truncation was taken at $\alpha_\text{max} = 50$ and $a_\text{max} = 1$. An animated version of this figure is available as Movie~5 in the supplementary material.}
    \label{padmtile}
\end{figure}

We can test our code on this geometry as well. To form a good physical picture of where the admittance singularities are, we may exploit the lack of coupling between symmetric and antisymmetric modes in this geometry to eliminate the plane-wave component from the solution altogether. To do this we make use of the alternative inlet condition mentioned in section \ref{pressurebc}. Figure \ref{padmtile} shows the pressure distribution lined up with a demonstration of the admittance singularity in this geometry. This was calculated with 50 spatial modes and 1 temporal mode (since we are testing just linearity here), for a duct of length $4.5X_\text{i}$ and decrease ratio of 4. As mentioned in section \ref{altnumdamp} we add a small imaginary part to the frequency to `dampen' any singularities. The frequency is then $(5 + 0.01\I)/X_\text{i}$, resulting in a value of $\Imag(\nu)$ high enough that the first root of $\K_\nu(\sigma)$ is within the duct domain, causing a node in the pressure, observable a little beyond $s = 0.5$.

\begin{figure}
    \centering
    \includegraphics[]{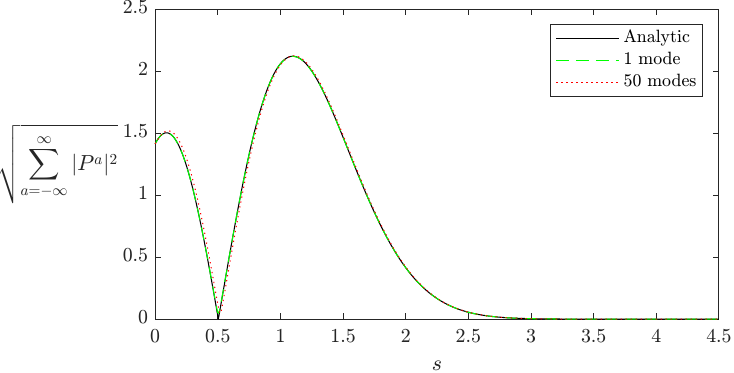}
    \caption{RMS pressure along the wall of the exponential horn for two different modal resolutions, compared with the analytic solution.}
    \label{testcasecomp.pdf}
\end{figure}

Figure \ref{testcasecomp.pdf} compares the results of a single-spatial-mode calculation with the 50-mode calculation (the same frequency, and therefore the same stabilising imaginary part, were used on the single-mode calculation). We find that the coupling is very minimal: while the 1-mode line matches the analytic solution exactly, the 50-mode line is a lot closer than it was in the case of the Webster plane-wave solution. We conclude from this that antisymmetric sources induce less coupling than symmetric plane-wave ones.

\subsection{Curvature and width variation combined}\label{sec:results-curvature-and-width}
We can use the inverse exponential horn to demonstrate mode coupling. In the linear case, if we send in an antisymmetric source and the duct is straight, only antisymmetric modes will be excited. This symmetry can be broken in multiple ways. If nonlinearity is included, higher-order temporal harmonics will be coupled, and this can include symmetric modes (a physical argument for this is that the nonlinear terms are quadratic in the pressure, so two odd pressures will multiply to create an even one). Alternatively symmetric modes of the same temporal order can be picked up if the originally straight duct is bent. Symmetric modes can include plane waves, which have the property of always being cut-on, so we see that for an antisymmetric source of a certain frequency, propagating waves will only `escape' the duct if a) they are allowed to steepen or b) the duct is bent.

\begin{figure}
    \centering
    \includegraphics{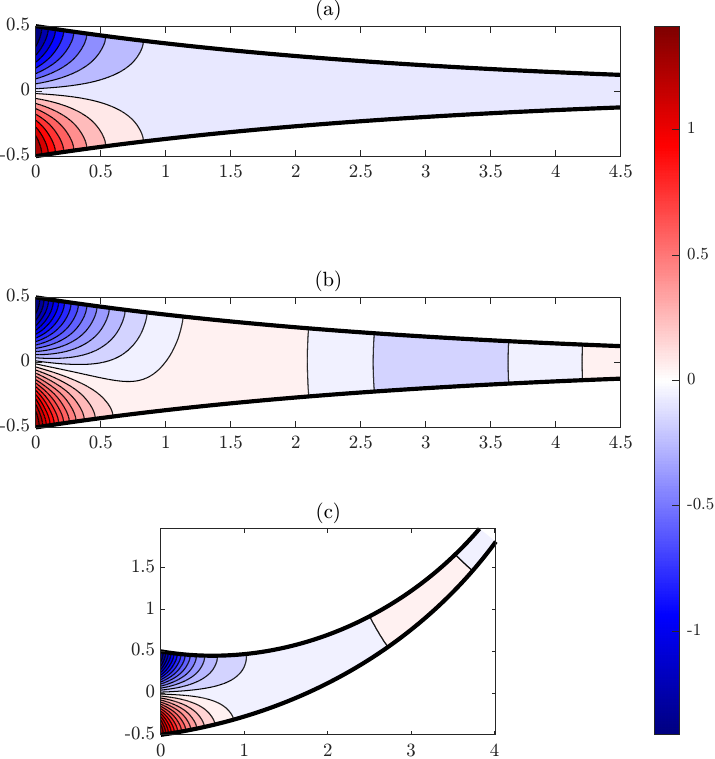}
    \caption{Pressure (normalised to the source amplitude) for an antisymmetric source of frequency $1/X_\text{i}$ in an inverse exponential horn for (a) linear, straight, (b) $M=0.05$, straight, and (c) linear, $\kappa = 0.2/X_\text{i}$. Truncation was taken at $\alpha_\text{max} = a_\text{max} = 10$. An animated version of this figure is available as Movie~6 in the supplementary material.}
    \label{planewavetunnelling.pdf}
\end{figure}

Figure \ref{planewavetunnelling.pdf} displays this phenomenon of `acoustic leakage', for the same inverse exponential horn as in figure \ref{padmtile}. An antisymmetric inlet source of frequency $1/X_\text{i}$ is used: this ensures that all non-plane waves are cut-off at the outlet. We see plane-wave tunnelling induced both for a straight duct with Mach number 0.05, or in the linear regime with a curvature of $0.2/X_\text{i}$. Truncation was taken at 10 spatial modes and 10 temporal modes for each of these calculations.

\subsection{Torsion}\label{sec:results-torsion}
\begin{figure}
    \centering
    \includegraphics{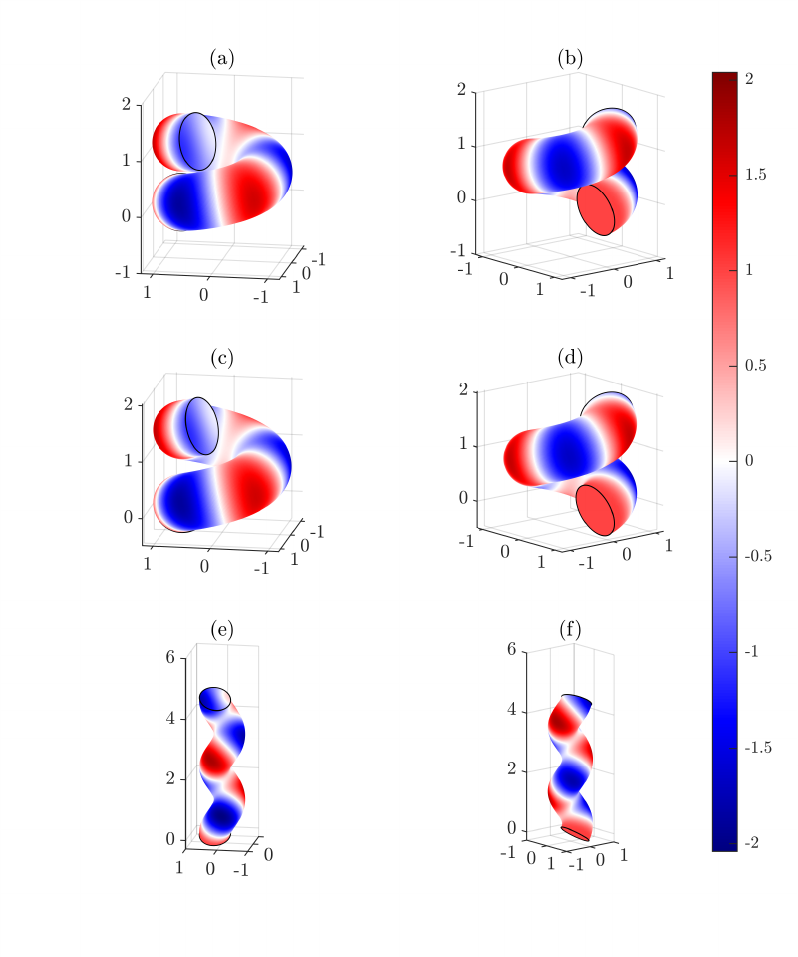}
    \caption{Linear pressure field (normalised by the Mach number) on the walls of a helical duct for the three torsions $\tau R = 0.16$, $0.20$ and $1.00$, with a plane piston source of frequency $\omega = 0.95\overline{\omega}_1^1 = 0.95 \times 1.8412/R$, viewed from two different angles in each case. Truncation was taken at $\alpha_\text{max} = a_\text{max} = 10$. An animated version of this figure is available as Movie~7 in the supplementary material.}
    \label{helixlin.pdf}
\end{figure}

\begin{figure}
    \centering
    \includegraphics{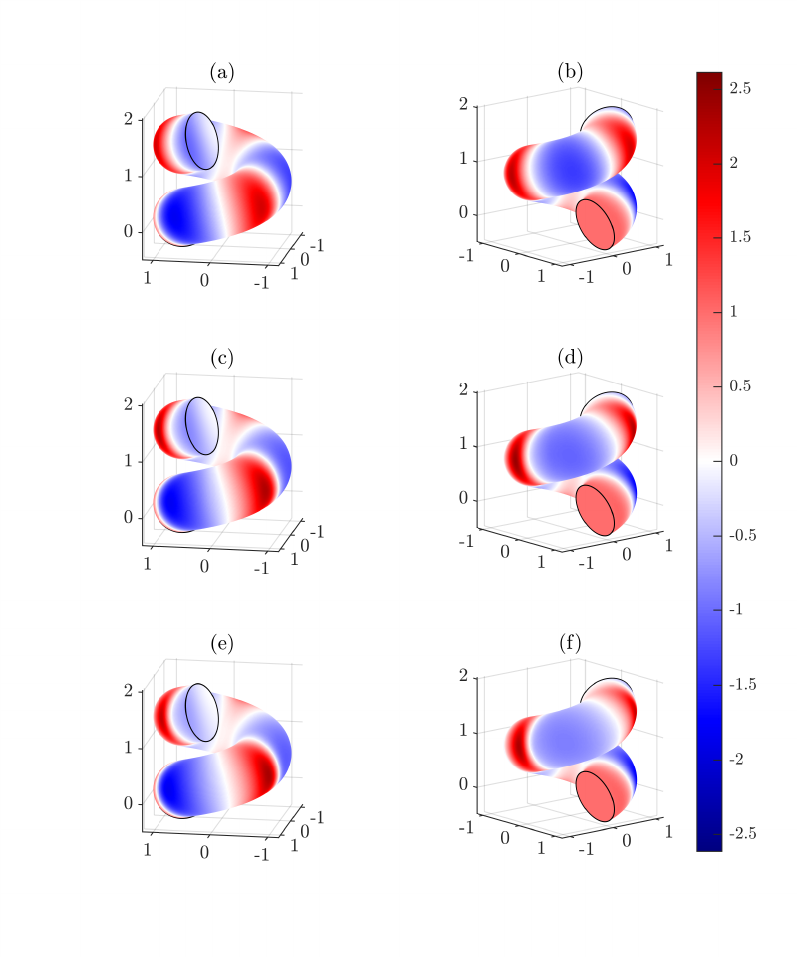}
    \caption{Pressure field (normalised by the Mach number) on the walls of a helical duct with torsion $\tau R = 0.20$, for $M = 0.05$ (a,b), 0.10 (c,d), and 0.15 (e,f), with a plane piston source of frequency $\omega = 0.95\overline{\omega}_1^1 = 0.95 \times 1.8412/R$. Truncation was taken at $\alpha_\text{max} = a_\text{max} = 10$. An animated version of this figure is available as Movie~8 in the supplementary material.}
    \label{tau0.20nonlin.pdf}
\end{figure}

Our framework allows us to investigate the effects of torsion on both linear and nonlinear acoustics. To this end, we examine helical ducts of constant wall radius $R$ and curvature $\kappa$, with various torsions $\tau$. For each helix, we have a duct centreline given by
\begin{equation}
    \pvect{q}(s) = (a\cos\tilde{s},a\sin\tilde{s},b\tilde{s}),\end{equation}
where $a$ is the helical radius, $2\uppi b$ is the \textit{pitch} of the helix, and $\tilde{s}$ is a transformed arclength. These are given in terms of $\kappa$ and $\tau$ by
\begin{align}
    a = \frac{\kappa}{\kappa^2 + \tau^2}, &&b = \frac{\tau}{\kappa^2 + \tau^2}, &&\tilde{s} = \sqrt{\kappa^2 + \tau^2}s.
\end{align}
The binormal is calculated as $\pvect{b} = \sqrt{\kappa^2 + \tau^2}(b\sin\tilde{s},-b\cos\tilde{s},a)$. At each point along the centreline, this vector points upward with a degree of backward tilt parallel to the centreline to make it perpendicular to the tangent vector. The angle between $\pvect{b}$ and the vertical is calculated with the dot product to be
\begin{equation}
    \cos\theta_{\pvect{b}} = \pvect{b}\cdot\pvect{e}_z = a\sqrt{\kappa^2 + \tau^2},
\end{equation}
so if we wish to know the vertical distance from a point $\tilde{s}_0$ to the duct wall above, we consider the hypotenuse of a right-angled triangle with adjacent side of length $R$ and an angle of $\theta_{\pvect{b}}$ between. This distance is therefore
\begin{equation}
    \frac{R}{\cos\theta_{\pvect{b}}} = \frac{R\sqrt{\kappa^2 + \tau^2}}{\kappa}.
\end{equation}
We also know from the definition of the helix that the vertical distance between point $\tilde{s}_0$ and point $\tilde{s}_0 + 2\uppi$ (where the centreline next passes over it) is $2\uppi b$. Therefore, for non-self-intersection, we require that $2\uppi b$ be greater than twice the vertical centreline-to-wall distance calculated above. This constraint yields the following condition for the helical duct
\begin{equation}
    0 > (\tau R)^6 + 3(\kappa R)^2(\tau R)^4 + \left(3(\kappa R)^2 - \uppi^2\right)(\kappa R)^2(\tau R)^2 + (\kappa R)^6.
\end{equation}
All of the helical ducts we consider have curvature given by $\kappa R = 2/3$. We consider three torsions: $\tau R$ = 0.16 (just large enough to avoid self-intersection), $\tau R = 0.2$, and $\tau R = 1$. All are compared in the linear case in figure \ref{helixlin.pdf}, while the second is held fixed for varying Mach number in figure \ref{tau0.20nonlin.pdf}. We use a frequency of $\omega = 0.95\overline{\omega}_1^1 = 0.95 \times 1.8412/R$ and truncation for all helical calculations is taken at 10 spatial modes and 10 temporal modes.

Each of these examples shows how torsional coupling causes a planar source to become non-planar at the duct outlet. It is also clear from figure \ref{helixlin.pdf} that (for all of the torsions we consider) the highest peaks and lowest troughs in the pressure occur on the outside of the helix, which is consistent with what was observed both in two- and three-dimensional planar bends. Figure \ref{tau0.20nonlin.pdf} shows that when the amplitude is high enough, the peaks and troughs then lose their even spacing as the wave steepens along the outside of the helix.  The effect of the $\theta_0(s)$ twisting due to the helical coordinate system is less visible in the static images shown in figures~\ref{helixlin.pdf} and~\ref{tau0.20nonlin.pdf}, but is more evident in the animated versions of them shown in Movies~7 and~8 in the supplementary material; the wave at the outlet is seen to spin as it propagates due to the helical nature of the duct.

\subsection{Comparison of two- and three-dimensions}\label{sec:results-3Dvs3D}
\begin{figure}
    \centering
    \includegraphics{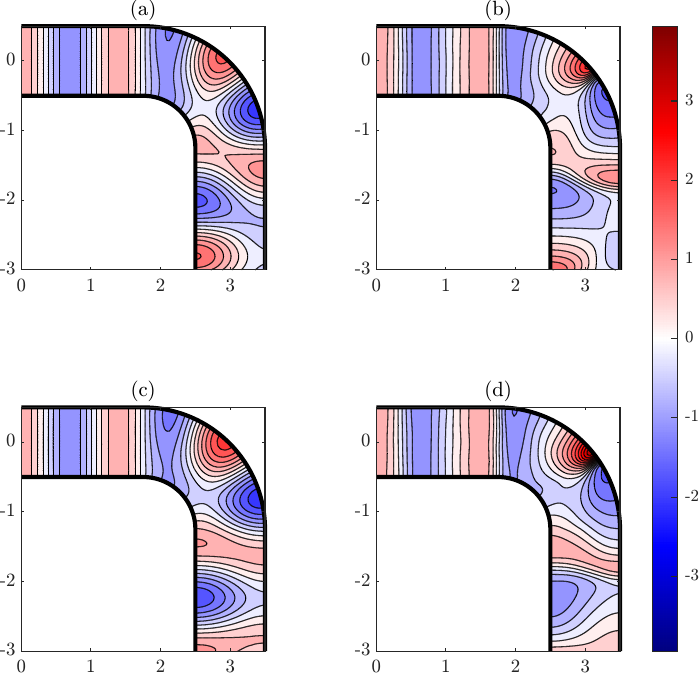}
    \caption{Forward-going pressure (normalised to the source amplitude) in an extended bend with a plane piston source of $9/2$, for (a) linear, two dimensions, (b) nonlinear, two dimensions, (c) linear, three dimensions and (d) nonlinear, three dimensions. Truncation was taken at $\alpha_\text{max} = a_\text{max} = 10$. An animated version of this figure is available as Movie~9 in the supplementary material.}
    \label{2D3Dcomp.pdf}
\end{figure}

\begin{figure}
    \centering
    \includegraphics{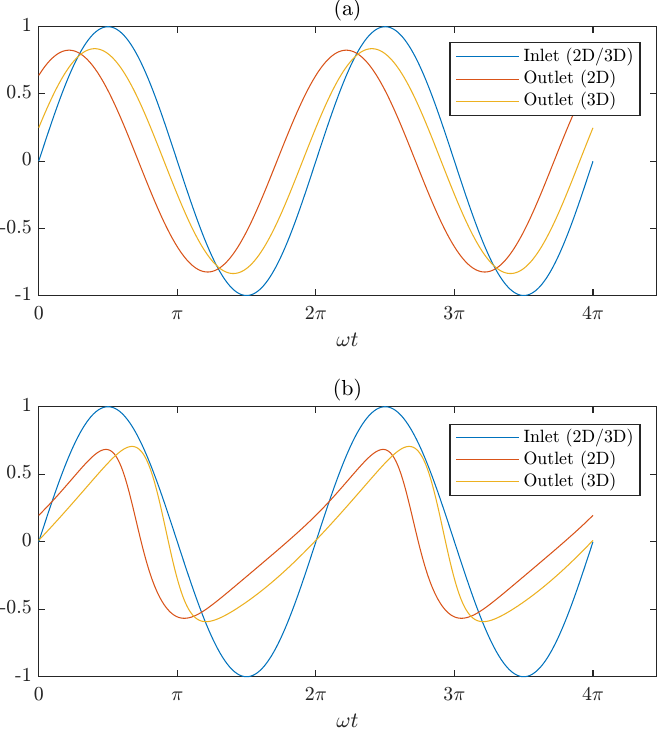}
    \caption{Forward-going pressure (normalised to the source amplitude), averaged over the inlet, the outlet in two dimensions, and the outlet in three dimensions, in the linear regime (a) and for $M = 0.10$ (b). Truncation was taken at $\alpha_\text{max} = a_\text{max} = 10$.}
    \label{inletoutletpressures.pdf}
\end{figure}

Width variation should not, by itself, produce significantly different effects between two- and three-dimensions, since the same symmetry is being broken in each case. Torsion is only possible in three dimensions, so does not invite comparison. Therefore, we will concentrate here on the difference between the effects of curvature in two- and three-dimensions. We consider a duct of width $X_\text{i} = 1$ in two dimensions, and radius $R_\text{i} = 0.5$ in three. In each case, there are straight sections of length $7/4$ either side of a right angle bend with bend radius $5/4$. We prescribe a piston source of frequency $9/2$: figure \ref{2D3Dcomp.pdf} compares the pressure fields in the linear case with those of Mach number $0.1$. In figure \ref{inletoutletpressures.pdf}, we plot the spatially-averaged outlet pressures against time, as a means of demonstrating the bend's effect on a piston source. Truncation was taken at 10 spatial modes and 10 temporal modes for each calculation.

We note that while the two cases seem to steepen equally, the effective acoustic length of the bend differs between them. The three-dimensional outlet pressure is consistently `ahead' of the two-dimensional outlet pressure for both linear and nonlinear regimes, suggesting that the effective bend length is longer in two dimensions. It may also be observed that the two-dimensional calculations exhibit more transverse oscillations than in three dimensions: this is unsurprising, since that is the only possible plane of oscillation in two dimensions, whereas in three dimensions there is an entire circular surface of oscillation at each $s$.

\subsection{Effective acoustic length of a bend}\label{sec:results-bend-length}

\begin{figure}
    \centering
    \includegraphics{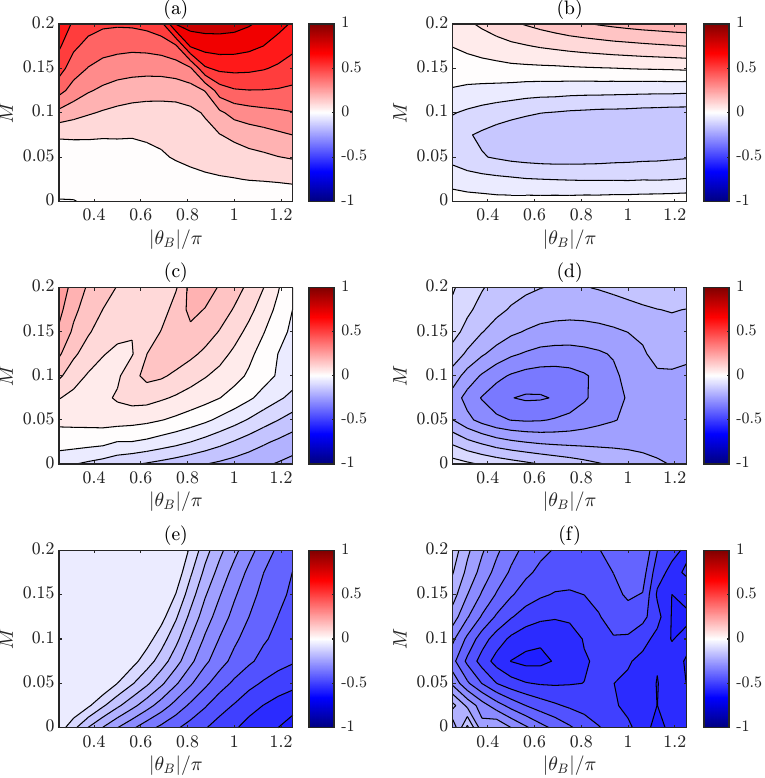}
    \caption{Upstroke (left) and downstroke (right) values of $B$, for $\omega X = 3$ (a,b), $3/2$ (c,d) and $3/4$ (e,f). Truncation was taken at $\alpha_\text{max} = a_\text{max} = 10$.}
    \label{curvaturemachtiles.pdf}
\end{figure}

We can also examine the competing effects of curvature and nonlinearity. We calculate the cross-sectional average pressure two wavelengths along the centreline from the inlet, for a range of different bend angles and Mach numbers, while keeping the arclength along the centre of the bend constant. The boundary condition at the inlet ensures that the average pressure across the inlet will pass upwardly through zero at $t = \uppi/\omega$, and downwardly at $t = \uppi/2\omega$. In a straight duct, the average pressure across a surface two wavelengths along would match this, but curvature causes a deviation of these upstroke and downstroke crossings, which we call $\Delta t$. We then define a \textit{bend correction factor} $B := 2c\Delta t/\Delta x$, where $\Delta x$ is the difference in arclength between the outer wall of the duct and the centreline for that particular bend angle. This parameter is equal to 1 for a wave travelling along the outside of the bend, and -1 for the inside. We can also choose whether to use the deviation time $\Delta t$ for the upstroke of the wave, or the downstroke. Figure \ref{curvaturemachtiles.pdf} shows upstroke and downstroke values of $B$ for three duct radii. Each data point on these contour plots is the result of a calculation with 10 spatial modes and 10 temporal modes.

From the upstroke plots we see that for greater bend angles (and correspondingly tighter bends), nonlinearity plays a bigger role in forcing the wave around the outside of the bend, whereas for smaller bend angles, the bend angle takes over as the determining parameter. The downstroke plots are less conclusive, but they do exhibit the emergence of a local minimum in $B$ as the radius is varied.

\section{Conclusions}\label{sec:conclusion}

We have presented a numerically tractable mathematical framework for solving for weakly-nonlinear acoustics in curved ducts of varying width, along with a number of numerical examples of its use.  The framework unifies both two- and three-dimensional governing equations, and allows for torsion in three dimensions.  The framework is numerically tractable in that it may be written in terms of large matrices and tensors that are invariant along the duct, unlike previous work in two dimensions where the large matrices and tensors varied with the duct geometry and therefore needed computing at each point along the duct~\citep{mctavish+brambley-2019}.  The method follows the multi-modal method of~\citet{felix1,felix2}, solving first for the admittance (a generalized ratio of acoustic velocity to acoustic pressure) throughout the duct from the outlet to the inlet, and then using this admittance to calculate the acoustic pressure and velocity from the inlet to the outlet.  This has the advantage that the effects of the duct geometry are encoded in the admittance and may be analyzed independently of the particular acoustic source used.
The unification of two and three dimensions allows for comparison between the two cases (such as was shown in figure~\ref{2D3Dcomp.pdf}).  

The method has been validated for a range of duct geometries where there are known analytic solutions or published numerical results.  Nonlinear steepening is validated against the \citet{fubini}, \citet{blackstock}, and \citet{fay} solutions for a straight duct in section~\ref{sec:results-straight}. Width variation in the linear case is validated against the \citet{webster} solution for an exponential horn in section~\ref{sec:results-horn}; this test case highlights the inaccuracy of approximating the full sound field by a plane wave in \citeauthor{webster}'s solution, which can be reproduced in our framework by restricting our solution to a single spatial mode or relaxed by allowing for many more spatial modes.  Curvature is validated in two dimensions in section~\ref{sec:results-bend}, in the linear regime against a result of~\citet{felix1} and in the nonlinear regime against a result for the same geometry of~\citet{mctavish+brambley-2019}.  The three dimensional equivalent in section~\ref{sec:results-bend3D} is validated against a result of~\citet{felix2} in the linear regime, and against an unpublished result of~\citet{mctavphd} in the nonlinear regime.  This extensive validation gives confidence that the combination of width variation, curvature, and nonlinearity is correctly modelled both in the mathematical framework and in its numerical implementation.

A number of duct geometries have then been investigated numerically.  These include a demonstration of wave leakage, where the inclusion of either slight curvature or slight nonlinearity can cause waves to propagate through a duct in which they would otherwise have been totally reflected (section~\ref{sec:results-curvature-and-width}).  Examples with torsion in three dimensions (section~\ref{sec:results-torsion}) show plane waves becoming localized on the outside of the duct curve.  The unification of two- and three-dimensional equations allows for the comparison of the two in section~\ref{sec:results-3Dvs3D}, showing that between two and three dimensions, wave steepening appears comparable but the effective acoustic length of a bend is different.

The effective length of a bend is investigated further in two dimensions in section~\ref{sec:results-bend-length}.  By considering the time lag in a wave propagating around a bend, a \emph{bend correction factor} $B$ is introduced, taking the value $+1$ for a wave propagating around the outside of the bend and $-1$ for a wave propagating around the inside of the bend.  These results are presented in figure~\ref{curvaturemachtiles.pdf}, and suggest that for higher curvature bends nonlinearity plays a bigger role in forcing wave propagation around the outside of the bend, whereas for lower curvature bends, the curvature takes over as the determining parameter.  This tool could potentially be used in the future to investigate whether resonant frequencies are likely to change when the sound amplitude is varied.

An obvious motivation of this work was the potential application to sound in brass instruments, where the brassy sound is caused by nonlinear wave steepening within the instrument.  The use of the admittance (both the usual linear admittance and its weakly nonlinear extension) makes the present framework well suited to this, as a duct's resonances could be investigated by solving for the admittance without needing to model or specify a sound source at the inlet.  Moreover, the bend correction factor of section~\ref{sec:results-bend-length} could be used to investigate whether instruments are likely to become sharper or flatter when played at louder or quieter volumes, and potentially even whether an instrument could be designed to have a stable pitch independently of the volume at which it is played.

The numerical solution for the admittance, and then for the pressure and velocity, is relatively standard, and future work could consider other methods of numerical solution.  As described in section~\ref{sec:numerics}, the infinite series of ODEs is truncated to a finite system by specifying a maximum number of temporal and spatial modes.  Since this truncation prevents the cascade of energy to higher modes and leads to artificial energy accumulating in the highest nontruncated mode, an artificial viscosity is added to help dissipate energy at higher modes.  This artificial viscosity is akin to the real molecular viscosity of the gas.  However, the dominant energy loss mechanism for acoustics in a duct is friction with the walls, which has been previously modelled using a fractional derivative approach~\citep[see, e.g.,][and references therein]{rendon2010}; a similar technique could be investigated for inclusion in the present model.  The now finite set of ODEs is then integrated here using a standard Runge--Kutta method, either with a fixed (RK4) or adaptive (RK45) method.  More sophisticated numerical methods that take note of the Riccati-style nature of the equations being integrated are also possible.  For example, \citet{pagneux2010} used a Magnus--M\"obius scheme in the linear case to avoid singularities of the admittance caused by the presence of pressure nodes in the duct.  Since the extension of Magnus schemes to nonlinear equations is possible~\citep{iserles2006}, potential future research might investigate the use of a weakly-nonlinear Magnus--M\"obius scheme for at least the admittance integration.

In order to calculate the admittance within the duct, a known admittance must be applied, for example at the duct outlet.  In all of the examples presented here, a non-reflecting admittance representing an infinite straight duct of constant width was used as the outlet admittance, as discussed in section~\ref{sec:invariant-admittances}.  One obvious outlet admittance for a musical instrument would be that of a duct exiting into an infinite space; such a situation may be modelled in the linear case either exactly using a Wiener--Hopf technique~\citep{munt-1977} or approximately using an outer duct with absorbing walls~\citep[e.g.][]{felix2018}, but the authors are not aware of a nonlinear, or even weakly-nonlinear, equivalent.  Both the open duct exit admittance and the directivity pattern of the far-field radiation would be interesting avenues of future research.

Other possibilities for future research might involve the inclusion of a mean flow within the duct~\citep[e.g.][]{mangin}, ducts with acoustically lined walls~\citep[e.g.][]{bi}, the investigation of other nonlinear effects such as resonant triads~\citep[e.g.][]{bustamante}, or other modal representations that might yield faster numerical convergence~\citep[e.g.][]{mercier}.

\begin{acknowledgements}
\noindent\textbf{Supplementary material.}
Animations of figures~\ref{McTavBramb_fig7}, \ref{FP2002_fig2}, \ref{McTav_fig3.6.pdf}, \ref{mctavfig5}, \ref{padmtile}, and \ref{planewavetunnelling.pdf}--\ref{2D3Dcomp.pdf} are available in the supplementary movies as Movies~1--9.
\textsc{Matlab} source code to generate the results given here is also available in supplementary material.

\vspace{1ex}
\noindent\textbf{Acknowledgements.}
For the purpose of open access, the authors have applied a Creative Commons Attribution (CC
BY) licence to any Author Accepted Manuscript version arising from this submission.

\vspace{1ex}
\noindent\textbf{Funding.}
F.J.\ was funded through the Warwick Mathematics Institute Centre for Doctoral Training, and gratefully acknowledges the support of the University of Warwick and the UK Engineering and Physical Sciences Research Council (EPSRC grant EP/W523793/1).
E.J.B.\ gratefully acknowledges the support of the UK Engineering and Physical Sciences Research Council (EPSRC grant EP/V002929/1).

\vspace{1ex}
\noindent\textbf{Declaration of interests.}
The authors report no conflict of interest.
\end{acknowledgements}

\appendix
\section{Spatial projection in more detail} \label{spatprodapp}
\subsection{Two-dimensional spatial projection}\label{app:A1}
The goal here is to project equations \eqref{equ:2Dgov} onto the basis of spatial modes~\eqref{2Dmodedef}. This is achieved by multiplying each equation by a mode $\psi_\alpha$ and integrating over a duct cross-section. If the expansions \eqref{2Dspatialmodedefs} are then employed, the acoustic mode coefficients may then be factored out of the integral, which becomes a matrix with rows in $\alpha$ and columns in the expansion dummy variable $\beta$, dependent only on the local geometry. However, the Neumann boundary condition \eqref{2DNeumann} prevents us from expanding any first-derivatives of acoustical quantities on the boundary, which means we may not expand second-derivatives of acoustical quantities within integrals, and must instead use integration-by-parts to remove the $x$-derivatives and then expand once it is safe to so.

If we start with the left-hand side of \eqref{2Delimmass}, we get
\begin{multline}
\label{projecting2Dmass}
        \int_{X_-}^{X_+}\psi_\alpha\left\{\frac{\partial U^a}{\partial s} - \I a\omega\left[h_s\left(1 + \frac{1}{a^2\omega^2}\frac{\partial^2}{\partial x^2}\right) - \frac{\kappa}{a^2\omega^2}\frac{\partial}{\partial x}\right]P^a\right\}\intd x
        = \underbrace{\frac{\intd }{\intd s}\left(\int_{X_-}^{X_+}\psi_\alpha U^a\intd x\right)}_\text{\fbox{1}}
        \\
        - \underbrace{\left[X_+'\left(\psi_\alpha U^a\right)\bigg|_{X_+} - X_-'\left(\psi_\alpha U^a\right)\bigg|_{X_-}\right]}_\text{\fbox{2}}
        - \underbrace{\frac{\I a\omega}{a^2\omega^2}\left[\psi_\alpha h_s\frac{\partial P^a}{\partial x} - \frac{\partial(\psi_\alpha h_s)}{\partial x}P^a - \kappa\psi_\alpha P^a\right]_{X_-}^{X_+}}_\text{\fbox{3}}
        \\
        - \int_{X_-}^{X_+}\underbrace{\frac{\partial\psi_\alpha}{\partial s}U^a}_\text{\fbox{4}}
        + \underbrace{\I a\omega\left(h_s\psi_\alpha + \frac{1}{a^2\omega^2}\frac{\partial^2(\psi_\alpha h_s)}{\partial x^2} + \frac{\kappa}{a^2\omega^2}\frac{\partial\psi_\alpha}{\partial x}\right)P^a}_\text{\fbox{5}}\intd x.
\end{multline}
Terms \fbox{1}, \fbox{2} and \fbox{4} are the result of bringing the partial $s$-derivative on the $U^a$ outside of the integral, whereas terms \fbox{3} and \fbox{5} are the result of removing all of the $x$-derivatives from the $P^a$ via integration-by-parts (in particular, the second-derivative has been removed by a double i-b-p to produce the first two constituent terms of \fbox{3}). Term \fbox{1} is an exact $s$-derivative of the projection of $U^a$ onto the basis, i.e.
\begin{equation}
    \frac{\intd }{\intd s}\left(\int_{X_-}^{X_+}\psi_\alpha U^a\intd x\right) = \frac{\intd }{\intd s}\left(\int_{X_-}^{X_+}\psi_\alpha\sum_\beta U_\beta^a\psi_\beta\intd x\right) = \frac{\intd }{\intd s}\left(\sum_\beta\delta_{\alpha\beta}U_\beta^a\right) = \frac{\intd U_\alpha^a}{\intd s},
\end{equation}
by mode orthogonality. Term \fbox{2} combines with the first term of \fbox{3} to form the left-hand side of the hard-walled boundary condition \eqref{2Dnopenetration}, leaving a term that is $O(M^2)$, while the second two terms of \fbox{3} cancel with one another due to the Neumann condition on the duct modes. Finally, \fbox{5}'s second term, by the Leibniz rule, makes -2 lots of \fbox{5}'s third term added to a term involving the second-derivative of a spatial mode, which by \eqref{2DHelmholtzeq} may be turned into multiplication by a squared eigenvalue. Thus, expanding into spatial modes, we have
\begin{align}
        \int_{X_-}^{X_+}\psi_\alpha&\left\{\frac{\partial U^a}{\partial s} - \I a\omega\left[h_s\left(1 + \frac{1}{a^2\omega^2}\frac{\partial^2}{\partial x^2}\right) - \frac{\kappa}{a^2\omega^2}\frac{\partial}{\partial x}\right]P^a\right\}\intd x \notag\\
        =& \sum_\beta\left(\delta_{\alpha\beta}\frac{\intd }{\intd s} - \int_{X_-}^{X_+}\frac{\partial\psi_\alpha}{\partial s}\psi_\beta\intd x\right)U_\beta^a + \frac{1}{\I a\omega}\left[h_s\psi_\alpha \frac{\partial Q^a}{\partial x}\right]_{X_-}^{X_+} \notag\\
        &- \I a\omega\sum_\beta\left[\left(1 - \frac{\lambda_\alpha^2}{a^2\omega^2X^2}\right)\int_{X_-}^{X_+}h_s\psi_\alpha\psi_\beta\intd x - \frac{\kappa}{a^2\omega^2}\int_{X_-}^{X_+}\frac{\partial\psi_\alpha}{\partial x}\psi_\beta\intd x\right]P_\beta^a.
\end{align}
Apart from the term involving $Q^a$ (which will be dealt with later), we have integrals, with two subscripts, summing over their second subscripts with acoustical quantities that have only one subscript. In other words, this is standard matrix-vector multiplication, so we may drop the sigmas in employment of the summation convention. Furthermore, we will define a shorthand for any mode-integral matrices: the integral of two modes is written as $\Psi$, with Greek-letter subscripts denoting the mode-numbers, while bracketed subscripts denote derivatives of the modes in question (curly brackets are $s$-derivatives and square brackets are $x$-derivatives, with a square bracket around all subscripts meaning that this is a boundary term). Finally, any non-mode functions present in the integral appear in square brackets after the subscripts. This notation is the same as was used in \cite{mctavish+brambley-2019}.  For example,
\begin{align}
    &\Psi_{\{\alpha\}[\beta]}[f(s,x)] &&\text{would mean}&& \int_{X_-}^{X_+}\frac{\partial\psi_\alpha}{\partial s}\frac{\partial\psi_\beta}{\partial x}f(s,x)\intd x.
\end{align}
With this notation in hand, the projected left-hand side becomes
\begin{multline}
    \left(\delta_{\alpha\beta}\frac{\intd }{\intd s} - \Psi_{\{\alpha\}\beta}\right)U_\beta^a + \frac{1}{\I a\omega}\left[h_s\psi_\alpha \frac{\partial Q^a}{\partial x}\right]_{X_-}^{X_+} \\
    - \I a\omega\left[\left(\delta_{\alpha\gamma} - \frac{\gmat{\Lambda}_{\alpha\gamma}^2}{a^2\omega^2X^2}\right)\Psi_{\gamma\beta}[h_s] - \frac{\kappa}{a^2\omega^2}\Psi_{[\alpha]\beta}\right]P_\beta^b = O(M^2).
\end{multline}
where $\gmat{\Lambda}$ is a matrix with entries given by $\gmat{\Lambda}_{\alpha\beta} = \lambda_{\alpha}\delta_{\alpha\beta}$ (no sum).

We now turn to the right-hand side of \eqref{2Delimmass}. Its first term projects very easily, giving
\begin{equation}
    \int_{X_-}^{X_+}-\I a\omega\cnon h_s\psi_\alpha\sum_bP^{a-b}P^b\intd x = -\I a\omega\cnon\Psi_{\alpha\beta\gamma}[h_s]\sum_bP_\beta^{a - b}P_\gamma^b,
\end{equation}
where the $\Psi$ matrix notation has been naturally extended to third-rank tensors, which contract with acoustical quantities on subscripts 2 and 3. The next term on the right-hand side becomes
\begin{multline}
        \I a\omega\int_{X_-}^{X_+}\psi_\alpha\left[h_s\left(1 - \frac{1}{a^2\omega^2}\frac{\partial^2}{\partial x^2}\right) + \frac{\kappa}{a^2\omega^2}\frac{\partial}{\partial x}\right]Q^a\intd x
        \\
        = \frac{\I a\omega}{a^2\omega^2}\bigg[-\psi_\alpha h_s\frac{\partial Q^a}{\partial x} + \frac{\partial(\psi_\alpha h_s)}{\partial x}Q^a + \kappa\psi_\alpha Q^a\bigg]_{X_-}^{X_+}
        \\
        + \I a\omega\int_{X_-}^{X_+}Q^a\left[\left(1 - \frac{1}{a^2\omega^2}\frac{\partial^2}{\partial x^2}\right)(h_s\psi_\alpha) - \frac{\kappa}{a^2\omega^2}\frac{\partial\psi_\alpha}{\partial x}\right]\intd x.
\end{multline}
This is more-or-less the same manipulation as took place on the left-hand side, only now with $Q^a$ in the place of $P^a$ and some sign changes. After the same cancellations have been made, and we switch to the $\Psi$ notation (which requires the terms making up $Q^a$ to be expanded into spatial modes), we get
\begin{multline}
\label{Q^a_expansion2D}
    \I a\omega\int_{X_-}^{X_+}\psi_\alpha\left[h_s\left(1 - \frac{1}{a^2\omega^2}\frac{\partial^2}{\partial x^2}\right) + \frac{\kappa}{a^2\omega^2}\frac{\partial}{\partial x}\right]Q^a\intd x = \frac{1}{\I a\omega}\left[h_s\psi_\alpha\frac{\partial Q^a}{\partial x}\right]_{X_-}^{X_+} \\
    + \I a\omega\bigg[\left(\delta_{\alpha\delta} + \frac{\gmat{\Lambda}_{\alpha\delta}^2}{a^2\omega^2X^2}\right)\Psi_{\delta\beta\gamma}[h_s] + \frac{\kappa}{a^2\omega^2}\Psi_{[\alpha]\beta\gamma}\bigg]\frac{1}{2}\sum_bP_\beta^{a - b}P_\gamma^b - U_\beta^{a - b}U_\gamma^b \\
    + \I a\omega\bigg[\left(\delta_{\alpha\delta} + \frac{\gmat{\Lambda}_{\alpha\delta}^2}{a^2\omega^2X^2}\right)\Psi_{\delta[\beta][\gamma]}[h_s] + \frac{\kappa}{a^2\omega^2}\Psi_{[\alpha][\beta][\gamma]}\bigg]\sum_b\frac{P_\beta^{a - b}P_\gamma^b}{2(a - b)b\omega^2}.
\end{multline}
The first term here is exactly the leftover boundary term from the left-hand side; this will therefore cancel when they are combined, meaning the projected mass conservation equation is
\begin{equation}
\label{2Dprojectedmass}
    \begin{aligned}
        \left(\delta_{\alpha\beta}\frac{\intd }{\intd s}\right.&\left.{}\!\!- \Psi_{\{\alpha\}\beta}\right)U_\beta^a - \I a\omega\left[\left(1 - \frac{\lambda_\alpha^2}{a^2\omega^2X^2}\right)\Psi_{\alpha\beta}[h_s] - \frac{\kappa}{a^2\omega^2}\Psi_{[\alpha]\beta}\right]P_\beta^b \\
        =&\, \I a\omega\sum_b\Bigg\{-\cnon\Psi_{\alpha\beta\gamma}[h_s]P_\beta^{a - b}P_\gamma^b \\
        &+ \bigg[\left(\delta_{\alpha\delta} + \frac{\gmat{\Lambda}_{\alpha\delta}^2}{a^2\omega^2X^2}\right)\Psi_{\delta\beta\gamma}[h_s] + \frac{\kappa}{a^2\omega^2}\Psi_{[\alpha]\beta\gamma}\bigg]\frac{P_\beta^{a - b}P_\gamma^b - U_\beta^{a - b}U_\gamma^b}{2} \\
        &+ \bigg[\left(\delta_{\alpha\delta} + \frac{\gmat{\Lambda}_{\alpha\delta}^2}{a^2\omega^2X^2}\right)\Psi_{\delta[\beta][\gamma]}[h_s] + \frac{\kappa}{a^2\omega^2}\Psi_{[\alpha][\beta][\gamma]}\bigg]\frac{P_\beta^{a - b}P_\gamma^b}{2(a - b)b\omega^2}\Bigg\}.
    \end{aligned}
\end{equation}
Projecting \eqref{2Delimmom} proves easier, since there are no boundary-term cancellations; care is only needed when dealing with the second $x$-derivative on the right-hand side. The left-hand side requires no integration-by-parts this time, due to the absence of any $x$-derivatives, so we simply have
\begin{equation}
    \int_{X_-}^{X_+}\psi_\alpha\left(\frac{\partial P^a}{\partial s} - \I a\omega h_sU^a\right)\intd x = \left(\delta_{\alpha\beta}\frac{\intd }{\intd s} + \Psi_{\alpha\{\beta\}}\right)P_\beta^a - \I a\omega\Psi_{\alpha\beta}[h_s]U_\beta^a.
\end{equation}
The term involving $(a - b)$ on the right-hand side is dealt with very easily, and the final term is effectively a repeat of the final term in \eqref{Q^a_expansion2D}. This leaves only the term with the $\partial^2/\partial x^2$, which must be integrated-by-parts before an expansion of the pressure into spatial modes may take place
\begin{multline}
        \int_{X_-}^{X_+}\I\omega h_s\psi_\alpha\sum_b\left[-bU^{a - b}\frac{1}{b^2\omega^2}\frac{\partial^2P^b}{\partial x^2}\right]\intd x = -\I\omega\left[h_s\psi_\alpha\sum_b\frac{b}{b^2\omega^2}U^{a - b}\frac{\partial P^b}{\partial x}\right]_{X_-}^{X_+} \\
        + \int_{X_-}^{X_+} \I\omega\sum_b\left[\frac{b}{b^2\omega^2}\frac{\partial\left(h_s\psi_\alpha U^{a - b})\right)}{\partial x}\sum_\gamma P_\gamma^b\frac{\partial \psi_\gamma}{\partial x}\right]\intd x.
\end{multline}
We know from the hard-walled boundary condition that the $\partial P^b/\partial x$ in the boundary term must turn into a $U^b$ (i.e. something expansible), and having now (legally) expanded the $P^b$ inside the integral, we may now simply reverse the integration-by-parts (and employ the summation convention over $\gamma$), giving
\begin{multline}
        \int_{X_-}^{X_+}\I\omega h_s\psi_\alpha\sum_b\left(-bU^{a - b}\frac{1}{b^2\omega^2}\frac{\partial^2P^b}{\partial x^2}\right)\intd x = \left[X_+\Psi_{[\alpha\beta\gamma]}^+ - X_-\Psi_{[\alpha\beta\gamma]}^-\right]\sum_bU_\beta^{a - b}U_\gamma^b \\
        + \I\omega\sum_b\left[\frac{b}{b^2\omega^2}h_s\psi_\alpha U^{a - b}P_\gamma^b\frac{\partial\psi_\gamma}{\partial x}\right]_{X_-}^{X_+} + \I\omega\sum_b\frac{b\gmat{\Lambda}_{\delta\gamma}^2}{b^2\omega^2X^2}\Psi_{\alpha\beta\delta}[h_s]U_\beta^{a - b}P_\gamma^b.
\end{multline}
The $\partial\psi_\gamma/\partial x$ in the second boundary term causes it to vanish, by the Neumann condition on spatial modes. Thus, this term has been dealt with, and the projected momentum equation is
\begin{multline}
\label{2Dprojectedmom}
        \left(\delta_{\alpha\beta}\frac{\intd }{\intd s} + \Psi_{\alpha\{\beta\}}\right)P_\beta^a - \I a\omega\Psi_{\alpha\beta}[h_s]U_\beta^a = \left[X_+\Psi_{[\alpha\beta\gamma]}^+ - X_-\Psi_{[\alpha\beta\gamma]}^-\right]\sum_bU_\beta^{a - b}U_\gamma^b \\
        + \I\omega\sum_b\left\{\Psi_{\alpha\beta\delta}[h_s]\left[(a - b)\delta_{\delta\gamma} - b\left(\delta_{\delta\gamma} - \frac{\gmat{\Lambda}_{\delta\gamma}^2}{b^2\omega^2X^2}\right)\right] + \frac{1}{b\omega^2}\Psi_{\alpha[\beta][\gamma]}[h_s]\right\}U_\beta^{a - b}P_\gamma^b,
\end{multline}
where the `$+$' and `$-$' superscripts on boundary terms indicate that only a single side of the boundary is being evaluated.

In theory, we have now done all of the work required to turn these equations from PDEs to ODEs and the next step would be to put them into a computer and solve them. However, more work may now be done to these equations in order to make them numerically efficient to solve. At present, we have a countably infinite set of coupled vector ODEs in $s$ with matricial coefficients that also depend on $s$, meaning that a solver would need to update the value of each matrix at each point along the duct. This is missing a trick, however, since each of these matrices' $s$-dependence may be factored out into the form of $s$-dependent scalars such as $\kappa$, while the matrices themselves are constant, and thus only need to be defined before solving. This also has the advantage of added clarity, since it will clearly highlight which terms in the equations are responsible for different geometrical irregularities. In two dimensions, we have $s$-dependence in the integration limits $X_-$ and $X_+$, which is removed via the substitution $\xi = (x - X_-)/X$, so that the integration limits are now 0 and 1. Expanding $h_s$ and factoring the $\kappa$ out then just leaves $1/\sqrt{X}$ in front of the modes, which can also be factored out. If we apply this technique to one of the matrices from \eqref{2Dprojectedmom}, we get
\begin{multline}
        \Psi_{\alpha\beta}[h_s] = \int_{X_-}^{X_+}\frac{C_\alpha C_\beta}{X}h_s\cos\left(\lambda_\alpha\frac{x - X_-}{X}\right)\cos\left(\lambda_\beta\frac{x - X_-}{X}\right)\intd x \\
        = (1 - \kappa X_-)\delta_{\alpha\beta} - \kappa X\int_0^1C_\alpha C_\beta\xi\cos(\lambda_\alpha\xi)\cos(\lambda_\beta\xi)\intd \xi = (1 - \kappa X_-)\delta_{\alpha\beta} - \kappa X\Xi_{\alpha\beta}[\xi]
\end{multline}
where we have adopted a new notation $\Xi$ for the constant matrices, analogous to the $\Psi$ notation but now with different limits and a different integration variable. In order to deal with $s$-derivatives, we note that
\begin{equation}
    \frac{\partial\psi_\alpha}{\partial s} = -\frac{X'}{2X}\psi_\alpha - \frac{X'}{X}(x - X_-)\frac{\partial\psi_\alpha}{\partial x} - X_-'\frac{\partial\psi_\alpha}{\partial x} = - \frac{X'}{2X}\left(\psi_\alpha + 2\xi\frac{\partial\psi_\alpha}{\partial\xi}\right) - \frac{X_-'}{X}\frac{\partial\psi_\alpha}{\partial\xi},
\end{equation}
and a more convenient form may be found for some of the right-hand-side terms involving derivatives on multiple indices, by use of the Leibniz rule. The two relations necessary for this are
\begin{subequations}\begin{align}
        \Psi_{\alpha[\beta][\gamma]}[h_s] &= \frac{\lambda_\beta^2 + \lambda_\gamma^2 - \lambda_\alpha^2}{2X^2}\Psi_{\alpha\beta\gamma}[h_s] - \kappa\Psi_{[\alpha]\beta\gamma} + \frac{\kappa}{2}\Psi_{[\alpha\beta\gamma]},
\\
        \Psi_{[\alpha][\beta][\gamma]} &= \frac{\lambda_{\beta}^2 + \lambda_\gamma^2 - \lambda_\alpha^2}{2X^2}\Psi_{[\alpha]\beta\gamma} + \frac{\lambda_\alpha^2}{2X^2}\Psi_{[\alpha\beta\gamma]}.
\end{align}\end{subequations}%
In section \ref{spatprodandnot} we introduced a compact notation which computationally and algebraically simplified the equations. In the new notation, \eqref{2Dprojectedmass} and \eqref{2Dprojectedmom} can be rewritten as
\begin{subequations}\label{app/2Dvector}\begin{gather}
    \label{app/2Dvectormass}
        \begin{aligned}
            \Bigg[\frac{\intd }{\intd s} + \frac{X'}{2X}\mat{W} &+ \frac{X_-'}{X}\widetilde{\mat{A}}\Bigg]\mvect{u}^a - \I a\omega\Bigg[\left(\mat{I} - \frac{\gmat{\Lambda}^2}{a^2\omega^2X^2}\right)\bigg((1 - \kappa X_-)\mat{I} - \kappa X\mat{A}\bigg) - \frac{\kappa\widetilde{\mat{A}}}{a^2\omega^2X}\Bigg]\mvect{p}^a \\
            =& \frac{\I a\omega}{\sqrt{X}}\sum_b\Bigg\{- \cnon\bigg((1 - \kappa X_-)\mathcal{I} - \kappa X\mathcal{A}\bigg)\langle\mvect{p}^{a - b},\mvect{p}^b\rangle\\&
            +\Bigg[\left(\mat{I} + \frac{\gmat{\Lambda}^2}{a^2\omega^2X^2}\right)\bigg((1 - \kappa X_-)\mathcal{I} - \kappa X\mathcal{A}\bigg) + \frac{\kappa\widetilde{\mathcal{A}}}{a^2\omega^2X}\Bigg]\frac{\langle\mvect{p}^{a - b},\mvect{p}^b\rangle - \langle\mvect{u}^{a - b},\mvect{u}^b\rangle}{2} \\&
            + \left[\frac{\left(\mat{I} + \frac{\gmat{\Lambda}^2}{a^2\omega^2X^2}\right)\bigg((1 - \kappa X_-)\mathcal{I}^\lambda - \kappa X\mathcal{A}^\lambda\bigg) + \frac{\kappa\widetilde{\mathcal{A}}^\lambda}{a^2\omega^2X}}{2(a - b)b\omega^2X^2} \right]\langle\mvect{p}^{a - b},\mvect{p}^b\rangle\Bigg\},
        \end{aligned}
\displaybreak[0]\\[1ex]
    \label{app/2Dvectormom}
        \begin{aligned}
            \Bigg[\frac{\intd }{\intd s} - \frac{X'}{2X}\mat{W}^\T &- \frac{X_-'}{X}\widetilde{\mat{A}}^\T\Bigg]\mvect{p}^a - \I a\omega\Bigg[(1 - \kappa X_-)\mat{I} - \kappa X\mat{A}\Bigg]\mvect{u}^a \\
            =& \frac{1}{\sqrt{X}}\sum_b\Bigg\{\bigg(\frac{X_+'}{X}\overline{\mathcal{W}}^+ - \frac{X_-'}{X}\overline{\mathcal{W}}^-\bigg)\langle\mvect{u}^{a - b},\mvect{u}^b\rangle
            \\&
            + \I\omega\Bigg[\bigg((1 - \kappa X_-)\mathcal{I} - \kappa X\mathcal{A}\bigg)
            \left\langle\mat{I},(a - b)\mat{I} - b\left(\mat{I} - \frac{\gmat{\Lambda}^2}{b^2\omega^2X^2}\right)\right\rangle
            \\&\qquad\quad
            + \frac{(1 - \kappa X_-)\mathcal{I}^\lambda - \kappa X\mathcal{A}^\lambda}{b\omega^2X^2}\Bigg]\langle\mvect{u}^{a - b},\mvect{p}^b\rangle\Bigg\},
        \end{aligned}
    \end{gather}\end{subequations}%
where
\begin{subequations}\begin{align}
\mat{I} &= \delta_{\alpha\beta},
&
\mat{W}_{\alpha\beta} &= \delta_{\alpha\beta} + 2\Xi_{[\alpha]\beta}[\xi],
\\
 \mat{A}_{\alpha\beta} &= \Xi_{\alpha\beta}[\xi],
 &
 \widetilde{\mat{A}}_{\alpha\beta} &= \Xi_{[\alpha]\beta},
 \\
 \mathcal{I}_{\alpha\beta\gamma} &= \Xi_{\alpha\beta\gamma},
 &
 \mathcal{A}_{\alpha\beta\gamma} &= \Xi_{\alpha\beta\gamma}[\xi],
 \\
 \widetilde{\mathcal{A}}_{\alpha\beta\gamma} &= \Xi_{[\alpha]\beta\gamma},
 &
 \overline{\mathcal{A}}_{\alpha\beta\gamma} &= \Xi_{[\alpha\beta\gamma]},
 \\
 \overline{\mathcal{W}}_{\alpha\beta\gamma}^+ &= \Xi_{[\alpha\beta\gamma]}^+,
 &
 \overline{\mathcal{W}}_{\alpha\beta\gamma}^- &= \Xi_{[\alpha\beta\gamma]}^-,
\end{align}
and
    \begin{align}
        \mathcal{I}^\lambda &= \mathcal{I}\frac{\langle\gmat{\Lambda}^2,\mat{I}\rangle + \langle\mat{I},\gmat{\Lambda}^2\rangle - \gmat{\Lambda}^2}{2},
 \\
 \mathcal{A}^\lambda &= \mathcal{A}\frac{\langle\gmat{\Lambda}^2,\mat{I}\rangle + \langle\mat{I},\gmat{\Lambda}^2\rangle - \gmat{\Lambda}^2}{2} + \widetilde{\mathcal{A}} - \frac{\overline{\mathcal{A}}}{2},
\\        \widetilde{\mathcal{A}}^\lambda &= \widetilde{\mathcal{A}}\frac{\langle\gmat{\Lambda}^2,\mat{I}\rangle + \langle\mat{I},\gmat{\Lambda}^2\rangle - \gmat{\Lambda}^2}{2} + \frac{\gmat{\Lambda}^2}{2}\overline{\mathcal{A}}.
\end{align}\end{subequations}%
The names of these matrices have been chosen to represent the geometric irregularities whose influence upon the equations they modulate, i.e. $\mat{W}$ for `width', and $\mat{A}$ for `annularity' (one matrix, $\widetilde{\mat{A}}$ is used for both, so a single name had to be picked). Calligraphic tensors are related to sans-serif matrices with the same name, i.e. $\mathcal{I}_{\alpha\beta0} = \mat{I}_{\alpha\beta}$.  Equations~\eqref{app/2Dvector} are equations~\eqref{equ:2Dvector} in the main text.

\subsection{Three-dimenensional spatial projection}\label{app:A2}
Projection of the governing equations is complicated by the extra dimension, though fortunately all boundary terms in $\theta$ disappear due to periodicity of the spatial modes. Thus, we only need to worry about boundary terms in $r$. To help with this, some quick formulae are introduced for integration-by-parts (over a duct cross-section $S$, with differential $\intd S = r\intd r\intd \theta$) involving the transverse gradient and FSND
\begin{subequations}
    \begin{equation}
    \label{transverselaplacianbyparts}
        \begin{aligned}
            \iint_Sf\pvect{\nabla}_\mathrm{t}^2g\intd S &= \int_0^{2\uppi}\left[rf\frac{\partial g}{\partial r}\right]_0^R\intd \theta - \iint_S\pvect{\nabla}_\mathrm{t}f\pvect{\cdot}\pvect{\nabla}_\mathrm{t}g\intd S \\
            &= \int_0^{2\uppi}\left[r\left(f\frac{\partial g}{\partial r} - \frac{\partial f}{\partial r}g\right)\right]_0^R\intd \theta + \iint_Sg\pvect{\nabla}_\mathrm{t}^2f\intd S,
        \end{aligned}
    \end{equation}
    \begin{equation}
        \iint_Sf\frac{\partial g}{\partial\pvect{n}}\intd S = \int_0^{2\uppi}\bigg[r\cos\phi fg\bigg]_0^R\intd \theta - \iint_Sg\frac{\partial f}{\partial\pvect{n}}\intd S.
    \end{equation}
\end{subequations}
Note that we introduce here a shorthand $\phi$ for use in this appendix, defined as $\phi := \theta - \theta_0(s)$. Using the formulae above, we can project the left-hand side of \eqref{3Delimmass}:
\begin{multline}
    \iint_S\psi_\alpha\left\{\frac{\partial U^a}{\partial s} - \I a\omega\left[h_s\left(1 + \frac{\pvect{\nabla}_\mathrm{t}^2}{a^2\omega^2}\right) - \frac{\kappa}{a^2\omega^2}\frac{\partial}{\partial\pvect{n}}\right]P^a\right\}\intd S = \underbrace{\frac{\intd }{\intd s}\left(\iint\psi_\alpha U^a\intd S\right)}_\text{\fbox{1}} \\
    - \underbrace{\int_0^{2\uppi}R'\left(r\psi_\alpha U^a\right)\bigg|_R}_\text{\fbox{2}} + \underbrace{\frac{\I a\omega}{a^2\omega^2}\left[r\psi_\alpha h_s\frac{\partial P^a}{\partial r} - r\frac{\partial(\psi_\alpha h_s)}{\partial r}P^a - \kappa r\cos\phi\psi_\alpha P^a\right]_0^R}_\text{\fbox{3}}\intd \theta \\
    -\iint_S\underbrace{\frac{\partial\psi_\alpha}{\partial s}U^a}_\text{\fbox{4}} + \underbrace{\I a\omega\left(h_s\psi_\alpha + \frac{\pvect{\nabla}_\mathrm{t}^2}{a^2\omega^2}(h_s\psi_\alpha) + \frac{\kappa}{a^2\omega^2}\frac{\partial\psi_\alpha}{\partial\pvect{n}}\right)P^a}_\text{\fbox{5}}\intd S.
\end{multline}
As in \eqref{projecting2Dmass}, we have brought the $s$-derivative on the $U^a$ outside of the integral, and removed all of the transverse derivatives from the pressure via integration-by-parts, so that all derivatives now act on the $\psi_\alpha$ mode. Once again term \fbox{2} combines with the first bit of term \fbox{3} to form the linear part of the boundary condition \eqref{3Dnopenetration}, while the rest of term \fbox{3} cancels, and term \fbox{5} changes after an application of the Leibniz rule, leaving
\begin{align}
    \iint_S\psi_\alpha&\left\{\frac{\partial U^a}{\partial s} - \I a\omega\left[h_s\left(1 + \frac{\pvect{\nabla}_\mathrm{t}^2}{a^2\omega^2}\right) - \frac{\kappa}{a^2\omega^2}\frac{\partial}{\partial\pvect{n}}\right]P^a\right\}\intd S \notag\\
    =& \left(\delta_{\alpha\beta}\frac{\intd }{\intd s} - \iint_S\frac{\partial\psi_\alpha}{\partial s}\psi_\beta\intd S\right)U_\beta^a + \frac{1}{\I a\omega}\left[rh_s\psi_\alpha\frac{\partial Q^a}{\partial r}\right]_0^R \notag\\ 
    &- \I a\omega\left[\left(1 - \frac{\lambda_\alpha^2}{a^2\omega^2R^2}\right)\iint_Sh_s\psi_\alpha\psi_\beta\intd S - \frac{\kappa}{a^2\omega^2}\iint_S\frac{\partial\psi_\alpha}{\partial\pvect{n}}\psi_\beta\intd S\right]P_\beta^a.
\end{align}
We can use the same shorthand for these matrix-integrals as we did earlier, introducing `$|\bullet|$' as a subscript notation for the FSND, and round brackets for $\theta$-derivatives, so that the projected left-hand side of \eqref{3Delimmass} becomes
\begin{multline}
    \left(\delta_{\alpha\beta}\frac{\intd }{\intd s} - \Psi_{\{\alpha\}\beta}[r]\right)U_\beta^a + \frac{1}{\I a\omega}\int_0^{2\uppi}\left[rh_s\psi_\alpha\frac{\partial Q^a}{\partial r}\right]_0^R\intd \theta \\
    - \I a\omega\left[\left(\delta_{\alpha\gamma} - \frac{\gmat{\Lambda}_{\alpha\gamma}^2}{a^2\omega^2R^2}\right)\Psi_{\gamma\beta}[rh_s] - \frac{\kappa}{a^2\omega^2}\Psi_{|\alpha|\beta}[r]\right]P_\beta^a = O(M^2)
\end{multline}
Turning now to the right-hand side of \eqref{3Delimmass}, we have an easy first term, followed by some derivatives of $Q^a$, which will removed by integration-by-parts as they just were with the pressure on the left-hand side
\begin{multline}
    \I a\omega\iint_S\psi_\alpha\left[h_s\left(1 - \frac{\pvect{\nabla}_\mathrm{t}^2}{a^2\omega^2}\right) + \frac{\kappa}{a^2\omega^2}\frac{\partial}{\partial\pvect{n}}\right]Q^a\intd S \\
    = \int_0^{2\uppi}\frac{\I a\omega}{a^2\omega^2}\bigg[-rh_s\psi_\alpha\frac{\partial Q^a}{\partial r} + r\frac{\partial(h_s\psi_\alpha)}{\partial r}Q^a + \kappa r\cos\phi\psi_\alpha Q^a\bigg]_0^R\intd \theta \\
    + \I a\omega\iint_SL^a\left[\left(1 - \frac{\pvect{\nabla}_\mathrm{t}^2}{a^2\omega^2}\right)(h_s\psi_\alpha) - \frac{\kappa}{a^2\omega^2}\frac{\partial\psi_\alpha}{\partial\pvect{n}}\right]\intd S
\end{multline}
Switching to the $\Psi$ notation and making some cancellations, we get
\begin{multline}
    \I a\omega\iint_S\psi_\alpha\bigg[h_s\left(1 - \frac{\pvect{\nabla}_\mathrm{t}^2}{a^2\omega^2}\right) + \frac{\kappa}{a^2\omega^2}\frac{\partial}{\partial\pvect{n}}\bigg]Q^a\intd S = \frac{1}{\I a\omega}\int_0^{2\uppi}\left[rh_s\psi_\alpha\frac{\partial Q^a}{\partial r}\right]_0^R\intd \theta \\ 
    + \I a\omega\Bigg\{\bigg[\left(\delta_{\alpha\delta} + \frac{\gmat{\Lambda}_{\alpha\delta}^2}{a^2\omega^2R^2}\right)\Psi_{\delta\beta\gamma}[rh_s] + \frac{\kappa}{a^2\omega^2}\Psi_{|\alpha|\beta\gamma}[r]\bigg]\sum_b\frac{P_\beta^{a - b}P_\gamma^b - U_\beta^{a - b}U_\gamma^b}{2} \\
    + \bigg[\left(\delta_{\alpha\delta} + \frac{\gmat{\Lambda}_{\alpha\delta}^2}{a^2\omega^2R^2}\right)\bigg(\Psi_{\delta[\beta][\gamma]}[rh_s] + \Psi_{\delta(\beta)(\gamma)}[h_s/r]\bigg) \\
    + \frac{\kappa}{a^2\omega^2}\bigg(\Psi_{|\alpha|[\beta][\gamma]}[r] + \Psi_{|\alpha|(\beta)(\gamma)}[1/r]\bigg)\bigg]\sum_b\frac{P_\beta^{a - b}P_\gamma^b}{2(a - b)b\omega^2}\Bigg\}.
\end{multline}
Combining this with the left-hand side of \eqref{3Delimmass}, the same leftover boundary term vanishes, to give the full projected mass conservation equation,
\begin{equation}
\label{3Dprojectedmass}
    \begin{aligned}
    \left(\delta_{\alpha\beta}\frac{\intd }{\intd s} \right.&\left.{}\!\!- \Psi_{\{\alpha\}\beta}[r]\right)U_\beta^a - \I a\omega\left[\left(1 - \frac{\lambda_\alpha^2}{a^2\omega^2R^2}\right)\Psi_{\alpha\beta}[rh_s] - \frac{\kappa}{a^2\omega^2}\Psi_{|\alpha|\beta}[r]\right]P_\beta^a \\
    =&\,  \I a\omega\sum_b\Bigg\{-\cnon\Psi_{\alpha\beta\gamma}[rh_s]P_\beta^{a - b}P_\gamma^b \\
    &+ \bigg[\left(\delta_{\alpha\delta} + \frac{\gmat{\Lambda}_{\alpha\delta}^2}{a^2\omega^2R^2}\right)\Psi_{\delta\beta\gamma}[rh_s] + \frac{\kappa}{a^2\omega^2}\Psi_{|\alpha|\beta\gamma}[r]\bigg]\frac{P_\beta^{a - b}P_\gamma^b - U_\beta^{a - b}U_\gamma^b}{2} \\
    &+ \bigg[\left(\delta_{\alpha\delta} + \frac{\gmat{\Lambda}_{\alpha\delta}^2}{a^2\omega^2R^2}\right)\bigg(\Psi_{\delta[\beta][\gamma]}[rh_s] + \Psi_{\delta(\beta)(\gamma)}[h_s/r]\bigg) \\
    &\quad\quad\quad\quad\quad\quad\quad+ \frac{\kappa}{a^2\omega^2}\bigg(\Psi_{|\alpha|[\beta][\gamma]}[r] + \Psi_{|\alpha|(\beta)(\gamma)}[1/r]\bigg)\bigg]\frac{P_\beta^{a - b}P_\gamma^b}{2(a - b)b\omega^2}\Bigg\}.
    \end{aligned}
\end{equation}
As before, the left-hand side of the momentum conservation equation \eqref{3Delimmom} has no boundary cancellations, so projection is very straightforward
\begin{equation}
    \iint_S\psi_\alpha\left(\frac{\partial P^a}{\partial s} - \I a\omega h_sU^a\right)\intd S = \left(\delta_{\alpha\beta}\frac{\intd }{\intd s} + \Psi_{\alpha\{\beta\}}[r]\right)P_\beta^a - \I a\omega\Psi_{\alpha\beta}[rh_s]U_\beta^a.
\end{equation}
Finally, the right-hand side of \eqref{3Delimmom} requires only that we deal carefully with the tranverse Laplacian, which produces a boundary term,
\begin{multline}
    \iint_S\I\omega h_s\psi_\alpha\sum_bU^{a - b}\left(-b\frac{\pvect{\nabla}_\mathrm{t}^2}{b^2\omega^2}P^b\right)\intd S = \I\omega\int_0^{2\uppi}\left[rh_s\psi_\alpha\sum_b\frac{-b}{b^2\omega^2}U^{a - b}\frac{\partial P^b}{\partial r}\right]_0^R\intd \theta \\
    - \I\omega\iint_S\sum_b\left(\frac{-b}{b^2\omega^2}\right)\pvect{\nabla}_\mathrm{t}(h_s\psi_\alpha U^{a - b})\pvect{\cdot}\pvect{\nabla}_\mathrm{t}(\psi_\gamma)P_\gamma^b\intd S.
\end{multline}
As took place in two dimensions, the boundary term here becomes a $U^{a - b}U^b$, which can be expanded, and the other term has its integration-by-parts reversed once more, leaving
\begin{multline}
    \iint_S\I\omega h_s\psi_\alpha\sum_bU^{a - b}\left(-b\frac{\pvect{\nabla}_\mathrm{t}^2}{b^2\omega^2}P^b\right)\intd S = R'\Psi_{[\alpha\beta\gamma]}\sum_bU_\beta^{a - b}U_\gamma^b \\
    - \I\omega\int_0^{2\uppi}\left[\sum_b\frac{-b}{b^2\omega^2}rh_s\psi_\alpha U^{a - b}\frac{\partial\psi_\gamma}{\partial r}P_\gamma^b\right]_0^R\intd \theta + \I\omega\sum_b\frac{b\gmat{\Lambda}_{\delta\gamma}^2}{b^2\omega^2R^2}\Psi_{\alpha\beta\delta}[h_s] U_\beta^{a - b}P_\gamma^b.
\end{multline}
The leftover boundary term here cancels due to the Neumann condition on the spatial modes, so the full projected version of \eqref{3Delimmom} is 
\begin{multline}
\label{3Dprojectedmom}
    \left(\delta_{\alpha\beta}\frac{\intd }{\intd s} + \Psi_{\alpha\{\beta\}}[r]\right)P_\beta^a - \I a\omega\Psi_{\alpha\beta}[rh_s]U_\beta^a = R'\Psi_{[\alpha\beta\gamma]}\sum_bU_\beta^{a - b}U_\gamma^b \\
    + \I\omega\sum_b\Bigg\{\Psi_{\alpha\beta\delta}[rh_s]\left[(a - b)\delta_{\delta\gamma} - b\left(\delta_{\delta\gamma} - \frac{\gmat{\Lambda}_{\delta\gamma}^2}{b^2\omega^2R^2}\right)\right] \\
    + \frac{\Psi_{\alpha[\beta][\gamma]}[rh_s] + \Psi_{\alpha(\beta)(\gamma)}[h_s/r]}{b\omega^2}\Bigg\}U_\beta^{a - b}P_\gamma^b.
\end{multline}
We now have the same task of rewriting these equations in a more numerically efficient way. The added complication in three dimensions is that these are double integrals, which means that they must be split into separate radial and azimuthal parts; nonetheless, everything may still be made independent of $s$ just as before. The notation for the split integrals is very similar to that of the $\Psi$ matrices, e.g.
\begin{equation}
    \Pi_{\alpha[\beta]}[f(x)] = C_\alpha C_\beta\int_{x = 0}^1\J_{p_\alpha}(\lambda_\alpha x)\frac{\partial}{\partial x}\left[\J_{p_\beta}(\lambda_\beta x)\right] f(x)\intd x,
\end{equation}
and
\begin{equation}
    \Phi_{\alpha(\beta)}[g(\phi)] = \frac{1}{\uppi}\int_{\theta = 0}^{2\uppi}\cos(p_\alpha\phi - \xi_\alpha\frac{\uppi}{2})\frac{\partial}{\partial\theta}\left[\cos(p_\beta\phi - \xi_\beta\frac{\uppi}{2})\right]g(\phi)\intd \theta.
\end{equation}
Once again, we need to know how $s$-derivatives of modes work: in three dimensions, the relevant equation is
\begin{equation}
    \frac{\partial\psi_\alpha}{\partial s} = -\frac{R'}{R}\left(\psi_\alpha + r\frac{\partial\psi_\alpha}{\partial r}\right) - \tau\frac{\partial\psi_\alpha}{\partial\theta},
\end{equation}
while the analogous improved forms of matrices with derivatives on multiple subscripts are
\begin{subequations}
    \begin{multline}
        \Psi_{\alpha[\beta][\gamma]}[rh_s] + \Psi_{\alpha(\beta)(\gamma)}[h_s/r] \\
        = \frac{\lambda_\beta^2 + \lambda_\gamma^2 - \lambda_\alpha^2}{2R^2}\Psi_{\alpha\beta\gamma}[rh_s] - \kappa\Psi_{[\alpha]\beta\gamma}[r] + \frac{\kappa}{2}\Psi_{[\alpha\beta\gamma]}[r\cos\phi],
    \end{multline}
    \begin{multline}
        \Psi_{|\alpha|[\beta][\gamma]}[r] + \Psi_{|\alpha|(\beta)(\gamma)}[1/r] \\
        = \frac{\lambda_\beta^2 + \lambda_\gamma^2 - \lambda_\alpha^2}{2R^2}\Psi_{|\alpha|\beta\gamma}[r] + \frac{\lambda_\alpha^2 - p_\alpha^2}{2R^2}\Psi_{[\alpha\beta\gamma]}[r\cos\phi] - \frac{1}{2R}\Psi_{[(\alpha)\beta\gamma]}[\sin\phi].
    \end{multline}
\end{subequations}
Equipped with the above, \eqref{3Dprojectedmass} and \eqref{3Dprojectedmom} become
\begin{subequations}\label{app/3Dvector}
    \begin{multline}
    \label{3Dvectormass}
        \left(\frac{\intd }{\intd s} + \frac{R'}{R}\mat{W} + \tau\mat{H}\right)\mvect{u}^a - \I a\omega\left[\left(\mat{I} - \frac{\gmat{\Lambda}^2}{a^2\omega^2R^2}\right)\left(\mat{I} - \kappa R\mat{A}\right) - \frac{\kappa\widetilde{\mat{A}}}{a^2\omega^2R}\right]\mvect{p}^a \\
        = \frac{\I a\omega}{\sqrt{\uppi}R}\sum_b\Bigg\{\left[\left(\mat{I} + \frac{\gmat{\Lambda}^2}{a^2\omega^2R^2}\right)\left(\mathcal{I} - \kappa R\mathcal{A}\right) + \frac{\kappa\widetilde{\mathcal{A}}}{a^2\omega^2 R}\right]\frac{\langle\mvect{p}^{a - b},\mvect{p}^b\rangle - \langle\mvect{u}^{a - b},\mvect{u}^b\rangle}{2} \\
        + \left[\frac{\left(\mat{I} + \frac{\gmat{\Lambda}^2}{a^2\omega^2R^2}\right)\left(\mathcal{I}^\lambda - \kappa R\mathcal{A}^\lambda\right) + \frac{\kappa\widetilde{\mathcal{A}}^\lambda}{a^2\omega^2R}}{2(a - b)b\omega^2R^2} - \cnon\left(\mathcal{I} - \kappa R\mathcal{A}\right)\right]\langle\mvect{p}^{a - b},\mvect{p}^b\rangle\Bigg\},
    \end{multline}
    \begin{multline}
    \label{3Dvectormom}
        \left(\frac{\intd }{\intd s} - \frac{R'}{R}\mat{W}^\T - \tau\mat{H}^\T\right)\mvect{p}^a - \I a\omega\bigg(\mat{I} - \kappa R\mat{A}\bigg)\mvect{u}^a = \frac{1}{\sqrt{\uppi}R}\sum_b\Bigg\{\frac{R'}{R}\overline{\mathcal{W}}\langle\mvect{u}^{a - b},\mvect{u}^b\rangle \\
        + \I\omega\left[\left(\mathcal{I} - \kappa R\mathcal{A}\right)\bigg\langle\mat{I},(a - b)\mat{I} - b\left(\mat{I} - \frac{\gmat{\Lambda}^2}{b^2\omega^2R^2}\right)\bigg\rangle + \frac{\mathcal{I}^\lambda - \kappa R\mathcal{A}^\lambda}{b\omega^2R^2}\right]\langle\mvect{u}^{a - b},\mvect{p}^b\rangle\Bigg\},
    \end{multline}
\end{subequations}
where
\begin{align}
    \gmat{\Lambda}_{\alpha\beta} &= \lambda_\alpha\delta_{\alpha\beta}, 
    &
    \mat{P}_{\alpha\beta} &= p_\alpha\delta_{\alpha\beta}, \\
    \mat{W}_{\alpha\beta} &= \delta_{\alpha\beta} + \Pi_{[\alpha]\beta}[x^2]\Phi_{\alpha\beta}, 
    &
    \mat{H}_{\alpha\beta} &= \Pi_{\alpha\beta}[x]\Phi_{(\alpha)\beta}, \\
    \mat{A}_{\alpha\beta} &= \Pi_{\alpha\beta}[x^2]\Phi_{\alpha\beta}[\cos\phi],
    &
    \widetilde{\mat{A}}_{\alpha\beta} &= \Pi_{[\alpha]\beta}[x]\Phi_{\alpha\beta}[\cos\phi]  - \Pi_{\alpha\beta}\Phi_{(\alpha)\beta}[\sin\phi], \\
    \mathcal{I}_{\alpha\beta\gamma} &= \Pi_{\alpha\beta\gamma}[x]\Phi_{\alpha\beta\gamma}, 
    &
    \overline{\mathcal{W}}_{\alpha\beta\gamma} &= \Pi_{[\alpha\beta\gamma]}\Phi_{\alpha\beta\gamma}, \\
    \mathcal{A}_{\alpha\beta\gamma} &= \Pi_{\alpha\beta\gamma}[x^2]\Phi_{\alpha\beta\gamma}[\cos\phi],
    &
    \widetilde{\mathcal{A}}_{\alpha\beta\gamma} &= \Pi_{[\alpha]\beta\gamma}[x]\Phi_{\alpha\beta\gamma}[\cos\phi]  - \Pi_{\alpha\beta\gamma}\Phi_{(\alpha)\beta\gamma}[\sin\phi], \\
    \overline{\mathcal{A}}_{\alpha\beta\gamma} &= \Pi_{[\alpha\beta\gamma]}\Phi_{\alpha\beta\gamma}[\cos\phi],
    &
    \overline{\mathcal{A}}_{\alpha\beta\gamma}^* &= \Pi_{[\alpha\beta\gamma]}\Phi_{(\alpha)\beta\gamma}[\sin\phi],
\end{align}
and the eigenvalue-modified matrices $\mathcal{I}^\lambda$, $\mathcal{A}^\lambda$ and $\widetilde{\mathcal{A}}^\lambda$ are defined by
\begin{align}
    \mathcal{I}^\lambda &= \mathcal{I}\frac{\langle\gmat{\Lambda}^2,\mat{I}\rangle + \langle\mat{I},\gmat{\Lambda}^2\rangle - \gmat{\Lambda}^2}{2}, \\
    \mathcal{A}^\lambda &= \mathcal{A}\frac{\langle\gmat{\Lambda}^2,\mat{I}\rangle + \langle\mat{I},\gmat{\Lambda}^2\rangle - \gmat{\Lambda}^2}{2} + \widetilde{\mathcal{A}} - \frac{\overline{\mathcal{A}}}{2}, \\
    \widetilde{\mathcal{A}}^\lambda &= \widetilde{\mathcal{A}}\frac{\langle\gmat{\Lambda}^2,\mat{I}\rangle + \langle\mat{I},\gmat{\Lambda}^2\rangle - \gmat{\Lambda}^2}{2} + \frac{(\gmat{\Lambda}^2 - \mat{P}^2)\overline{\mathcal{A}} - \overline{\mathcal{A}}^*}{2}.
\end{align}
As before, notation has been chosen such that curvature acts through \textit{annularity} matrices ($\mat{A}$ etc), \textit{width} variation through $\mat{W}$, and with the new quantity, torsion, through a \textit{helicity} matrix $\mat{H}$.  Equations~\eqref{app/3Dvector} are equations~\eqref{equ:3Dvector} in the main text.

\subsection{Evaluation of integrals}
Here the mode-integrals employed in the projection of the governing equations above are calculated explicitly, where possible. These are the expressions used when performing computations, in order to save time.
\allowdisplaybreaks
\subsubsection{Integrals from the two-dimensional formulation}
\begin{align}
    \Xi_{\alpha\beta} &= \delta_{\alpha\beta}, \\
    \Xi_{\alpha\beta}[\xi] &= \frac{\delta_{\alpha\beta}}{2} + C_\alpha C_\beta\frac{\left((-1)^{\alpha + \beta} - 1\right)(\alpha^2 + \beta^2)}{(\alpha^2 - \beta^2)^2\uppi^2}, \\
    \Xi_{[\alpha]\beta} &= \sqrt{2}\alpha^2 C_\beta\frac{(-1)^{\alpha + \beta} - 1}{\alpha^2 - \beta^2}, \\
    \Xi_{[\alpha]\beta}[\xi] &= \delta_{\alpha\beta}\frac{(1 - \delta_{\alpha0})}{2} + (1 - \delta_{\alpha\beta})\frac{\sqrt{2}\alpha^2C_\beta(-1)^{\alpha + \beta}}{\alpha^2 - \beta^2}.\displaybreak[0]\\
    \Xi_{\alpha\beta\gamma} &= \frac{C_\alpha C_\beta}{2 C_\gamma}(\delta_{\alpha+\beta,\gamma} + \delta_{|\alpha - \beta|,\gamma}), \\
    \Xi_{\alpha\beta\gamma}[\xi] &= \frac{\Xi_{\alpha\beta\gamma}}{2} + \frac{C_\alpha C_\beta C_\gamma\left((-1)^{\alpha + \beta + \gamma} - 1\right)}{4\uppi^2}\Bigg(\frac{1}{(\alpha + \beta + \gamma)^2}\notag \\
    &\quad\quad\quad+ \frac{1}{(\alpha + \beta - \gamma)^2} + \frac{1}{(\alpha - \beta + \gamma)^2} + \frac{1}{(\alpha - \beta + \gamma)^2}\Bigg), \\
    \Xi_{[\alpha]\beta\gamma} &= \frac{\alpha C_\beta C_\gamma\left((-1)^{\alpha + \beta + \gamma} - 1\right)}{2\sqrt{2}}\Bigg(\frac{1}{\alpha + \beta - \gamma} + \frac{1}{\alpha - \beta + \gamma} + \frac{1}{\alpha - \beta - \gamma}\Bigg), \\
    \Xi_{[\alpha\beta\gamma]} &= C_\alpha C_\beta C_\gamma\left((-1)^{\alpha + \beta + \gamma} - 1\right), \\
    \Xi_{[\alpha\beta\gamma]}^+ &= C_\alpha C_\beta C_\gamma(-1)^{\alpha + \beta + \gamma}, \\
    \Xi_{[\alpha\beta\gamma]}^- &= C_\alpha C_\beta C_\gamma.
\end{align}
\subsubsection{Integrals from the three-dimensional formulation}
\begin{align}
    \Pi_{\alpha\beta} &= C_\alpha C_\beta\int_0^1 J_{p_\alpha}(\lambda_\alpha x)J_{p_\beta}(\lambda_\beta x)\mathrm{d}x, \\
    \Pi_{\alpha\beta}[x] &= C_\alpha C_\beta\int_0^1xJ_{p_\alpha}(\lambda_\alpha x)J_{p_\beta}(\lambda_\beta x)\mathrm{d}x, \\
    \Pi_{\alpha\beta}[x^2] &= C_\alpha C_\beta\int_0^1 x^2J_{p_\alpha}(\lambda_\alpha x)J_{p_\beta}(\lambda_\beta x)\mathrm{d}x, \\
    \Pi_{[\alpha]\beta}[x] &= C_\alpha C_\beta\int_0^1 \left[p_\alpha J_{p_\alpha}(\lambda_\alpha x) - \lambda_\alpha  xJ_{p_\alpha + 1}(\lambda_\alpha x)\right]J_{p_\beta}(\lambda_\beta x)\mathrm{d}x, \\
    \Pi_{[\alpha]\beta}[x^2] &= C_\alpha C_\beta\int_0^1 x\left[p_\alpha J_{p_\alpha}(\lambda_\alpha x) - \lambda_\alpha  xJ_{p_\alpha + 1}(\lambda_\alpha x)\right]J_{p_\beta}(\lambda_\beta x)\mathrm{d}x, \\
    \Phi_{\alpha\beta} &= (\delta_{p_\alpha + p_\beta,0} +  \delta_{p_\alpha p_\beta})\delta_{\xi_\alpha\xi_\beta}, \\
    \Phi_{(\alpha)\beta} &= -p_\alpha(-1)^{\xi_\alpha}\delta_{p_\alpha p_\beta}\delta_{\xi_\alpha + \xi_\beta,1}, \\
    \Phi_{\alpha\beta}[\cos\phi] &= \frac{1}{2}\left[(-1)^{\xi_\alpha}\delta_{p_\alpha + p_\beta,1} + \delta_{|p_\alpha - p_\beta|,1}\right]\delta_{\xi_\alpha\xi_\beta}, \\
    \Phi_{\alpha\beta}[\sin\phi] &= \frac{1}{2}\left[\delta_{p_\alpha + p_\beta,1} + (-1)^{\xi_\beta}(p_\alpha - p_\beta)\delta_{|p_\alpha - p_\beta|,1}\right]\delta_{\xi_\alpha + \xi_\beta,1}, \\
    \Phi_{(\alpha)\beta}[\sin\phi] &= -\frac{p_\alpha}{2}\left[\delta_{p_\alpha1}\delta_{p_\beta0}\delta_{\xi_\alpha0} + (p_\alpha - p_\beta)\delta_{|p_\alpha - p_\beta|,1}\right]\delta_{\xi_\alpha\xi_\beta}.
\end{align}
The linear coefficient matrices in three dimensions always consist of a combination of a $\Pi$ with a $\Phi$, which rarely simplifies further, although the notable cases are the identity,
\begin{equation}
    \mathsf{I}_{\alpha\beta} = \Pi_{\alpha\beta}[x]\Phi_{\alpha\beta} = \delta_{\alpha\beta},
\end{equation}
and the helicity matrix,
\begin{equation}
    \mathsf{H}_{\alpha\beta} = \Pi_{\alpha\beta}[x]\Phi_{(\alpha)\beta}  = -p_\alpha(-1)^{\xi_\alpha}\delta_{\xi_\alpha+\xi_\beta,1}\delta_{p_\alpha p_\beta}\delta_{q_\alpha q_\beta}.
\end{equation}
Likewise,
\begin{align}
    \Pi_{\alpha\beta\gamma} &= C_\alpha C_\beta C_\gamma \int_0^1 J_{p_\alpha}(\lambda_\alpha x)J_{p_\beta}(\lambda_\beta x)J_{p_\gamma}(\lambda_\gamma x)\mathrm{d}x, \\
    \Pi_{\alpha\beta\gamma}[x] &= C_\alpha C_\beta C_\gamma \int_0^1 xJ_{p_\alpha}(\lambda_\alpha x)J_{p_\beta}(\lambda_\beta x)J_{p_\gamma}(\lambda_\gamma x)\mathrm{d}x, \\
    \Pi_{\alpha\beta\gamma}[x^2] &= C_\alpha C_\beta C_\gamma \int_0^1 x^2J_{p_\alpha}(\lambda_\alpha x)J_{p_\beta}(\lambda_\beta x)J_{p_\gamma}(\lambda_\gamma x)\mathrm{d}x, \\
    \Pi_{[\alpha]\beta\gamma}[x] &= C_\alpha C_\beta C_\gamma \int_0^1 \bigg(p_\alpha J_{p_\alpha}(\lambda_\alpha x) - \lambda_\alpha xJ_{p_\alpha + 1}(\lambda_\alpha x)\bigg)J_{p_\beta}(\lambda_\beta x)J_{p_\gamma}(\lambda_\gamma x)\mathrm{d}x, \\
    \Pi_{[\alpha\beta\gamma]}[x] &= C_\alpha C_\beta C_\gamma J_{p_\alpha}(\lambda_\alpha )J_{p_\beta}(\lambda_\beta)J_{p_\gamma}(\lambda_\gamma), \\
    \Phi_{\alpha\beta\gamma} &= \frac{1}{2}\bigg(\delta_{p_\alpha + p_\beta + p_\gamma,0} + (-1)^{\xi_\alpha}\delta_{p_\beta + p_\gamma,p_\alpha} + (-1)^{\xi_\beta}\delta_{p_\alpha + p_\gamma,p_\beta}\notag \\
    &\quad\quad\quad\quad\quad + (-1)^{\xi_\gamma}\delta_{p_\alpha + p_\beta,p_\gamma}\bigg)\bigg(\delta_{\xi_\alpha + \xi_\beta + \xi_\gamma,0} - \delta_{\xi_\alpha + \xi_\beta + \xi_\gamma,2}\bigg), \\
    \Phi_{\alpha\beta\gamma}[\cos\phi] &= \frac{1}{4}\bigg(\delta_{p_\alpha + p_\beta + p_\gamma,1} + (-1)^{\xi_\alpha}\delta_{|p_\beta + p_\gamma - p_\alpha|,1} + (-1)^{\xi_\beta}\delta_{|p_\alpha + p_\gamma - p_\beta|,1}\notag \\
    &\quad\quad\quad\quad\quad + (-1)^{\xi_\gamma}\delta_{|p_\alpha + p_\beta - p_\gamma|,1}\bigg)\bigg(\delta_{\xi_\alpha + \xi_\beta + \xi_\gamma,0} - \delta_{\xi_\alpha + \xi_\beta + \xi_\gamma,2}\bigg), \\
    \Phi_{(\alpha)\beta\gamma}[\sin\phi] &= -\frac{p_\alpha}{4}\Bigg(\delta_{p_\alpha + p_\beta + p_\gamma,1} + \delta_{|p_\beta + p_\gamma - p_\alpha|,1}(-1)^{p_\beta + p_\gamma - p_\alpha + \xi_\alpha}\notag \\
    &\quad + \delta_{|p_\alpha + p_\gamma - p_\beta|,1}(-1)^{p_\alpha + p_\gamma - p_\beta + \xi_\beta} + \delta_{|p_\alpha + p_\beta - p_\gamma|,1}(-1)^{p_\alpha + p_\beta - p_\gamma + \xi_\gamma}\bigg]\Bigg)\notag \\
    &\quad\quad\quad\quad\quad\quad\quad\quad\quad\quad\quad\quad\quad\quad\quad\quad\quad\times\bigg(\delta_{\xi_\alpha + \xi_\beta + \xi_\gamma,0} - \delta_{\xi_\alpha + \xi_\beta + \xi_\gamma,2}\bigg).
\end{align}

\section{Definitions of block matrices}
\label{blockdetailapp}
In this appendix we detail the entries of the block matrix $\mat{L}^a$ and block tensor $\mathcal{N}^{ab}$ from equation \eqref{LandNintro}. In terms of square blocks $\mat{L}^a =: \begin{pmatrix}
        \mat{L}_1^a &\mat{L}_2^a \\ \mat{L}_3^a &\mat{L}_4^a
    \end{pmatrix}$, we have a two-dimensional $\mat{L}^a$
\begin{equation}
    \mat{L}^a = \begin{pmatrix}
        -\frac{X'}{2X}\mat{W} - \frac{X_-'}{X}\widetilde{\mat{A}} &\I a\uomega\left[\left(\mat{I} - \frac{\gmat{\Lambda}^2}{a^2\omega^2X^2}\right)\left(\mat{I} - \kappa X\mat{A}\right) - \frac{\kappa\widetilde{\mat{A}}}{a^2\omega^2X}\right] \\
        \I a\uomega\bigg(\mat{I} - \kappa X\mat{A}\bigg) &\frac{X'}{2X}\mat{W}^T + \frac{X_-'}{X}\widetilde{\mat{A}}^T
    \end{pmatrix},
\end{equation}
and a three-dimensional $\mat{L}^a$
\begin{equation}
    \mat{L}^a = \begin{pmatrix}
        -\frac{R'}{R}\mat{W} - \tau\mat{H} &\I a\uomega\left[\left(\mat{I} - \frac{\gmat{\Lambda}^2}{a^2\omega^2R^2}\right)\left(\mat{I} - \kappa R\mat{A}\right) - \frac{\kappa\widetilde{\mat{A}}}{a^2\omega^2R}\right] \\
        \I a\uomega\bigg(\mat{I} - \kappa R\mat{A}\bigg) &\frac{R'}{R}\mat{W}^T + \tau\mat{H}^T
    \end{pmatrix},
\end{equation}
and in terms of `cubic blocks'
\begin{align}
    &\mathcal{N}_{\alpha,\beta,0:\alpha_\text{max}}^{ab} = \begin{pmatrix}
        \mathcal{N}_1^{ab} &\mathcal{N}_2^{ab} \\ \mathcal{N}_3^{ab} &\mathcal{N}_4^{ab}
    \end{pmatrix}, && \mathcal{N}_{\alpha,\beta,\alpha_\text{max}+1:2\alpha_\text{max}+1}^{ab} = \begin{pmatrix}
        \mathcal{N}_5^{ab} &\mathcal{N}_6^{ab} \\ \mathcal{N}_7^{ab} &\mathcal{N}_8^{ab}
    \end{pmatrix},
\end{align}
we have, in two dimensions
\begin{align}
    \mathcal{N}_1^{ab} &= -\frac{\I a\uomega}{2\sqrt{X}}\left[\left(\mat{I} + \frac{\gmat{\Lambda}^2}{a^2\omega^2X^2}\right)\left(\mathcal{I} - \kappa X\mathcal{A}\right) + \frac{\kappa\widetilde{\mathcal{A}}}{a^2\omega^2 X}\right], \\
    \mathcal{N}_3^{ab} &= \frac{1}{\sqrt{X}}\bigg(\frac{X_+'}{X}\overline{\mathcal{W}}^+ - \frac{X_-'}{X}\overline{\mathcal{W}}^-\bigg), \\
    \mathcal{N}_6^{ab} &= -\mathcal{N}_1^{ab} + \frac{\I a\uomega}{\sqrt{X}}\left[\frac{\left(\mat{I} + \frac{\gmat{\Lambda}^2}{a^2\omega^2X^2}\right)\left(\mathcal{I}^\lambda - \kappa X\mathcal{A}^\lambda\right) + \frac{\kappa\widetilde{\mathcal{A}}^\lambda}{a^2\omega^2X}}{2(a - b)b\omega^2X^2} - \cnon\left(\mathcal{I} - \kappa R\mathcal{A}\right)\right], \\
    \mathcal{N}_7^{ab} &= \frac{\I \uomega}{\sqrt{X}}\left\{\left(\mathcal{I} - \kappa X\mathcal{A}\right)\bigg\langle\mat{I},(a - b)\mat{I} - b\left(\mat{I} - \frac{\gmat{\Lambda}^2}{b^2\omega^2X^2}\right)\bigg\rangle + \frac{\mathcal{I}^\lambda - \kappa X\mathcal{A}^\lambda}{b\omega^2X^2}\right\},
\end{align}
and in three dimensions
\begin{align}
    \mathcal{N}_1^{ab} &= -\frac{\I a\uomega}{2\sqrt{\uppi}R}\left[\left(\mat{I} + \frac{\gmat{\Lambda}^2}{a^2\omega^2R^2}\right)\left(\mathcal{I} - \kappa R\mathcal{A}\right) + \frac{\kappa\widetilde{\mathcal{A}}}{a^2\omega^2 R}\right], \\
    \mathcal{N}_3^{ab} &= \frac{R'}{\sqrt{\uppi}R^2}\overline{\mathcal{W}}, \\
    \mathcal{N}_6^{ab} &= -\mathcal{N}_1^{ab} + \frac{\I a\uomega}{\sqrt{\uppi}R}\left[\frac{\left(\mat{I} + \frac{\gmat{\Lambda}^2}{a^2\omega^2R^2}\right)\left(\mathcal{I}^\lambda - \kappa R\mathcal{A}^\lambda\right) + \frac{\kappa\widetilde{\mathcal{A}}^\lambda}{a^2\omega^2R}}{2(a - b)b\omega^2R^2} - \cnon\left(\mathcal{I} - \kappa R\mathcal{A}\right)\right], \\
    \mathcal{N}_7^{ab} &= \frac{\I \uomega}{\sqrt{\uppi}R}\left\{\left(\mathcal{I} - \kappa R\mathcal{A}\right)\bigg\langle\mat{I},(a - b)\mat{I} - b\left(\mat{I} - \frac{\gmat{\Lambda}^2}{b^2\omega^2R^2}\right)\bigg\rangle + \frac{\mathcal{I}^\lambda - \kappa R\mathcal{A}^\lambda}{b\omega^2R^2}\right\},
\end{align}
while in both two and three dimensions $\mathcal{N}_2^{ab} = \mathcal{N}_4^{ab} = \mathcal{N}_5^{ab} = \mathcal{N}_8^{ab} = 0$.
\section{Characteristic admittances of non-straight ducts}
\label{charadmapp}
This appendix deals with the non-straight-duct versions of section \ref{straightductcharadmderivation}.
\subsection{Curved-duct characteristic admittances}
When we introduce curvature, the straight-duct operator from equation \eqref{straightductoperator} retains its block structure
\begin{equation}
    \breve{\mat{L}}^a = \mat{L}^a\bigg|_{\text{no}~\tau,~X'~\text{or}~R'} = \begin{pmatrix}
        \mat{0} &\breve{\mat{L}}_2^a \\
        \breve{\mat{L}}_3^a &\mat{0}
    \end{pmatrix},
\end{equation}
with the difference that $\breve{\mat{L}}_2^a$ and $\breve{\mat{L}}_3^a$ are no longer diagonal. They do remain symmetric, however (by inspection in $\breve{\mat{L}}_3^a$'s case, and by a complicated integration-by-parts in $\breve{\mat{L}}_2^a$'s). The eigenvalues $\breve{\gamma}^a$ now satisfy (from equation \eqref{dettheorem})
\begin{equation}
\label{curvedductchareq}
    0 = \det\left[\breve{\mat{L}}^a - \breve{\gamma}^a\begin{pmatrix}
        \mat{I} &\mat{0} \\ \mat{0} &\mat{I}
    \end{pmatrix}\right] = \det\left((\breve{\gamma}^{a})^2\mat{I} - \breve{\mat{L}}_3^a\breve{\mat{L}}_2^a\right).
\end{equation}
so they will exhibit the same properties as before (purely real/imaginary and being mirrored either side of the origin). We once again have $\breve{\mat{L}}_3^a\breve{\mat{L}}_2^a\breve{\mvect{c}}_{\alpha,2}^{a\pm} = (\breve{\gamma}_\alpha^a)^2\breve{\mvect{c}}_{\alpha,2}^{a\pm}$, so a single lower-eigenvector matrix $\breve{\mat{C}}^a$ still exists, but this time it is non-diagonal. Substituting the admittance into the lower block this time, we get
\begin{align}
    &\breve{\mat{L}}_3^a\breve{\mat{Y}}^{a\pm}\breve{\mat{C}}^a = \pm\breve{\mat{C}}^a\breve{\gmat{\Gamma}}^a,&& \text{so}&& \breve{\mat{Y}}^{a\pm} = \pm(\breve{\mat{L}}_3^a)^{-1}\breve{\mat{C}}^a\breve{\gmat{\Gamma}}^a(\breve{\mat{C}}^a)^{-1}  =: \pm\breve{\mat{Y}}^a.
\end{align}
The nonlinear characteristic admittance takes a bit more work this time: the term multiplying $\breve{\mathcal{Y}}^{ab\pm}$ in the nonlinear admittance equation is no longer diagonal. We have already diagonalised $\breve{\mat{L}}_3^a\breve{\mat{Y}}^{a\pm}$, but we also need to diagonalise $\breve{\mat{Y}}^{a\pm}\breve{\mat{L}}_3^a$. To do this we note that \eqref{linadmeq} is satisfied by $(\mat{Y}^a)^\T$ as well as $\mat{Y}^a$, so with symmetric $\breve{\mat{L}}_3^a$, we have
\begin{equation}
    \breve{\mat{Y}}^{a\pm}\breve{\mat{L}}_3^a = (\breve{\mat{L}}_3^a\breve{\mat{Y}}^{a\pm})^\T = \pm\left(\breve{\mat{C}}^a\right)^{-\T}\breve{\gmat{\Gamma}}^a\left(\breve{\mat{C}}^a\right)^\T.
\end{equation}
With both diagonalisations, we now have
\begin{multline}
    \breve{\mathcal{Y}}_{\alpha\beta\gamma}^{ab\pm} = \pm\sum_{\zeta,\delta,\epsilon}\left(\breve{\mat{C}}^a\right)_{\zeta\alpha}^{-1}\left(\breve{\mat{C}}^{a - b}\right)_{\delta\beta}^{-1}\left(\breve{\mat{C}}^b\right)_{\epsilon\gamma}^{-1} \\
    \times\frac{\left[\left(\breve{\mat{C}}^a\right)^{\T}\left[\breve{\mathcal{N}}_1^{ab}\langle\breve{\mat{Y}}^{a - b},\breve{\mat{Y}}^b\rangle + \breve{\mathcal{N}}_6^{ab} - \breve{\mat{Y}}^a\breve{\mathcal{N}}_7^{ab}\langle\breve{\mat{Y}}^{a - b},\mat{I}\rangle\right]\left\langle\breve{\mat{C}}^{a - b},\breve{\mat{C}}^b\right\rangle\right]_{\zeta\delta\epsilon}}{\breve{\gamma}_\zeta^a + \breve{\gamma}_\delta^{a - b} + \breve{\gamma}_\epsilon^b}
\end{multline}
\subsection{Torsional-duct characteristic admittances}
The final parameter that we can switch on is torsion, and this is only possible in the three-dimensional case. When we do so, the matrix $\mat{L}^a$ becomes
\begin{equation}
    \widetilde{\mat{L}}^a = \mat{L}^a\bigg|_{\text{no}~R'} = \begin{pmatrix}
        \widetilde{\mat{L}}_1^a &\widetilde{\mat{L}}_2^a \\
        \widetilde{\mat{L}}_3^a &\widetilde{\mat{L}}_1^a
    \end{pmatrix},
\end{equation}
where $\widetilde{\mat{L}}_2^a$ and $\widetilde{\mat{L}}_3^a$ remain real and symmetric, and $\widetilde{\mat{L}}_1^a$ is a real antisymmetric matrix (apparent by direct calculation of the matrix $\mat{H}$). Since a linear combination of the identity and an antisymmetric matrix is non-singular, we know that the characteristic equation reads
\begin{equation}
    0 = \det\left(\left[\widetilde{\mat{L}}_1^a - \widetilde{\gamma}^{a}\mat{I}\right]^2 - \left[\widetilde{\mat{L}}_1^a - \widetilde{\gamma}^{a}\mat{I}\right]\widetilde{\mat{L}}_3^a\left[\widetilde{\mat{L}}_1^a - \widetilde{\gamma}^{a}\mat{I}\right]^{-1}\widetilde{\mat{L}}_2^a\right),
\end{equation}
but also, from the alternative expression in \eqref{dettheorem}, that we have
\begin{equation}
    0 = \det\left(\left[\widetilde{\mat{L}}_1^a - \widetilde{\gamma}^{a}\mat{I}\right]^2 - \widetilde{\mat{L}}_2^a\left[\widetilde{\mat{L}}_1^a - \widetilde{\gamma}^{a}\mat{I}\right]^{-1}\widetilde{\mat{L}}_3^a\left[\widetilde{\mat{L}}_1^a - \widetilde{\gamma}^{a}\mat{I}\right]\right).
\end{equation}
The expression on the right will have the same determinant if we take its transpose. Doing so reverses the signs on every term except those with a $\widetilde{\gamma}^a\mat{I}$, demonstrating that the eigenvalue mirroring property remains, even if the eigenvalues are no longer necessarily on the real/imaginary axes. We may still partition them into two matched groups, so we define
\begin{equation}
    \{\widetilde{\gamma}_\alpha^a\}_{\alpha = 0}^\infty = \{\widetilde{\gamma}^a : \mathrm{Re}(\widetilde{\gamma}^a) < 0\}\cup\{\widetilde{\gamma}^a : \mathrm{Re}(\widetilde{\gamma}^a) = 0,\mathrm{Im}(\widetilde{\gamma}^a) > 0\},
\end{equation}
a set that we use to construct the positive characteristic admittance. These correspond to purely-forward-propagating waves, as well as forward-decaying waves that may propagate either forwards or backwards.

We must also now keep track of eigenvalue signs when defining eigenvectors. If we define the diagonal $\widetilde{\gmat{\Gamma}}^a$ matrix, and a general acoustic eigenvector $(\widetilde{\mvect{c}}_{\alpha,1}^{a\pm},\widetilde{\mvect{c}}_{\alpha,2}^{a\pm})$, there will exist matrices $\widetilde{\mat{C}}^{a\pm}$, constructed from an ordering of either the `$+$' lower eigenvectors or the `$-$'. The lower-block admittance equation is then
\begin{align}
    &(\widetilde{\mat{L}}_3^a\widetilde{\mat{Y}}^{a\pm} + \widetilde{\mat{L}}_1^a)\widetilde{\mat{C}}^{a\pm} = \pm\widetilde{\mat{C}}^{a\pm}\widetilde{\gmat{\Gamma}}^{a},&&\text{so} && \widetilde{\mat{Y}}^{a\pm} = \left(\widetilde{\mat{L}}_3^a\right)^{-1}\left[-\widetilde{\mat{L}}_1^a \pm \widetilde{\mat{C}}^{a\pm}\widetilde{\gmat{\Gamma}}^{a}\left(\widetilde{\mat{C}}^{a\pm}\right)^{-1}\right],
\end{align}
from which we deduce that the torsional entries $\widetilde{\mat{L}}_1^a$ have caused a symmetry-breaking between the positive and negative characteristic admittances. When it comes to computing the nonlinear admittances, we once again must take a transpose
\begin{equation}
    \widetilde{\mat{Y}}^{a\pm}\widetilde{\mat{L}}_3^a - \widetilde{\mat{L}}_1^a = \left(\widetilde{\mat{L}}_3^a\widetilde{\mat{Y}}^{a\pm} + \widetilde{\mat{L}}_1^a\right)^\T = \pm\left(\widetilde{\mat{C}}^{a\pm}\right)^{-\T}\widetilde{\gmat{\Gamma}}^{a}\left(\widetilde{\mat{C}}^{a\pm}\right)^\T,
\end{equation}
meaning that the nonlinear characteristic admittances are given by
\begin{multline}
    \widetilde{\mathcal{Y}}_{\alpha\beta\gamma}^{ab\pm} = \pm\sum_{\zeta,\delta,\epsilon}\left(\widetilde{\mat{C}}^{a\pm}\right)_{\zeta\alpha}^{-1}\left(\widetilde{\mat{C}}^{(a - b)\pm}\right)_{\delta\beta}^{-1}\left(\widetilde{\mat{C}}^{b\pm}\right)_{\epsilon\gamma}^{-1} \\
    \times\frac{\left[\left(\widetilde{\mat{C}}^{a\pm}\right)^{\T}\left[\widetilde{\mathcal{N}}_1^{ab}\langle\widetilde{\mat{Y}}^{(a - b)\pm},\widetilde{\mat{Y}}^{b\pm}\rangle + \widetilde{\mathcal{N}}_6^{ab} - \widetilde{\mat{Y}}^{a\pm}\widetilde{\mathcal{N}}_7^{ab}\langle\widetilde{\mat{Y}}^{(a - b)\pm},\mat{I}\rangle\right]\left\langle\widetilde{\mat{C}}^{(a - b)\pm},\widetilde{\mat{C}}^{b\pm}\right\rangle\right]_{\zeta\delta\epsilon}}{\widetilde{\gamma}_\zeta^a + \widetilde{\gamma}_\delta^{a - b} + \widetilde{\gamma}_\epsilon^b}.
\end{multline}

\section{Viscosity in more detail}
\label{numviscapp}
\subsection{Numerical viscosity calculation}
A more detailed derivation of the numerical viscosity term in equation \eqref{truncationerrorpostcalc} is presented here. Starting from equation \eqref{truncationerrorprecalc}, the summand may be split into partial fractions as
\begin{equation}
    \frac{a}{2(a - b)b} = \frac{1}{2(a - b)} + \frac{1}{2b}
\end{equation}
and when summed from $-\infty$ to $\infty$ this becomes $-1/a$. This can be folded into the other sum, so that the error becomes
\begin{equation}
    E^a = \frac{\I\omega\cnon \sqrt{A_\text{cs}}M^2e^{\I a\omega s}}{2(1 + \sigma)^2}\Bigg(\sum_{\substack{b = a - \sgn(a)a_{\text{max}},\\b \neq a}}^{\sgn(a)a_{\text{max}}}\frac{1}{a - b} + \sum_{\substack{b = a - \sgn(a)a_{\text{max}},\\b \neq 0}}^{\sgn(a)a_{\text{max}}}\frac{1}{b}\Bigg).
\end{equation}
The first sum is split as
\begin{equation}
    \sum_{\substack{b = a - \sgn(a)a_{\text{max}},\\b \neq a}}^{\sgn(a)a_{\text{max}}}\frac{1}{a - b} = \Bigg(\sum_{b = a - \sgn(a)a_{\text{max}}}^{2a - \sgn(a)(a_\text{max} + 1)} + \sum_{\substack{2a - \sgn(a)a_\text{max},\\ b \neq a}}^{\sgn(a)a_{\text{max}}}\Bigg)\frac{1}{a - b},
\end{equation}
with the second term vanishing due to symmetric limits and an odd summand about $b = a$. The second sum is treated similarly
\begin{equation}
    \sum_{\substack{b = a - \sgn(a)a_{\text{max}},\\b \neq 0}}^{\sgn(a)a_{\text{max}}}\frac{1}{b} = \Bigg(\sum_{\substack{b = a - \sgn(a)a_{\text{max}},\\b \neq 0}}^{\sgn(a)a_{\text{max}} - a} + \sum_{\sgn(a)(a_{\text{max}} + 1) - a}^{\sgn(a)a_{\text{max}}}\Bigg)\frac{1}{b},
\end{equation}
and this time the first term vanishes. Both of the remaining sums, with the correct substitution, become
\begin{equation}
    \sum_{b = a - \sgn(a)a_{\text{max}}}^{2a - \sgn(a)(a_\text{max} + 1)}\frac{1}{a - b} = \sum_{\sgn(a)(a_{\text{max}} + 1) - a}^{\sgn(a)a_{\text{max}}}\frac{1}{b} = \sum_{b = 0}^{|a| - 1}\frac{\sgn(a)}{a_\text{max} - b},
\end{equation}
so then the error becomes
\begin{equation}
    E^a = \frac{\I\omega\cnon \sqrt{A_\text{cs}}M^2e^{\I a\omega s}}{(1 + \sigma)^2}\sum_{b = 0}^{|a| - 1}\frac{\sgn(a)}{a_\text{max} - b} = \frac{|a|\omega\cnon MP_0^a}{1 + \sigma}\sum_{b = 0}^{|a| - 1}\frac{1}{a_\text{max} - b}.
\end{equation}
The sum has upper and lower bounds in the form of integrals:
\begin{equation}
    \int_0^{|a| - 1}\frac{\intd x}{a_\text{max} - x} < \sum_{b = 0}^{|a| - 1}\frac{1}{a_\text{max} - b} < \int_0^{|a| - 1}\frac{\intd x}{a_\text{max} - (x + 1)}.
\end{equation}
We choose the lower bound as the approximation to the sum since the upper bound's integrand is singular at its upper limit. Thus, evaluating the integral and including the viscous scale factor $\nu_0$, we arrive at \eqref{truncationerrorpostcalc}.

\subsection{Physical viscosity calculation}
If we instead consider physical viscosity in one dimension, the (single) momentum equation picks up an extra term
\begin{equation}
    -\I a\omega U^a + \frac{\partial P^a}{\partial s} - \left(\frac{4\mu}{3} + \zeta\right)\frac{\partial^2U^a}{\partial s^2} = \frac{\partial Q^a}{\partial s},
\end{equation}
while the mass conservation equation is unchanged. We still wish to use the Riccati method, which requires our two equations to be first-order in $s$; this can be achieved by substituting the $O(M^2)$ expansion of $\partial U^a/\partial s$ from the mass conservation equation here. That substitution results in
\begin{equation}
    \begin{aligned}
        -\I a\omega U^a + \left(1 - \I a\omega\left(\frac{4\mu}{3} + \zeta\right)\right)\frac{\partial P^a}{\partial s} = \frac{\partial Q^a}{\partial s} - \I a\omega\left(\frac{4\mu}{3} + \zeta\right)\frac{\partial}{\partial s}\left(\sum_b\cnon P^{a - b}P^b - Q^a\right).
    \end{aligned}
\end{equation}
The $s$-derivatives of $O(M^2)$ terms may now have linear terms substituted into them. For notational ease, we define a modified frequency for each temporal index, and absorb both viscosities into a single constant 
\begin{align}
    \Omega^a = \frac{\I a\omega}{1 - \I a\omega\Tilde{\mu}}, && \Tilde{\mu} = \frac{4\mu}{3} + \zeta,
\end{align}
so then we have
\begin{equation}
    \frac{\partial P^a}{\partial s} - \Omega^aU^a = \frac{\partial Q^a}{\partial s} -\Tilde{\mu}\cnon\Omega^a\frac{\partial}{\partial s}\sum_bP^{a - b}P^b,
\end{equation}
Defining the one-dimensional admittances in the usual way, we end up with the following two equations for them
\begin{align}
    \frac{\mathrm{d}\mathsf{Y}^a}{\mathrm{d}s} &= -\Omega^a(\mathsf{Y}^a)^2 + \I a\omega, \\
    \frac{\intd\mathcal{Y}^{ab}}{\intd s} &= -\mathcal{Y}^{ab}\bigg(\Omega^a\mat{Y}^a + \Omega^{a - b}\mat{Y}^{a - b} + \Omega^b\mat{Y}^b\bigg)\notag \\
    &\quad-\frac{1}{\sqrt{\uppi}R}\Bigg\{\left(\Omega^{a - b}\mat{Y}^{a - b} + \Omega^b\mat{Y}^b\right)\left(\frac{\mat{Y}^a}{2} + \Tilde{\mu}\cnon\Omega^a\right)\notag \\
    &\quad+ \frac{i\omega}{2}\left(a\gamma + a\mat{Y}^{a - b}\mat{Y}^b - b\mat{Y}^a\mat{Y}^{a - b} - (a - b)\mat{Y}^a\mat{Y}^b\right)\Bigg\}.
\end{align}
Since we are working in one dimension, the admittances must be characteristic, and so we have the solutions
\begin{align}
    \overline{\mat{Y}}^{a\pm} &= \pm\overline{\mat{Y}}^a = \pm\exp\left(-\frac{\I\uppi}{4}\left[\sgn(1 - \I a\omega\Tilde{\mu}) - 1\right]\right)\sqrt{|1 - \I a\omega\Tilde{\mu}|}, \\
    \overline{\mathcal{Y}}^{ab\pm} &= \mp\frac{1}{\sqrt{\uppi}R}\Bigg\{\left(\Omega^{a - b}\overline{\mat{Y}}^{a - b} + \Omega^b\overline{\mat{Y}}^b\right)\left(\frac{\overline{\mat{Y}}^a}{2} \pm \Tilde{\mu}\cnon\Omega^a\right)\notag \\
    &\quad+ \frac{\I\omega}{2}\left(a\gamma + a\overline{\mat{Y}}^{a - b}\overline{\mat{Y}}^b - b\overline{\mat{Y}}^a\overline{\mat{Y}}^{a - b} - (a - b)\overline{\mat{Y}}^a\overline{\mat{Y}}^b\right)\Bigg\}\bigg/\left(\Omega^a\overline{\mat{Y}}^a + \Omega^{a - b}\overline{\mat{Y}}^{a - b} + \Omega^b\overline{\mat{Y}}^b\right).
\end{align}
Assuming $\Tilde{\mu}$ is small, first-order expansions of these quantities may be found, though the nonlinear admittance only appears in $O(M^2)$ terms to begin with, so its second term here will eventually be $O(M^2\mu)$, and thus discarded. Hence, we have
\begin{equation}
    \frac{\intd P_0^a}{\intd s} \approx \I a\omega\left[P_0^a - \frac{\I a\omega\cnon}{2\sqrt{\uppi} R}\sum_bP_0^{a - b}P_0^b\right] -\frac{a^2\omega^2\Tilde{\mu}}{2}P_0^a,
\end{equation}
suggesting that in order to stabilise a truncated sawtooth wave the physical viscosity would need to be around $\Tilde{\mu} \approx 2\cnon M/a_\text{max}\omega$.

\section{Inverse exponential horn in more detail}
\label{inverseexpapp}
\begin{figure}
    \centering
    \includegraphics[]{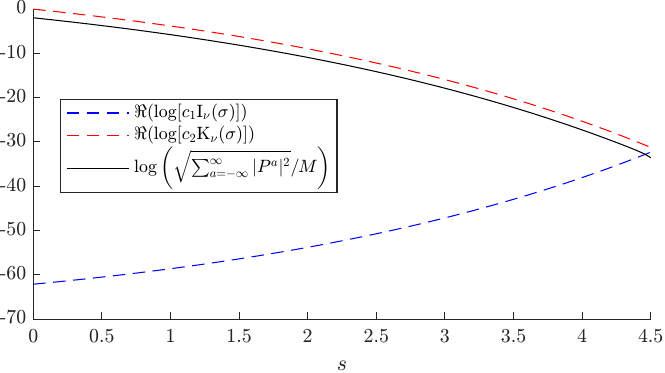}
    \caption{Figure comparing the (real parts of the) logarithmically plotted modified Bessel functions ($\Ib_\nu(\sigma)$ in blue, $\K_\nu(\sigma)$ in red), along an inverse exponential horn of length $4.5X_i$ and width decrease ratio of 4, for a frequency of 0.5/$X_i$. RMS pressure (according to this analytical solution) normalised by Mach number is logarithmically plotted in black.}
    \label{modbesselplotting}
\end{figure}

This appendix contains a more detailed presentation of the analytical solution to the inverse exponential horn's first antisymmetric mode equation, discussed earlier in section \eqref{inverseexpresults}. By use of the substitution $\sigma = \uppi\e^{2|m|s}/2|m|X_\text{i}$, we reduced the problem to the modified Bessel equation
\begin{equation}
    \sigma^2\frac{\intd^2p}{\intd \sigma^2} + \sigma\frac{\intd p}{\intd\sigma} - (\nu^2 + \sigma^2)p = 0,
\end{equation}
where $\nu$ is defined by $\nu^2 = 1 - \omega^2/4m^2$, and takes purely real values or purely imaginary values, depending on whether $\omega$ exceeds $2|m|$. If $\nu$ is not a real integer, then the solution is given by
\begin{equation}
    p(\sigma) = c_1\Ib_\nu(\sigma) + c_2\K_{\nu}(\sigma),
\end{equation}
The velocity was eliminated in the derivation of equation \eqref{antisympressureeq}, but can be retrieved from the fourth row of equation \eqref{2modetruncation}, since we have
\begin{align}
    \frac{\intd p}{\intd s} = \I\omega~u + \frac{X'}{X}p &&\implies &&u(\sigma) = \frac{2|m|}{\I\omega}\left[\sigma p'(\sigma) + p(\sigma)\right],
\end{align}
so we have
\begin{equation}
    u(\sigma) = \frac{2|m|}{\I\omega}\left[c_1\Tilde{\Ib}_\nu(\sigma) + c_2\Tilde{\K}_{\nu}(\sigma)\right],\quad \text{where}\quad \Tilde{\Ib}_\nu(\sigma) := \sigma\Ib_\nu'(\sigma) + \Ib_\nu(\sigma).
\end{equation}
and $\Tilde{\K}_\nu(\sigma)$ is defined similarly. We then have an admittance given by
\begin{equation}
    Y(\sigma) = \frac{u(\sigma)}{p(\sigma)} = \frac{2|m|}{\I\omega}\frac{c_1\Tilde{\Ib}_\nu(\sigma) + c_2\Tilde{\K}_{\nu}(\sigma)}{c_1\Ib_\nu(\sigma) + c_2\K_{\nu}(\sigma)}.
\end{equation}
Applying the boundary condition $Y(\sigma_\text{o}) = \overline{Y}_\text{o}$, we get
\begin{equation}
    Y(\sigma) = \frac{2|m|}{\I\omega}\frac{\left[\Tilde{\K}_{\nu}(\sigma_\text{o}) - \frac{\I\omega \overline{Y}_\text{o}}{2|m|}\K_{\nu}(\sigma_\text{o})\right]\Tilde{\Ib}_\nu(\sigma) - \left[\Tilde{\Ib}_\nu(\sigma_\text{o}) - \frac{\I\omega \overline{Y}_\text{o}}{2|m|}\Ib_{\nu}(\sigma_\text{o})\right]\Tilde{\K}_{\nu}(\sigma)}{\left[\Tilde{\K}_{\nu}(\sigma_\text{o}) - \frac{\I\omega \overline{Y}_\text{o}}{2|m|}\K_{\nu}(\sigma_\text{o})\right]\Ib_\nu(\sigma) - \left[\Tilde{\Ib}_\nu(\sigma_\text{o}) - \frac{\I\omega \overline{Y}_\text{o}}{2|m|}\Ib_{\nu}(\sigma_\text{o})\right]\K_{\nu}(\sigma)},
\end{equation}
where the characteristic admittance $\overline{Y}_\text{o}$ is given by plugging the definition of $X_\text{o}$ into equation \eqref{straightcharadm}
\begin{equation}
    \overline{Y}_\text{o} = \overline{Y}(\sigma_\text{o}) = \exp\left\{-\frac{\I\uppi}{4}\left[\sgn\left(1 - \frac{4m^2\sigma_\text{o}^2}{\omega^2}\right) -1 \right]\right\}\sqrt{\left|1 - \frac{4m^2\sigma_\text{o}^2}{\omega^2}\right|}.
\end{equation}
Specifying the total pressure at the inlet, we match our overall remaining constant with equation \eqref{pistonprojection} to get
\begin{equation}
    p(\sigma) = \frac{M\sqrt{X_\text{i}}}{2\I}\frac{\left[\Tilde{\K}_{\nu}(\sigma_\text{o}) - \frac{\I\omega \overline{Y}_\text{o}}{2|m|}\K_{\nu}(\sigma_\text{o})\right]\Ib_\nu(\sigma) - \left[\Tilde{\Ib}_\nu(\sigma_\text{o}) - \frac{\I\omega \overline{Y}_\text{o}}{2|m|}\Ib_{\nu}(\sigma_\text{o})\right]\K_{\nu}(\sigma)}{\left[\Tilde{\K}_{\nu}(\sigma_\text{o}) - \frac{\I\omega \overline{Y}_\text{o}}{2|m|}\K_{\nu}(\sigma_\text{o})\right]\Ib_\nu(\sigma_\text{i}) - \left[\Tilde{\Ib}_\nu(\sigma_\text{o}) - \frac{\I\omega \overline{Y}_\text{o}}{2|m|}\Ib_{\nu}(\sigma_\text{o})\right]\K_{\nu}(\sigma_\text{i})}.
\end{equation}
The boundary condition at the duct outlet $\sigma_\text{o}$ can be rewritten as a vector orthogonality condition in two dimensions, i.e.
\begin{equation}
    \begin{pmatrix}
        c_1 \\ c_2
    \end{pmatrix}\cdot \left[\sigma_\text{o}\begin{pmatrix}
        \Ib_\nu'(\sigma_\text{o}) \\ \K_\nu'(\sigma_\text{o})
    \end{pmatrix} + \left(1 - \frac{\I\omega\overline{Y}_\text{o}}{2|m|}\right)\begin{pmatrix}
        \Ib_\nu(\sigma_\text{o}) \\ \K_\nu(\sigma_\text{o})
    \end{pmatrix} \right] = 0.
\end{equation}
$\Ib_\nu(\sigma)$ and $\Ib_\nu'(\sigma)$ are exponentially-growing functions, while $\K_\nu(\sigma)$ and $\K_\nu'(\sigma)$ exponentially decay. Unscaled, these functions will therefore be of very different orders of magnitude at the outlet. This relation, however, effectively sets the vector $(c_1,c_2)$ to be `asymptotically orthogonal' to the Bessel function vector, meaning that the two terms in both the velocity and the pressure are instead of comparable order at the outlet. If we then move backwards from the outlet, the $\K$ functions will grow rapidly while the $\Ib$ functions decay, so for most of the duct the $\K$ functions will dominate. For higher frequencies, $\nu$ can become imaginary, causing $\K$ to oscillate towards the inlet: these oscillations will pass through zero, resulting in pressure nodes, which cause singularities in the admittance We note that for the inverse exponential horn, the lowest-mode admittance singularities correspond approximately to roots of $\K_{\I\sqrt{\omega^2/4m^2 - 1}}(\uppi\e^{2|m|s}/2|m|X_\text{i})$. Figure \ref{modbesselplotting} shows these relationships in more detail, for a frequency low enough that oscillations do not yet occur.

Another feature of this geometry (or any geometry with width variation) is that \textit{turning points} can occur, i.e. points where a mode can go from being cut-on to cut-off. These are significant because they correspond mathematically to the characteristic admittance (for that particular mode, which in a straight duct is a scalar) passing from the real axis to the imaginary axis, and therefore necessarily passing through zero. Since the splitting operators then involve the inverse of a matrix with a zero eigenvalue, this results in singularities in the forward and backward-going pressures at this point. By inverting equation \eqref{cutofffreqdef} for $\omega = \overline{\omega}_\alpha^a, \lambda_\alpha = \alpha\uppi$, and $X = X_\text{i}\e^{2|m|s}$, we note that the location of the turning point for mode $(a,\alpha)$ is
\begin{equation}
    s_\text{tp} = \frac{1}{2|m|}\log\left(\frac{a\omega X_\text{i}}{\alpha\uppi}\right).
\end{equation}

\bibliography{biblio.bib}

\end{document}